\documentclass[11pt,english,onecolumn]{IEEEtran}
\usepackage[T1]{fontenc}
\usepackage[latin9]{inputenc}
\usepackage{units}
\usepackage{amsthm}
\usepackage{amsmath}
\usepackage{amssymb}
\usepackage{mathdots}
\usepackage{graphicx}
\usepackage{setspace}
\PassOptionsToPackage{version=3}{mhchem}
\usepackage{mhchem}
\usepackage{esint}
\PassOptionsToPackage{normalem}{ulem}
\usepackage{ulem}
\makeatletter
\theoremstyle{plain}
\newtheorem{thm}{\protect\theoremname}
\theoremstyle{remark}
\newtheorem{rem}[thm]{\protect\remarkname}
\theoremstyle{plain}
\newtheorem{lem}[thm]{\protect\lemmaname}
\theoremstyle{plain}
\newtheorem{prop}[thm]{\protect\propositionname}
\theoremstyle{plain}
\newtheorem{fact}[thm]{\protect\factname}
\theoremstyle{plain}
\newtheorem{cor}[thm]{\protect\corollaryname}

\usepackage{hyperref}
\usepackage{breqn}
  
\DeclareMathOperator*{\argmax}{arg\,max} \DeclareMathOperator*{\argmin}{arg\,min}

\allowdisplaybreaks

\global\long\def\s[#1]{\textnormal{\scriptsize #1}}
\global\long\def\st[#1]{\textnormal{\tiny #1}}

\global\long\def\P{\mathbb{P}}
\global\long\def\E{\mathbb{E}}
\global\long\def\I{\mathbb{I}}

\global\long\def\teq{\triangleq}
\global\long\def\dfn{\teq}

\global\long\def\trre[#1,#2]{\overset{{\scriptstyle (#2)}}{#1}} % transition explained with reason
\global\long\def\bin[#1,#2]{\mathbb{B}[#1;#2]} %Binary representation of an element in a set
\global\long\def\dec[#1]{\mathbb{D}_{#1}} %Decimal representation of bits

\author{
\authorblockN{Nir Weinberger and Neri Merhav}

\authorblockA{Dept. of Electrical Engineering\\
   	    Technion - Israel Institute of Technology\\
Technion City, Haifa 3200004, Israel
} \\
\authorblockA{\{nirwein@campus, merhav@ee\}.technion.ac.il}
}
\makeatother

\usepackage{babel}
\providecommand{\corollaryname}{Corollary}
\providecommand{\factname}{Fact}
\providecommand{\lemmaname}{Lemma}
\providecommand{\propositionname}{Proposition}
\providecommand{\remarkname}{Remark}
\providecommand{\theoremname}{Theorem}

\begin{document}

\title{Channel Detection in Coded Communication}
\maketitle
\begin{abstract}
We consider the problem of block-coded communication, where in each
block, the channel law belongs to one of two disjoint sets. The decoder
is aimed to decode only messages that have undergone a channel from
one of the sets, and thus has to detect the set which contains the
prevailing channel. We begin with the simplified case where each of
the sets is a singleton. For any given code, we derive the optimum
detection/decoding rule in the sense of the best trade-off among the
probabilities of decoding error, false alarm, and misdetection, and
also introduce sub-optimal detection/decoding rules which are simpler
to implement. Then, various achievable bounds on the error exponents
are derived, including the exact single-letter characterization of
the random coding exponents for the optimal detector/decoder. We then
extend the random coding analysis to general sets of channels, and
show that there exists a universal detector/decoder which performs
asymptotically as well as the optimal detector/decoder, when tuned
to detect a channel from a specific pair of channels. The case of
a pair of binary symmetric channels is discussed in detail.\end{abstract}
\renewcommand\[{\begin{equation}}
\renewcommand\]{\end{equation}}
\thispagestyle{empty}

\section{Introduction\label{sec:Introduction}}

Consider communicating over a channel, for which the prevailing channel
law $P_{Y|X}$ ($X$ and $Y$ being the channel input and output,
respectively) is supposed to belong to a family of channels ${\cal W}$.
For example, ${\cal W}$ could be a singleton ${\cal W}=\{W\}$, or
some ball centered at $W$ with respect to (w.r.t.) a given metric
(say, total variation). This ball represents some uncertainty regarding
the channel, which may result, e.g., from estimation errors. The receiver
would also like to examine an alternative hypothesis, in which the
channel $P_{Y|X}$ is not in ${\cal W}$, and belongs to a different
set ${\cal V}$, disjoint from ${\cal W}$. Such a detection procedure
will be useful, for example, in the following cases:
\begin{enumerate}
\item \emph{Time-varying channels}: In many protocols, communication begins
with a channel estimation phase, and later on, at the data transmission
phase, the channel characteristics are tracked using adaptive algorithms
\cite[Chapters 8 and 9]{dig_com_lee_messer}. However, it is common,
that apart from its slow variation, the channel may occasionally also
change abruptly, for some reason. Then, the tracking mechanism totally
fails, and it is necessary to initialize communication again with
a channel estimation phase. The detection of this event is usually
performed at high communication layers, e.g., by inspecting the output
data bits of the decoder, and verifying their correctness in some
way. This procedure could be aided, or even replaced, by identifying
a distinct change in the channel as part of the decoding. Note that
this problem is a block-wise version of the change-point detection
problem from sequential analysis \cite{lorden1971procedures,generalized_change_point}
(see, also \cite{chandar2015quickest} and referenced therein for
a recent related work).
\item \emph{Arbitrarily varying channels in blocks}:\emph{ }In the same
spirit, consider a bursty communication system, where within each
burst, the underlying channel may belong to either one of two sets,
resulting from two very distinctive physical conditions. For example,
a wireless communication signal may occasionally be blocked by some
large obstacle which results in low channel gain compared to the case
of free-space propagation, or it may experience strong interference
from other users \cite{Tse_book}. The receiver should then decide
if the current channel enables reliable decoding.  
\item \emph{Secure decoding}: In channels that are vulnerable to intrusions,
the receiver would like to verify that an authorized transmitter has
sent the message. In these cases, the channel behavior could serve
as a proxy for the identity of the transmitter. For example, a channel
with a significantly lower or larger signal-to-noise ratio (SNR) than
predicted by the geographical distance between the transmitter and
receiver, could indicate a possible attempt to intrude the system.
The importance of identifying such cases is obvious, e.g., if the
messages are used to control a sensitive equipment at the receiver
side.
\item \emph{Multiple access channels with no collisions}: Consider a slotted
sparse multiple access channel, for which two transmitters are sending
messages to a common receiver only in a very small portion of the
available slots%
\footnote{For simplicity, assume that each codeword occupies exactly a single
slot.%
}, via \emph{different} channels. Thus, it may be assumed that at each
slot, at most one transmitter is active. The receiver would like to
identify the sender with high reliability. As might be dictated by
practical considerations, the same codebook is used by both transmitters
and the receiver identifies the transmitter via a short header, which
is common to all codewords of the same transmitter.%
\footnote{Also, if senders simply use different codebooks, then the detection
performance would be related to the error probability of the codebook
which is comprised from joining the two codebooks. The random coding
exponents for the case that the codebook of each transmitter is chosen
independently from the codebook of the other user can be obtained
by slightly modifying the results of \cite{bin_index}.%
} The receiver usually identifies the transmitter based on the received
header only. Of course, this header is an undesired overhead, and
so it is important to maximize the detection performance for any given
header. To this end, the receiver can also use the codeword sent,
and identify the transmitter using the different channel.
\end{enumerate}
Thus, beyond the ordinary task of decoding the message, the receiver
would also like to detect the event $P_{Y|X}\in{\cal V}$, or, in
other words, perform \emph{hypothesis testing} between the null hypothesis
$P_{Y|X}\in{\cal W}$ and the alternative hypothesis $P_{Y|X}\in{\cal V}$.
For example, if the channel quality is gauged by a single parameter,
say, the crossover probability of a binary symmetric channel (BSC),
or the SNR of an additive white Gaussian noise channel (AWGN), then
${\cal W}$ and ${\cal V}$ could be two disjoint intervals of this
parameter. 

This problem of joint detection/decoding belongs to a larger class
of hypothesis testing problems, in which after performing the test,
another task should be performed, depending on the chosen hypothesis.
For example, in \cite{Moustakides11,Moustakides12}, the problem of
joint hypothesis testing and Bayesian estimation was considered, and
in \cite{Joint_detection_source_coding} the subsequent task is lossless
source coding. A common theme for all the problems in this class,
is that separately optimizing the detection and the task is sub-optimal,
and so, joint optimization is beneficial.

In a more recent work \cite{CoN}, we have studied the related problem
of joint detection and decoding for sparse communication \cite{Wang10},
which is motivated by strongly asynchronous channels \cite{TCW08,TCW09}.
In these channels the transmitter is either completely silent or transmits
a codeword from a given codebook. The task of the detector/decoder
is to decide whether transmission has taken place, and if so, to decode
the message. Three figures of merit were defined in order to judge
performance: (i) the probability of \textit{false alarm }\textit{\emph{(FA)}}
- i.e., deciding that a message has been sent when actually, the transmitter
was silent and the channel output was pure noise, (ii) the probability
of \textit{misdetection} (MD) - that is, deciding that the transmitter
was silent when it actually transmitted some message, and (iii) the
probability of \textit{inclusive error} (IE) - namely, not deciding
on the correct message sent, namely, either misdetection of erroneous
decoding. We have then found the optimum detector/decoder that minimizes
the IE probability subject to given constraints on the FA and the
MD probabilities for a given codebook, and also provided single-letter
expressions for the exact random coding exponents. While this is a
\emph{joint} detector/decoder, we have also observed that an \emph{asymptotic
separation principle} holds, in the following sense: A detector/decoder
which achieves the optimal exponents may be comprised of an optimal
detector in the Neyman-Pearson sense for the FA and MD probabilities,
followed by ordinary maximum likelihood (ML) decoding.

In this paper, we study the problem of joint channel detection between
two disjoint sets of memoryless channels ${\cal W},{\cal V}$, and
decoding. We mainly consider discrete alphabets, but some of the results
are easily adapted to continuous alphabets. We begin by considering
the case of simple hypotheses, namely ${\cal W}=\{W\}$ and ${\cal V}=\{V\}$.
As in \cite{CoN}, we measure the performance of the detector/decoder
by its FA, MD and IE probabilities, derive the optimal detector/decoder,
and show that here too, an asymptotic separation principle holds.
Due to the numerical instability of the optimal detector, we also
propose two simplified detectors, each of which suits better a different
rate range. Then, we discuss a plethora of lower bounds on the achievable
exponents: For the optimal detector/decoder, we derive single-letter
expressions for the \emph{exact} random coding exponents, as well
as expurgated bounds which improve the bounds at low rates. The exact
random coding exponents are also derived for the simplified detectors/decoders.
In addition, we also derive Gallager/Forney-style random coding and
expurgated bounds, which are simpler to compute, and can be directly
adapted to continuous channels. However, as we show in a numerical
example, the Gallager/Forney-style exponents may be strictly loose
when compared to the exact exponents, even in simple cases. Thus,
using the refined analysis technique which is based on type class
enumeration (see, e.g., \cite{Merhav09,SBM11} and references therein)
and provides the \textit{exact} random coding exponents is beneficial
in this case. Afterwards, we discuss a generalization to composite
hypotheses, i.e., ${\cal W},{\cal V}$ that are not singletons. Finally,
we discuss in detail the archetype example for which $W,V$ are a
pair BSCs. 

The detection problem addressed in \cite{CoN} can be seen to be a
special case of the problem studied here, for which the the output
of the channel $V$ is completely independent of its input, and plays
the role of noise. It turns out that the optimal detector/decoder
and its properties for the problem studied here are straightforward
generalizations of \cite{CoN}, and thus we will discuss them rather
briefly and only cite the relevant results from \cite{CoN}. However,
there is a substantial difference in the analysis of the random coding
detection exponents in \cite{CoN}, compared to the analysis here.
In \cite{CoN}, the discrimination is between the codebook and noise.
The detector compares a likelihood which depends on the codebook with
a likelihood function that depends on the noise. So, when analyzing
the performance of random coding, the random choice of codebook only
affects the distribution of the likelihood of the `codebook hypothesis'.
By contrast, here, since we would like to detect the channel, the
random choice of codebook affects the likelihood of \emph{both} hypotheses,
and consequently, the two hypotheses may be highly dependent. One
consequence of this situation, is that to derive the random coding
exponents, it is required to analyze the joint distribution of type
class enumerators (cf. Subsection \ref{sub:A-Tight-Random}), and
not just rely on their marginal distributions. The expurgated and
Gallager/Forney-style exponents, as well as the simplified detectors/decoders
are studied here for the first time.

The outline of the rest of the paper is as follows. In Section \ref{sec:Notation-Conventions},
we establish notation conventions and provide some preliminaries,
and in Section \ref{sec:Problem-Formulation}, we formulate the problem
of detecting between two channels. In Section \ref{sec:Joint-Detector/Decoders},
we derive the optimum detector/decoder and discuss some of its properties,
and also introduce sub-optimal detectors/decoders. In Section \ref{sec:Achievable-Error-Exponents },
we present our main results regarding various single-letter achievable
exponents. In Section \ref{sec:Composite-Detection}, we discuss the
problem of detection of composite hypotheses. Finally, in Section
\ref{sec:Examples}, we exemplify the results for a pair of BSCs.
We defer most of the proofs to the appendices.

\section{Notation Conventions and Preliminaries\label{sec:Notation-Conventions}}

Throughout the paper, random variables will be denoted by capital
letters, specific values they may take will be denoted by the corresponding
lower case letters, and their alphabets, similarly as other sets,
will be denoted by calligraphic letters. Random vectors and their
realizations will be denoted, respectively, by capital letters and
the corresponding lower case letters, both in the bold face font.
Their alphabets will be superscripted by their dimensions. For example,
the random vector $\mathbf{X}=(X_{1},\ldots,X_{n})$, ($n$ - positive
integer) may take a specific vector value $\mathbf{x}=(x_{1},\ldots,x_{n})$
in ${\cal X}^{n}$, the $n$-th order Cartesian power of ${\cal X}$,
which is the alphabet of each component of this vector.

A joint distribution of a pair of random variables $(X,Y)$ on ${\cal X}\times{\cal Y}$,
the Cartesian product alphabet of ${\cal X}$ and ${\cal Y}$, will
be denoted by $Q_{XY}$ and similar forms, e.g. $\tilde{Q}_{XY}$.
Since usually $Q_{XY}$ will represent a joint distribution of $X$
and $Y$, we will abbreviate this notation by omitting the subscript
$XY$, and denote, e.g, $Q_{XY}$ by $Q$. The $X$-marginal (\textbf{$Y$}-marginal),
induced by $Q$ will be denoted by $Q_{X}$ (respectively, $Q_{Y}$),
and the conditional distributions will be denoted by $Q_{Y|X}$ and
$Q_{X|Y}$. In accordance with this notation, the joint distribution
induced by $Q_{X}$ and $Q_{Y|X}$ will be denoted by $Q=Q_{X}\times Q_{Y|X}$.

For a given vector $\mathbf{x}$, let $\hat{Q}_{\mathbf{x}}$ denote
the empirical distribution, that is, the vector $\{\hat{Q}_{\mathbf{x}}(x),~x\in{\cal X}\}$,
where $\hat{Q}_{\mathbf{x}}(x)$ is the relative frequency of the
letter $x$ in the vector $\mathbf{x}$. Let ${\cal T}(P_{X})$ denote
the type class%
\footnote{The blocklength will not be displayed since it will be understood
from the context.%
} associated with $P_{X}$, that is, the set of all sequences $\{\mathbf{x}\}$
for which $\hat{Q}_{\mathbf{x}}=P_{X}$. Similarly, for a pair of
vectors $(\mathbf{x},\mathbf{y})$, the empirical joint distribution
will be denoted by $\hat{Q}_{\mathbf{xy}}$. 

The mutual information of a joint distribution $Q$ will be denoted
by $I(Q)$, where $Q$ may also be an empirical joint distribution.
The information divergence between $Q_{X}$ and $P_{X}$ will be denoted
by $D(Q_{X}\|P_{X})$, and the conditional information divergence
between the empirical conditional distribution $Q_{Y|X}$ and $P_{Y|X}$,
averaged over $Q_{X}$, will be denoted by $D(Q_{Y|X}\|P_{Y|X}|Q_{X})$.
Here too, the distributions may be empirical. 

The probability of an event ${\cal A}$ will be denoted by $\P\{{\cal A}\}$,
and the expectation operator will be denoted by $\E\{\cdot\}$. Whenever
there is room for ambiguity, the underlying probability distribution
$Q$ will appear as a subscript, i.e., $\P_{Q}\{\cdot\}$ and $\E_{Q}\{\cdot\}$.
The indicator function will be denoted by $\I\{\cdot\}$. Sets will
normally be denoted by calligraphic letters.  The complement of a
set ${\cal A}$ will be denoted by $\overline{{\cal A}}$. Logarithms
and exponents will be understood to be taken to the natural base.
The notation $[t]_{+}$ will stand for $\max\{t,0\}$. We adopt the
standard convention that when a minimization (respectively, maximization)
problem is performed on an empty set the result is $\infty$ (respectively,
$-\infty$). 

For two positive sequences, $\{a_{n}\}$ and $\{b_{n}\}$, the notation
$a_{n}\doteq b_{n}$ will mean asymptotic equivalence in the exponential
scale, that is, $\lim_{n\to\infty}\frac{1}{n}\log(\frac{a_{n}}{b_{n}})=0$,
and similar standard notations $\dot{\leq}$ and $\dot{\geq}$ will
also be used. When $a_{n}$ is a sequence of conditional probabilities,
i.e, $a_{n}=\P\left({\cal A}_{n}|{\cal B}_{n}\right)$ for some pair
of sequence of events $\{{\cal A}_{n}\}_{n=1}^{\infty}$ and $\{{\cal B}_{n}\}_{n=1}^{\infty}$,
the notation $\P({\cal A}_{n}|{\cal B}_{n})\doteq b_{n}$ will mean
\[
\lim_{l\to\infty}\frac{1}{n_{l}}\log\left(\frac{a_{n_{l}}}{b_{n_{l}}}\right)=0,
\]
where $\{n_{l}\}_{l=1}^{\infty}$ is the sequence of blocklengths
such that $\P({\cal B}_{n_{l}})>0$. We shall use the notation $a_{n}\doteq e^{-n\infty}$
when $a_{n}$ decays super-exponentially to zero.

Throughout the sequel, we will make a frequent use of the fact that
$\sum_{i=1}^{k_{n}}a_{n}(i)\doteq\max_{1\le i\le k_{n}}a_{n}(i)$
as long as $\{a_{n}(i)\}$ are positive and $k_{n}\doteq1$. Accordingly,
for $k_{n}$ sequences of positive random variables $\{A_{n}(i)\}$,
all defined on a common probability space, and a deterministic sequence
$b_{n}$, 
\begin{eqnarray}
\P\left\{ \sum_{i=1}^{k_{n}}A_{n}(i)\ge b_{n}\right\}  & \doteq & \P\left\{ \max_{1\le i\le k_{n}}A_{n}(i)\ge b_{n}\right\} \\
 & = & \P\bigcup_{i=1}^{k_{n}}\left\{ A_{n}(i)\ge b_{n}\right\} \\
 & \doteq & \sum_{i=1}^{k_{n}}\P\left\{ A_{n}(i)\ge b_{n}\right\} \\
 & \doteq & \max_{1\le i\le k_{n}}\P\left\{ A_{n}(i)\ge b_{n}\right\} ,\label{eq: union rule}
\end{eqnarray}
provided that $b_{n}'\doteq b_{n}$ implies $\P\{A_{n}(i)\ge b_{n}'\}\doteq\P\{A_{n}(i)\ge b_{n}\}$.%
\footnote{Consider the case where $b_{n}\doteq e^{bn}$ ($b$ being a constant,
independent of $n$) and the exponent of $\P\{A_{n}(i)\ge e^{bn}\}$
is a continuous function of $b$.%
} In simple words, summations and maximizations are equivalent and
can be both ``pulled out outside'' $\P\{\cdot\}$ without changing
the exponential order, as long as $k_{n}\doteq1$. The equalities
in \eqref{eq: union rule} will be termed henceforth `the \emph{union
rule}' (UR). By the same token, 
\begin{eqnarray}
\P\left\{ \sum_{i=1}^{k_{n}}A_{n}(i)\le b_{n}\right\}  & \doteq & \P\left\{ \max_{1\le i\le k_{n}}A_{n}(i)\le b_{n}\right\} \\
 & = & \P\bigcap_{i=1}^{k_{n}}\{A_{n}(i)\le b_{n}\},\label{eq: intersection rule}
\end{eqnarray}
and these equalities will be termed henceforth `the \emph{intersection
rule}' (IR). 

The natural candidate for $k_{n}$ is the number of joint types possible
for a given block length $n$, and this fact, along with all other
rules of the \emph{method of types }\cite{csiszar2011information}\emph{
}will be used extensively henceforth, without explicit reference.

\section{Problem Formulation\label{sec:Problem-Formulation}}

Consider a discrete memoryless channel (DMC), characterized by a finite
input alphabet ${\cal X}$, a finite output alphabet ${\cal Y}$,
and a given matrix of single-letter transition probabilities $\{P_{Y|X}(y|x)\}_{x\in{\cal X},y\in{\cal Y}}$.
Let ${\cal C}_{n}=\{\mathbf{x}_{1},\mathbf{x}_{2}\ldots,\mathbf{x}_{M}\}\subset{\cal X}^{n}$,
denote a codebook for blocklength $n$ and rate $R$, for which the
transmitted codeword is chosen with a uniform probability distribution
over the $M=\left\lceil e^{nR}\right\rceil $ codewords. The conditional
distribution $P_{Y|X}$ may either satisfy $P_{Y|X}=W$ (the \emph{null
hypothesis}), or $P_{Y|X}=V$ (the \emph{alternative hypothesis}).
It is required to design a detector/decoder which is oriented to decode
messages only arriving via the channel $W$. Formally, such a detector/decoder
$\phi$ is a partition of ${\cal Y}^{n}$ into $M+1$ regions, denoted
by $\{{\cal R}_{m}\}_{m=0}^{M}$.%
\footnote{The decoder $\phi$ naturally depends on the blocklength via the codebook
${\cal C}_{n}$, but this will be omitted.%
} If $\mathbf{y}\in{\cal R}_{m}$ for some $1\leq m\leq M$ then the
$m$-th message is decoded. If $\mathbf{y}\in{\cal R}_{0}$ (the \emph{rejection
region}) then the channel $V$ is identified, and no decoding takes
place. 

For a codebook ${\cal C}_{n}$ and a given detector/decoder $\phi$,
the probability of \emph{false alarm} (FA) is given by

\begin{equation}
P_{\st[FA]}({\cal C}_{n},\phi)\dfn\frac{1}{M}\sum_{m=1}^{M}W({\cal R}_{0}|\mathbf{x}_{m}),\label{eq: FA probability}
\end{equation}
the probability of \emph{misdetection} (MD) is given by 
\begin{equation}
P_{\st[MD]}({\cal C}_{n},\phi)\dfn\frac{1}{M}\sum_{m=1}^{M}V(\overline{{\cal R}_{0}}|\mathbf{x}_{m}),\label{eq: MD probability}
\end{equation}
and the probability of \emph{inclusive error} (IE) is defined as 
\begin{equation}
P_{\st[IE]}({\cal C}_{n},\phi)\dfn\frac{1}{M}\sum_{m=1}^{M}W\left(\overline{{\cal R}_{m}}|\mathbf{x}_{m}\right).\label{eq: IE probability}
\end{equation}
Thus, the IE event is the total error event, namely, when the correct
codeword $\mathbf{x}_{m}$ is not decoded either because of a FA or
an ordinary erroneous decoding.%
\footnote{This definition is conventional in related problems. For example,
in Forney's error/erasure setting\emph{ }\cite{Forney68}, one of
the events defined and analyzed is the total error event, which is
comprised of a union of an undetected error event and an erasure event.%
} The probability of decoding to an erroneous\emph{ }codeword, excluding
the rejection region, is termed the \emph{exclusive error} (EE) probability
and is defined as 
\begin{equation}
P_{\st[EE]}({\cal C}_{n},\phi)\dfn P_{\st[IE]}({\cal C}_{n},\phi)-P_{\st[FA]}({\cal C}_{n},\phi).
\end{equation}
When obvious from context, we will omit the notation of the dependence
of these probabilities on ${\cal C}_{n}$ and $\phi$.

For a given code ${\cal C}_{n}$, we are interested in achievable
trade-offs between $P_{\st[FA]}$, $P_{\st[MD]}$ and $P_{\st[IE]}$.
Consider the following problem: 
\begin{alignat}{1}
 & \mbox{minimize}~~~P_{\st[IE]}\nonumber \\
 & \mbox{subject to}~~P_{\st[FA]}\le\epsilon_{\st[FA]}\nonumber \\
 & ~~~~~~~~~~~~~P_{\st[MD]}\le\epsilon_{\st[MD]}\label{eq: IE- FA and MD tradeoff}
\end{alignat}
where $\epsilon_{\st[FA]}$ and $\epsilon_{\st[MD]}$ are given prescribed
quantities, and it is assumed that these two constraints are not contradictory.
Indeed, there is some tension between $P_{\st[MD]}$ and $P_{\st[FA]}$
as they are related via the Neyman-Pearson lemma \cite[Theorem 11.7.1]{Cover:2006:EIT:1146355}.
For a given $\epsilon_{\st[FA]}$, the minimum achievable $P_{\st[MD]}$
is positive, in general. It is assumed then that the prescribed value
of $\epsilon_{\st[MD]}$ is not smaller than this minimum. In the
problem under consideration, it makes sense to relax the tension between
the two constraints to a certain extent, in order to allow some freedom
to minimize $P_{\st[IE]}$ under these constraints. While this is
true for any \emph{finite} blocklength, as we shall see (Proposition
\ref{prop:Full tension decoder is asym optimal}), an asymptotic separation
principle holds, and the optimal detector in terms of exponents has
full tension between the FA and MD exponents. The optimal detector/decoder
for the problem \eqref{eq: IE- FA and MD tradeoff} will be denoted
by $\phi^{*}$.
\begin{rem}
Naturally, one can use the detector/decoder $\phi^{*}$ for messages
sent via $V$. The detection performance for this detector/decoder
would simply be obtained by exchanging the meaning of FA with MD.
\end{rem}
Our goal is to find the optimum detector/decoder for the problem \eqref{eq: IE- FA and MD tradeoff},
and then analyze the achievable exponents associated with the resulting
error probabilities.

\section{Joint Detectors/Decoders\label{sec:Joint-Detector/Decoders}}

In this section, we discuss the optimum detector/decoder for the problem
\eqref{eq: IE- FA and MD tradeoff}, and some of its properties. We
will also derive an asymptotically optimal version, and discuss simplified
decoders, whose performance is close to optimal in some regimes.

\subsection{The Optimum Detector/Decoder\label{sub:The-Optimum-Detector/Decoder}}

Let $a,b\in\mathbb{R}$, and define the detector/decoder $\phi^{*}=\{{\cal R}_{m}^{*}\}_{m=0}^{M}$,
where: 
\begin{equation}
{\cal R}_{0}^{*}\dfn\left\{ \mathbf{y}:~a\cdot\sum_{m=1}^{M}W(\mathbf{y}|\mathbf{x}_{m})+\max_{m}W(\mathbf{y}|\mathbf{x}_{m})\leq b\cdot\sum_{m=1}^{M}V(\mathbf{y}|\mathbf{x}_{m})\right\} ,\label{eq: R0 optimal}
\end{equation}
and 
\begin{equation}
{\cal R}_{m}^{*}\dfn\overline{{\cal R}_{0}^{*}}\bigcap\left\{ \mathbf{y}:\:\max_{m}W(\mathbf{y}|\mathbf{x}_{m})\geq\max_{k\neq m}W(\mathbf{y}|\mathbf{x}_{k})\right\} ,\label{Rm optimal}
\end{equation}
where ties are broken arbitrarily. 
\begin{lem}
\label{lem: Optimal detector/decoder}Let a codebook ${\cal C}_{n}$
be given, let $\phi^{*}$ be as above, and let $\phi$ be any other
partition of ${\cal Y}^{n}$ into $M+1$ regions. If $P_{\st[FA]}({\cal C}_{n},\phi)\leq P_{\st[FA]}({\cal C}_{n},\phi^{*})$
and $P_{\st[MD]}({\cal C}_{n},\phi)\leq P_{\st[MD]}({\cal C}_{n},\phi^{*})$
then $P_{\st[IE]}({\cal C}_{n},\phi)\geq P_{\st[IE]}({\cal C}_{n},\phi^{*})$.\end{lem}
\begin{IEEEproof}
\noindent The proof is almost identical to the proof of \cite[Lemma 1]{CoN}
and thus omitted. 
\end{IEEEproof}
Note that this detector/decoder is optimal (in the Neyman-Pearson
sense) for any \emph{given }blocklength $n$ and codebook ${\cal C}_{n}$.
Thus, upon a suitable choice of the coefficients $a$ and $b$, its
solves the problem \eqref{eq: IE- FA and MD tradeoff} \emph{exactly}.
As common, to assess the achievable performance, we resort to large
blocklength analysis of error exponents. For a given sequence of codes
${\cal C}\teq\{{\cal C}_{n}\}_{n=1}^{\infty}$ and a detector/decoder
$\phi$, the FA exponent is defined as
\begin{equation}
E_{\st[FA]}\left({\cal C},\phi\right)\dfn\liminf_{n\to\infty}-\frac{1}{n}\log P_{\st[FA]}\left({\cal C}_{n},\phi\right),\label{eq: FA exponent definition given codebook}
\end{equation}
and the MD exponent $E_{\st[MD]}\left({\cal C},\phi\right)$ and the
IE exponent $E_{\st[IE]}\left({\cal C},\phi\right)$ are defined similarly.
The asymptotic version of \eqref{eq: IE- FA and MD tradeoff} is then
stated as finding the detector/decoder which achieves the largest
$E_{\st[IE]}$ under constraints on $E_{\st[FA]}$ and $E_{\st[MD]}$.
To affect these error exponents, the coefficients $a,b$ in \eqref{eq: R0 optimal}
need to exponentially increase/decrease as a functions of $n$. Denoting
$a\dfn e^{n\alpha}$ and $b\dfn e^{n\beta}$, the rejection region
of Lemma \ref{lem: Optimal detector/decoder} becomes 
\begin{equation}
{\cal R}_{0}^{*}=\left\{ \mathbf{y}:~e^{n\alpha}\cdot\sum_{m=1}^{M}W(\mathbf{y}|\mathbf{x}_{m})+\max_{m}W(\mathbf{y}|\mathbf{x}_{m})\leq e^{n\beta}\cdot\sum_{m=1}^{M}V(\mathbf{y}|\mathbf{x}_{m})\right\} .\label{eq: asymptotical rejection region}
\end{equation}
For $\alpha\geq0$, the ML term on the right-hand side (r.h.s.) of
\eqref{eq: asymptotical rejection region} is negligible w.r.t. the
left-hand side (l.h.s.), and the obtained rejection region is asymptotically
equivalent to
\begin{equation}
{\cal R}_{0}'\dfn\left\{ \mathbf{y}:~e^{n\alpha}\cdot\sum_{m=1}^{M}W(\mathbf{y}|\mathbf{x}_{m})\leq e^{n\beta}\cdot\sum_{m=1}^{M}V(\mathbf{y}|\mathbf{x}_{m})\right\} \label{eq: asymptotical rejection region full tension}
\end{equation}
which corresponds to an ordinary Neyman-Pearson test between the hypotheses
that the channel is $W$ or $V$. Thus, unlike the fixed blocklength
case, asymptotically, we obtain a complete tension between the FA
and MD probabilities. Also, comparing \eqref{eq: asymptotical rejection region full tension},
and \eqref{eq: asymptotical rejection region}, we may observe that
the term $\max_{m}W(\mathbf{y}|\mathbf{x}_{m})$ in ${\cal R}_{0}^{*}$
is added in favor of the alternative hypothesis $W$. So, in case
of a tie in the ordinary Neyman-Pearson test \eqref{eq: asymptotical rejection region full tension},
the optimal detector/decoder will actually decide in favor of $W$. 

As the next proposition shows, the above discussion implies that there
is no loss in error exponents when using the detector/decoder $\phi'$,
whose rejection region is as in \eqref{eq: asymptotical rejection region full tension},
and if $\mathbf{y}\notin{\cal R}_{0}'$ then ordinary ML decoding
for $W$ is used, as in \eqref{Rm optimal}. This implies an \emph{asymptotic
separation principle} between detection and decoding: the optimal
detector can be used without considering the subsequent decoding,
and the optimal decoder can be used without considering the preceding
detection. As a result, asymptotically, there is only a single degree
of freedom to control the exponents. Thus, when analyzing error exponents
in Section \ref{sec:Achievable-Error-Exponents }, we will assume
that $\phi'$ is used, and since \eqref{eq: asymptotical rejection region full tension}
depends on the difference $\alpha-\beta$ only, we will set henceforth
$\beta=0$ for $\phi'$. The parameter $\alpha$ will be used to control
the trade-off between the FA and MD exponents, just as in ordinary
hypothesis testing.
\begin{prop}
\label{prop:Full tension decoder is asym optimal}For any given sequence
of codes ${\cal C}=\{{\cal C}_{n}\}_{n=1}^{\infty}$, and given constraints
on the FA and MD exponents, the detector/decoder $\phi'$ achieves
the same IE exponent as $\phi^{*}$. \end{prop}
\begin{IEEEproof}
Assume that the coefficients $\alpha,\beta$ of $\phi^{*}$ (in \eqref{eq: asymptotical rejection region})
are tuned to satisfy constraints on the FA and MD exponents, say $\overline{E}_{\st[FA]}$
and $\overline{E}_{\st[MD]}$. Let us consider replacing $\phi^{*}$
by $\phi'$, with the same $\alpha,\beta$. Now, given that the $m$th
codeword was transmitted, the conditional IE probability \eqref{eq: IE probability}
is the union of the FA event and the event
\[
\left\{ W(\mathbf{Y}|\mathbf{x}_{m})<\max_{k\neq m}W(\mathbf{Y}|\mathbf{x}_{k})\right\} ,
\]
namely, an ordinary ML decoding error. The union bound then implies
\[
P_{\st[IE]}({\cal C}_{n},\phi)\leq P_{\st[O]}^{*}({\cal C}_{n})+P_{\st[FA]}({\cal C}_{n},\phi)
\]
where $P_{\st[O]}^{*}({\cal C}_{n})$ is the ordinary decoding error
probability, assuming the ML decoder tuned to $W$. As the union bound
is asymptotically exponentially\emph{ tight} for a union of two events,
then
\begin{align}
P_{\st[IE]}\left({\cal C}_{n},\phi^{*}\right) & \doteq P_{\st[O]}\left({\cal C}_{n},\phi^{*}\right)+P_{\st[FA]}\left({\cal C}_{n},\phi^{*}\right)\\
 & \doteq\max\left\{ P_{\st[O]}\left({\cal C}_{n},\phi^{*}\right),P_{\st[FA]}\left({\cal C}_{n},\phi^{*}\right)\right\} ,\label{eq: IE probability is minimum betweeen ordinary and FA}
\end{align}
or 
\begin{equation}
E_{\st[IE]}\left({\cal C},\phi^{*}\right)=\min\left\{ E_{\st[O]}\left({\cal C},\phi^{*}\right),E_{\st[FA]}\left({\cal C},\phi^{*}\right)\right\} .\label{eq: IE exponent is minimum between ordinary and FA}
\end{equation}
Now, the ordinary decoding error probability is the same for $\phi^{*}$
and $\phi'$ and so the first term in \eqref{eq: IE probability is minimum betweeen ordinary and FA}
is the same for both detectors/decoders. Also, given any constraint
on the MD exponent, the detector defined by ${\cal R}_{0}'$ achieves
the maximal FA exponent, and so
\[
E_{\st[FA]}\left({\cal C},\phi^{*}\right)\leq E_{\st[FA]}({\cal C},\phi').
\]
In light of \eqref{eq: IE exponent is minimum between ordinary and FA},
this implies that $\phi'$ satisfies the MD and FA constraints, and
at the same time, achieves an IE exponent at least as large as that
of $\phi^{*}$.
\end{IEEEproof}
The achievable exponent bounds will be proved by random coding over
some ensemble of codes. Letting over-bar denote an average w.r.t.
some ensemble, we will define the random coding exponents, as
\[
E_{\st[FA]}\left(\phi\right)\dfn\lim_{l\to\infty}-\frac{1}{n_{l}}\log\overline{P_{\st[FA]}}\left({\cal C}_{n_{l}},\phi\right),
\]
where $\{n_{l}\}_{l=1}^{\infty}$ is a sub-sequence of blocklengths.
When we assume a fixed composition ensemble with distribution $P_{X}$,
this sub-sequence will simply be the blocklengths such that ${\cal T}(P_{X})$
is not empty, and when we will assume the independent identically
distributed (i.i.d.) ensemble, all blocklengths are valid. To comply
with definition \eqref{eq: FA exponent definition given codebook},
one can obtain codes which are good for \emph{all} sufficiently large
blocklength by slightly modifying the input distribution. The MD exponent
$E_{\st[MD]}\left(\phi\right)$ and the IE exponent $E_{\st[IE]}\left(\phi\right)$
are defined similarly, where the three exponents share the \emph{same}
sequence of blocklengths. 

Now, if we provide random coding exponents for the FA, MD and ordinary
decoding exponents, then the existence of a good sequence of codes
can be easily shown. Indeed, Markov inequality implies that
\[
\P\left(\overline{P_{\st[FA]}}({\cal C}_{n_{l}},\phi)\geq\exp\left[-n_{l}\left(E_{\st[FA]}\left(\phi\right)-\delta\right)\right]\right)\leq e^{-n_{l}\frac{\delta}{2}},
\]
for all $l$ sufficiently large. Thus, with probability tending to
$1$, the chosen codebook will have FA probability not larger than
$\exp\left[-n\left(E_{\st[FA]}\left(\phi\right)-\delta\right)\right]$.
As the same can be said on the MD probability and the ordinary error
probability, then one can find a sequence of codebooks with simultaneously
good FA, MD and ordinary decoding error probabilities, and from \eqref{eq: IE exponent is minimum between ordinary and FA},
also good IE probability. For this reason, henceforth we will only
focus on the detection performance, namely the FA and MD exponents.
The IE exponent can be simply obtained by \eqref{eq: IE exponent is minimum between ordinary and FA}
and the known bounds of ordinary decoding, namely: (i) the standard
Csisz{\'a}r and K{\"o}rner random coding bounds \cite[Theorem 10.2]{csiszar2011information}
(and its tightness \cite[Problem 10.34]{csiszar2011information}%
\footnote{See also the extended version \cite[Appendix C]{SW_paper_extended},
which provides a simple proof to the tightness of the random coding
exponent of Slepian-Wolf coding \cite{slepian1973noiseless}. A very
similar method can show the tightness of the random coding exponent
of channel codes.%
}) and the expurgated bound \cite[Problem 10.18]{csiszar2011information}
for fixed composition ensembles, (ii) the random coding bound \cite[Theorem 5.6.2]{gallager1968information},
and the expurgated bound \cite[Theorem 5.7.1]{gallager1968information}
for the ensemble of i.i.d. codes. 

Beyond the fact that $\phi'$ is slightly a simpler detector/decoder
than $\phi^{*}$, it also enables to prove a very simple relation
between its FA and MD exponents. For the next proposition, we will
use the notation $\phi'_{\alpha}$ and ${\cal R}_{0,\alpha}'$ to
explicitly denote their dependence on $\alpha$. 
\begin{prop}
\label{prop: exponents balance} For any ensemble of codes such that
$E_{\st[FA]}({\cal C},\phi_{\alpha}')$ and $E_{\st[MD]}({\cal C},\phi_{\alpha}')$
are continuous in $\alpha$, the FA and MD exponents of $\phi'_{\alpha}$
satisfy
\begin{equation}
E_{\st[FA]}({\cal C},\phi_{\alpha}')=E_{\st[MD]}({\cal C},\phi_{\alpha}')+\alpha.\label{eq: FA exponent is MD exponent plus alpha}
\end{equation}
\end{prop}
\begin{IEEEproof}
For typographical convenience, let us assume that the sub-sequence
of blocklengths is simply $\mathbb{N}$. The detector/decoder $\phi'_{\alpha}$
is the one which minimizes the FA probability under an MD probability
constraint. Considering $e^{-n\alpha}\geq0$ as a positive Lagrange
multiplier, it is readily seen that for any given code, $\phi'_{\alpha}$
minimizes the following Lagrangian:
\begin{align}
L({\cal C}_{n},\phi,\alpha) & \dfn P_{\st[FA]}\left({\cal C}_{n},\phi\right)+e^{-n\alpha}P_{\st[MD]}\left({\cal C}_{n},\phi\right)\\
 & =\sum_{\mathbf{y}}\left\{ \frac{1}{M}\sum_{m=1}^{M}W(\mathbf{y}|\mathbf{x}_{m})\I\left\{ \mathbf{y}\in{\cal R}_{0}\right\} +e^{-n\alpha}\frac{1}{M}\sum_{m=1}^{M}V(\mathbf{y}|\mathbf{x}_{m})\I\left\{ \mathbf{y}\in\overline{{\cal R}_{0}}\right\} \right\} \label{eq: Lagrange formulation}
\end{align}
Hence,
\[
\overline{L({\cal C}_{n},\phi,\alpha)}\geq\overline{L({\cal C}_{n},\phi'_{\alpha},\alpha)}=\overline{P_{\st[FA]}}({\cal C}_{n},\phi'_{\alpha})+e^{-n\alpha}\overline{P_{\st[MD]}}({\cal C}_{n},\phi'_{\alpha}),
\]
or, after taking limits
\begin{align}
\lim_{n\to\infty}-\frac{1}{n}\log\overline{L({\cal C}_{n},\phi,\alpha)} & =\min\left\{ E_{\st[FA]}(\phi),E_{\st[MD]}(\phi)+\alpha\right\} .\\
 & \leq\lim_{n\to\infty}-\frac{1}{n}\log\overline{L({\cal C}_{n},\phi'_{\alpha},\alpha)}\\
 & =\min\left\{ E_{\st[FA]}(\phi'_{\alpha}),E_{\st[MD]}(\phi'_{\alpha})+\alpha\right\} .\label{eq: phi' maximizes Lagrangian}
\end{align}
Now, assume by contradiction that 
\begin{equation}
E_{\st[FA]}(\phi'_{\alpha})>E_{\st[MD]}(\phi'_{\alpha})+\alpha.\label{eq: inequality between FA and MD}
\end{equation}
Then, from continuity of the FA and MD exponents, one can expand ${\cal R}_{0,\alpha}'$
to some ${\cal R}_{0,\overline{\alpha}}'$ with $\overline{\alpha}<\alpha$
and obtain a decoder $\phi'_{\overline{\alpha}}$ for which 
\[
E_{\st[MD]}(\phi'_{\alpha})+\alpha<E_{\st[MD]}({\cal C},\phi'_{\overline{\alpha}})+\alpha=E_{\st[FA]}({\cal C},\phi'_{\overline{\alpha}})<E_{\st[FA]}({\cal C},\phi'_{\alpha}).
\]
Thus,
\[
\overline{L({\cal C}_{n},\phi'_{\overline{\alpha}},\alpha)}\geq\overline{L({\cal C}_{n},\phi'_{\alpha},\alpha)}
\]
which contradicts \eqref{eq: inequality between FA and MD}, and so
\[
E_{\st[FA]}({\cal C},\phi'_{\alpha})\leq E_{\st[MD]}({\cal C},\phi'_{\alpha})+\alpha.
\]
Similarly, it can be shown that reversed strict inequality in \eqref{eq: inequality between FA and MD}
contradicts the optimality of $\phi'_{\alpha}$, and so \eqref{eq: FA exponent is MD exponent plus alpha}
follows.\end{IEEEproof}
\begin{rem}
Consider the following related problem 
\begin{eqnarray}
 &  & \mbox{minimize}~~~P_{\st[EE]}\nonumber \\
 &  & \mbox{subject to}~~P_{\st[FA]}\le\epsilon_{\st[FA]}\nonumber \\
 &  & ~~~~~~~~~~~~~P_{\st[MD]}\le\epsilon_{\st[MD]}\label{eq: EE- FA and MD tradeoff}
\end{eqnarray}
and let $\phi^{**}$ be the optimal detector/decoder for the problem
\eqref{eq: EE- FA and MD tradeoff}. Now, as $P_{\st[IE]}=P_{\st[EE]}+P_{\st[FA]}$,
it may be easily verified that when $P_{\st[FA]}=\epsilon_{\st[FA]}$
for the optimal detector/decoder $\phi^{*}$ (of the problem \eqref{eq: IE- FA and MD tradeoff}),
then $\phi^{*}$ is also the optimal detector/decoder for the problem
\eqref{eq: EE- FA and MD tradeoff}. However, when $P_{\st[FA]}<\epsilon_{\st[FA]}$
for $\phi^{*}$, then $\phi^{**}$ is different, since it easy to
check that for the problem \eqref{eq: EE- FA and MD tradeoff}, the
constraint $P_{\st[FA]}\leq\epsilon_{\st[FA]}$ for $\phi^{**}$ must
be achieved with equality. To gain some intuition why \eqref{eq: EE- FA and MD tradeoff}
is more complicated than \eqref{eq: IE- FA and MD tradeoff}, see
the discussion in \cite[Section III]{CoN}.
\end{rem}

\subsection{Simplified Detectors/Decoders\label{sub:Simplified-Detector/Decoders}}

Unfortunately, the asymptotically optimal detector/decoder \eqref{eq: asymptotical rejection region full tension}
is very difficult to implement in its current form. The reason is
that the computation of $\sum_{m=1}^{M}W(\mathbf{y}|\mathbf{x}_{m})$
is usually intractable, as it is the sum of exponentially many likelihood
terms, where each likelihood term is exponentially small. This is
in sharp contrast to ordinary decoders, based on comparison of single
likelihood terms which can be carried out in the logarithmic scale,
rendering them numerically feasible. In a recent related work \cite{Simplified_erausre}
dealing with the optimal erasure/list decoder \cite{Forney68}, it
was observed that a much simplified decoder is asymptotically optimal.
For the detector/decoder discussed in this paper, this simplification
of \eqref{eq: asymptotical rejection region full tension} implies
that the rejection region
\[
{\cal R}_{0}''\dfn\left\{ \mathbf{y}:~e^{n\alpha}\cdot\max_{Q}\tilde{N}(Q|\mathbf{y})e^{nf_{W}(Q)}\leq e^{n\beta}\cdot\max_{Q}\tilde{N}(Q|\mathbf{y})e^{nf_{V}(Q)}\right\} ,
\]
is asymptotically optimal, where the \emph{type class enumerators
}are defined as
\begin{equation}
\tilde{N}(Q|\mathbf{y})\dfn\left|\left\{ \mathbf{x}\in{\cal C}_{n}:\;\hat{Q}_{\mathbf{x}\mathbf{y}}=Q_{XY}\right\} \right|.\label{eq: type enumertor including x1}
\end{equation}
While the above mentioned numerical problem does not arise in ${\cal R}_{0}''$,
there is still room for additional simplification which significantly
facilitates implementation, at the cost of degrading the performance,
perhaps only slightly. For zero rate, the type class enumerators cannot
increase exponentially, and so either $\tilde{N}(Q|\mathbf{y})=0$
or $\tilde{N}(Q|\mathbf{y})\doteq1$. Thus, for low rates, we propose
the use of a sub-optimal detector/decoder, which has the following
rejection region
\[
{\cal R}_{0,\st[L]}\dfn\left\{ \mathbf{y}:~e^{n\alpha}\cdot\max_{1\leq m\leq M}W(\mathbf{y}|\mathbf{x}_{m})<\max_{1\leq m\leq M}V(\mathbf{y}|\mathbf{x}_{m})\right\} .
\]
We will denote the resulting detector/decoder by $\phi_{\st[L]}$.
In this context, this is a \emph{generalized likelihood ratio test}
\cite{van2013detection}, in which the codeword is the `nuisance parameter'
for the detection problem. For high rates (close to the capacity of
the channel), the output distribution $\frac{1}{M}\sum_{m=1}^{M}W(\mathbf{y}|\mathbf{x}_{m})$
of a `good' code \cite{good_codes} tends to be close to a memoryless
distribution $\tilde{W}\dfn(P_{X}\times W)_{Y}$ for some distribution
$P_{X}$. Thus, for high rates, a possible approximation is a sub-optimal
detector/decoder, which has the following rejection region 
\[
{\cal R}_{0,\st[H]}\dfn\left\{ \mathbf{y}:~e^{n\alpha}\cdot\tilde{W}(\mathbf{y})<\tilde{V}(\mathbf{y})\right\} ,
\]
where $\tilde{V}\dfn(P_{X}\times W)_{Y}$. We will denote the resulting
detector/decoder by $\phi_{\st[H]}$.

As was recently demonstrated in \cite{Simplified_erausre}, while
$\phi_{\st[L]}$ and $\phi_{\st[H]}$ are much simpler to implement
than $\phi'$, they have the potential to cause only slight loss in
exponents compared to $\phi'$. Since the random coding performance
of $\phi_{\st[H]}$ is simply obtained by the standard analysis of
hypothesis testing between two memoryless hypotheses (cf. Subsection
\ref{sub:Exponents-Simplified-Decoders}), we will mainly focus on
$\phi_{\st[L]}$.

\section{Achievable Error Exponents\label{sec:Achievable-Error-Exponents }}

In this section, we derive various achievable exponents for the joint
detection/decoding problem \eqref{eq: IE- FA and MD tradeoff}, for
a given pair of DMCs $(W,V)$, at rate $R$. In Subsection \ref{sub:A-Tight-Random},
we derive the \emph{exact} random coding performance of the asymptotically
optimal detector/decoder $\phi'$. In Subsection \ref{sub:An-Expurgated-Bound},
we derive an improved bound for low rates using the expurgation technique.
In Subsection \ref{sub:Exponents-Simplified-Decoders}, we discuss
the exponents achieved by the sub-optimal detectors/decoders $\phi_{\st[L]}$
and $\phi_{\st[H]}$. In Subsection \ref{sub:Gallager-Forney-Style-Bounds},
we provide Gallager/Forney-style lower bounds on the exponents. While
these bounds can be loose and only lead to inferior exponents when
compared to Subsections \ref{sub:A-Tight-Random} and \ref{sub:An-Expurgated-Bound},
it is indeed useful to derive them since: (i) they are simpler to
compute, since they require solving at most two-dimensional optimization
problems%
\footnote{When there are no input constraints. When input constraints are given,
as e.g. in the power limited AWGN channel, it is required to solve
four-dimensional optimization problem (cf. \eqref{eq: E0'' AWGN}).%
}, irrespective of the input/output alphabet sizes, (ii) the bounds
are translated almost verbatim to memoryless channels with continuous
input/output alphabets, like the AWGN channel. For brevity, in most
cases the notation of the dependence on the problem parameters (i.e.
$R,P_{X},\alpha,W,V$) will be omitted, and will be reintroduced only
when necessary.

\subsection{Exact Random Coding Exponents\label{sub:A-Tight-Random}}

We begin with a sequence of definitions. Throughout, $\tilde{Q}$
will represent the joint type of the true transmitted codeword and
the output, and $\overline{Q}$ is some type of competing codewords.
We denote the \emph{normalized log-likelihood ratio} of a channel
$W$ by 
\begin{equation}
f_{W}(Q)\dfn\sum_{x\in{\cal X},y\in{\cal Y}}Q(x,y)\log W(y|x),\label{eq: log likelihood definition}
\end{equation}
with the convention $f_{W}(\hat{Q}_{\mathbf{x}\mathbf{y}})=-\infty$
if $W(\mathbf{y}|\mathbf{x})=0$. We define the set
\[
{\cal Q}_{W}\dfn\left\{ Q:\; f_{W}(Q)>-\infty\right\} 
\]
and for $\gamma\in\mathbb{R}$,
\begin{equation}
\mathbf{s}(\tilde{Q}_{Y},\gamma)\dfn\min_{Q\in{\cal Q}_{W}:\, Q{}_{Y}=\tilde{Q}_{Y}}I(Q)+\left[-\alpha-f_{W}(Q)+\gamma\right]_{+}.\label{eq: s bold lower case}
\end{equation}
Now, define the sets
\[
{\cal J}_{1}\dfn\left\{ \tilde{Q}:\; f_{W}(\tilde{Q})\leq-\alpha+f_{V}(\tilde{Q})\right\} ,
\]
\[
{\cal J}_{2}\dfn\left\{ \tilde{Q}:\;\mathbf{s}\left(\tilde{Q}_{Y},f_{V}(\tilde{Q})\right)\geq R\right\} ,
\]
the exponent
\begin{equation}
E_{A}\dfn\min_{\tilde{Q}\in\cap_{i=1}^{2}{\cal J}_{i}}D(\tilde{Q}{}_{Y|X}\|W|P_{X}),\label{eq: EA}
\end{equation}
the sets
\[
{\cal K}_{1}\dfn\left\{ (\tilde{Q},\overline{Q}):\;\overline{Q}_{Y}=\tilde{Q}_{Y}\right\} ,
\]
\[
{\cal K}_{2}\dfn\left\{ (\tilde{Q},\overline{Q}):\; f_{W}(\overline{Q})\leq-\alpha+f_{V}(\overline{Q})\right\} ,
\]
\[
{\cal K}_{3}\dfn\left\{ (\tilde{Q},\overline{Q}):\; f_{V}(\overline{Q})\geq\alpha+f_{W}(\tilde{Q})-\left[R-I(\overline{Q})\right]_{+}\right\} ,
\]
\[
{\cal K}_{4}\dfn\left\{ (\tilde{Q},\overline{Q}):\;\mathbf{s}\left(\tilde{Q}_{Y},f_{V}(\overline{Q})+\left[R-I(\overline{Q})\right]_{+}\right)\geq R\right\} ,
\]
and the exponent
\begin{equation}
E_{B}\dfn\min_{(\tilde{Q},\overline{Q})\in\cap_{i=1}^{4}{\cal K}_{i}}\left\{ D(\tilde{Q}{}_{Y|X}\|W|P_{X})+\left[I(\overline{Q})-R\right]_{+}\right\} .\label{eq: EB}
\end{equation}
In addition, let us define the \emph{type-enumeration detection random
coding exponent} as
\begin{equation}
E_{\st[TE]}^{\st[RC]}\left(R,\alpha,P_{X},W,V\right)\dfn\min\left\{ E_{A},E_{B}\right\} .\label{eq: Detection exponent RC enumeration}
\end{equation}

\begin{thm}
\label{thm: RC bound type-enumeration}Let a distribution $P_{X}$
and a parameter $\alpha\in\mathbb{R}$ be given. Then, there exists
a sequence of codes ${\cal C}=\{{\cal C}_{n}\}_{n=1}^{\infty}$ of
rate $R$ such that for any $\delta>0$
\begin{equation}
E_{\st[FA]}\left({\cal C},\phi^{*}\right)\geq E_{\st[TE]}^{\st[RC]}\left(R,\alpha,P_{X},W,V\right)-\delta,\label{eq: FA exponent RC type-enumeration thm}
\end{equation}
\begin{equation}
E_{\st[MD]}\left({\cal C},\phi^{*}\right)\geq E_{\st[TE]}^{\st[RC]}\left(R,\alpha,P_{X},W,V\right)-\alpha-\delta.\label{eq: MD exponent RC type-enumaration thm}
\end{equation}

\end{thm}
The main challenge in analyzing the random coding FA exponent, is
that the \emph{likelihoods} of both hypotheses, namely $\sum_{m=1}^{M}W(\mathbf{Y}|\mathbf{X}_{m})$
and $\sum_{m=1}^{M}V(\mathbf{Y}|\mathbf{X}_{m})$ are very correlated
due to the fact the once the codewords are drawn, they are common
for both likelihoods. This is significantly different from the situation
in \cite{CoN}, in which the likelihood $\sum_{m=1}^{M}W(\mathbf{Y}|\mathbf{X}_{m})$
was compared to a likelihood $Q_{0}(\mathbf{Y})$, of a completely
different distribution%
\footnote{In \cite{CoN}, $Q_{0}(\mathbf{Y})$ represented the hypothesis that
no codeword was transmitted and only noise was received.%
}.

We first make the following observation.
\begin{fact}
\label{fact: alternative expression for MD exponent}For the detector/decoder
$\phi'$ 
\begin{align}
P_{\st[FA]}({\cal C}_{n},\phi') & =\P_{W}\left(\mathbf{Y}\in{\cal R}_{0}'\right)\\
 & =\P_{W}\left(\frac{\sum_{m=1}^{M}W(\mathbf{Y}|\mathbf{x}_{m})}{\sum_{m=1}^{M}V(\mathbf{Y}|\mathbf{x}_{m})}\leq e^{-n\alpha}\right)
\end{align}
where $\P_{W}\left({\cal A}\right)$ is the probability of the event
${\cal A}$ under the hypothesis that the channel is $W$. Similarly,
\begin{align}
P_{\st[MD]}({\cal C}_{n},\phi') & =\P_{V}\left(\mathbf{Y}\not\in{\cal R}_{0}'\right)\\
 & =\P_{V}\left(\frac{\sum_{m=1}^{M}W(\mathbf{Y}|\mathbf{x}_{m})}{\sum_{m=1}^{M}V(\mathbf{Y}|\mathbf{x}_{m})}\geq e^{-n\alpha}\right)\\
 & =\P_{V}\left(\frac{\sum_{m=1}^{M}V(\mathbf{Y}|\mathbf{x}_{m})}{\sum_{m=1}^{M}W(\mathbf{Y}|\mathbf{x}_{m})}\leq e^{n\alpha}\right).
\end{align}
Thus, the random coding MD exponent can be obtained by replacing $\alpha$
with $-\alpha$, and $W$ with $V$ in the FA exponent, i.e.
\[
\lim_{l\to\infty}-\frac{1}{n_{l}}\log\overline{P_{\st[MD]}}({\cal C}_{n_{l}},\phi^{*})=E_{\st[TE]}^{\st[RC]}\left(R,-\alpha,P_{X},V,W\right)
\]
where $\{n_{l}\}$ is the sub-sequence of blocklengths such that ${\cal T}(P_{X})$
is not empty.
\end{fact}
Before rigorously proving Theorem \ref{thm: RC bound type-enumeration},
we make a short detour to present the \emph{type class enumerators
}concept\emph{ }\cite{Merhav09}, and also derive two useful lemmas.
Recall that when analyzing the performance of a randomly chosen code,
a common method is to first evaluate the error probability conditioned
on the transmitted codeword (assumed, without loss of generality,
to be $\mathbf{x}_{1}$) and the output vector $\mathbf{y}$, and
average only over $\{\mathbf{X}_{m}\}_{m=2}^{M}$. Afterwards, the
ensemble average error probability is obtained by averaging w.r.t.
the random choice of $(\mathbf{X}_{1},\mathbf{Y})$. We will assume
that the codewords are drawn randomly and uniformly from ${\cal T}(P_{X})$,
and so all joint types $Q$ mentioned henceforth will satisfy $Q_{X}=P_{X}$,
even if this is not explicitly displayed. 

To analyze the conditional error probability, it is useful \cite{Merhav09}
to define the \emph{type class enumerators}
\[
N(Q|\mathbf{y})\dfn\left|\left\{ \mathbf{x}\in{\cal C}_{n}\backslash\mathbf{x}_{1}:\;\hat{Q}_{\mathbf{x}\mathbf{y}}=Q\right\} \right|,
\]
which, for a given $\mathbf{y}$, count the number of codewords, excluding
$\mathbf{x}_{1}$, which have joint type $Q$ with $\mathbf{y}$.
As the codewords in the ensemble are drawn independently, $N(Q|\mathbf{y})$
is a binomial random variable pertaining to $M=\left\lceil e^{nR}\right\rceil $
trials and probability of success of the exponential order of $e^{-nI(Q)}$,
and consequently, $\E\left[N(Q|\mathbf{y})\right]\doteq\exp\left[n(R-I(Q))\right]$.
A more refined analysis, similar to the one carried in \cite[Subsection 6.3]{Merhav09},
shows that for any given $u\in\mathbb{R}$ 
\begin{equation}
\P\left\{ N(Q|\mathbf{y})\ge e^{nu}\right\} \doteq\exp\left\{ -e^{n[u]_{+}}\left(n\left[I(Q)-R+[u]_{+}\right]-1\right)\right\} .\label{eq: large deviations of type enumerator}
\end{equation}
Consequently, if $I(Q)<R$, $N(Q|\mathbf{y})$ concentrates double-exponentially
rapidly around its average $\doteq e^{n\left[R-I(Q)\right]}$, and
if $I(Q)>R$, then with probability tending to $1$ we have $N(Q|\mathbf{y})=0$,
and $\P\left\{ N(Q|\mathbf{y})\ge1\right\} \doteq e^{-n\left[I(Q)-R\right]}$,
as well as $\P\left\{ N(Q|\mathbf{y})\ge e^{nu}\right\} \doteq e^{-n\infty}$
for any $u>0$.

We now derive two useful lemmas. In the first lemma, we show that
if a single joint type $\overline{Q}$ is excluded from the possible
joint types for a randomly chosen codeword $\mathbf{X}_{l}$ and\textbf{
$\mathbf{y}$}, then the probability of drawing some other joint type
is not significantly different from its unconditional counterpart.
In the second lemma we characterize the behavior of the probability
of the intersection of events in which the type class enumerators
are upper bounded. 
\begin{lem}
\label{lem: conditional Ber}For any $Q\neq\overline{Q}$
\[
\P\left(\hat{Q}_{\mathbf{X}_{l}\mathbf{y}}=Q|\hat{Q}_{\mathbf{X}_{l}\mathbf{y}}\neq\overline{Q}\right)\doteq\P\left(\hat{Q}_{\mathbf{X}_{l}\mathbf{y}}=Q\right)\doteq e^{-nI(Q)}.
\]
\end{lem}
\begin{IEEEproof}
For any given $\overline{Q}$
\[
\P\left(\hat{Q}_{\mathbf{X}_{l}\mathbf{y}}=\overline{Q}\right)\doteq e^{-nI(\overline{Q})},
\]
and if $I(\overline{Q})=0$ then
\[
\P\left(\hat{Q}_{\mathbf{X}_{l}\mathbf{y}}=\overline{Q}\right)\to0,
\]
as $n\to\infty$, although sub-exponentially \cite[Problem 2.2]{csiszar2011information}.
Thus, for any $Q\neq\overline{Q}$, 
\begin{align}
\P\left(\hat{Q}_{\mathbf{X}_{l}\mathbf{y}}=Q|\hat{Q}_{\mathbf{X}_{l}\mathbf{y}}\neq\overline{Q}\right) & =\frac{\P\left(\hat{Q}_{\mathbf{X}_{l}\mathbf{y}}=Q,\hat{Q}_{\mathbf{X}_{l}\mathbf{y}}\neq\overline{Q}\right)}{\P\left(\hat{Q}_{\mathbf{X}_{l}\mathbf{y}}\neq\overline{Q}\right)}\\
 & =\frac{\P\left(\hat{Q}_{\mathbf{X}_{l}\mathbf{y}}=Q\right)}{1-\P\left(\hat{Q}_{\mathbf{X}_{l}\mathbf{y}}=\overline{Q}\right)}\\
 & \doteq e^{-nI(Q)}.
\end{align}
\end{IEEEproof}
\begin{lem}
\label{lem: behavior of intersection}Let a set ${\cal Q}$ of joint
types, a continuous function $J(Q)$ in ${\cal Q}$, and a type $\tilde{Q}_{Y}$
be given. Let $\{\hat{N}(Q|\mathbf{y})\}_{Q\in{\cal Q}}$ be a sequence
of sets of binomial random variables pertaining to $K_{n}$ trials
and probability of success $p_{n}$. Then, if $K_{n}\doteq e^{nR}$
and $p_{n}\doteq e^{-nI(Q)}$ 
\begin{equation}
\P\left(\bigcap_{Q\in{\cal Q}:\, Q_{Y}=\tilde{Q}_{Y}}\left\{ \hat{N}(Q|\mathbf{y})<e^{nJ(Q)}\right\} \right)\:\begin{cases}
=1-o(n), & \mathbf{S}(\tilde{Q}_{Y};J,{\cal Q})>R\\
\doteq e^{-n\infty}, & \mbox{otherwise}
\end{cases},\label{eq: exponential of intersection}
\end{equation}
where $\mathbf{y}\in{\cal T}(\tilde{Q}_{Y})$, and 
\begin{equation}
\mathbf{S}(\tilde{Q}_{Y};J,{\cal Q})\dfn\min_{Q\in{\cal Q}:\, Q{}_{Y}=\tilde{Q}_{Y}}I(Q)+\left[J(Q)\right]_{+}.\label{eq: S bold function}
\end{equation}
\end{lem}
\begin{IEEEproof}
A similar statement was proved in \cite[pp. 5086-5087]{CoN}, but
for the sake of completeness, we include its short proof. If there
exists at least one $Q\in{\cal Q}$ with $Q{}_{Y}=\tilde{Q}_{Y}$
for which $I(Q)<R$ and $R-I(Q)>J(Q)$, then this $Q$ alone is responsible
for a double exponential decay of the intersection probability, because
then the event in question would be a large deviations event whose
probability decays exponentially with $M=\left\lceil e^{nR}\right\rceil $,
thus double-exponentially with $n$, let alone the intersection over
all $Q\in{\cal Q}$. The condition for this to happen is $R>\mathbf{S}(\tilde{Q}_{Y};J,{\cal Q})$.
Conversely, if for every $Q\in{\cal Q}$ with $Q{}_{Y}=\tilde{Q}_{Y}$,
we have $I(Q)>R$ or $R-I(Q)<J(Q)$, i.e., $R<\mathbf{S}(\tilde{Q}_{Y};J,{\cal Q})$,
then the intersection probability is close to $1$, since the intersection
is over a sub-exponential number of events with very high probability.
Thus \eqref{eq: exponential of intersection} follows.\end{IEEEproof}
\begin{rem}
\label{rem: continuity of S}A natural choice for $\hat{N}(Q|\mathbf{y})$
is simply $N(Q|\mathbf{y})$. However, in what follows, we will need
to analyze a conditional version of the type enumerators, namely,
events of the form $\{N(Q|\mathbf{y})=N_{1}|N(\overline{Q}|\mathbf{y})=N_{2}\}$
for some $0\leq N_{1},N_{2}\leq M$. As Lemma \ref{lem: conditional Ber}
above hints, in some cases the conditional distribution of $N(Q|\mathbf{y})$
is asymptotically the same as the unconditional distribution. In this
respect, it should be noted that the result of Lemma \ref{lem: behavior of intersection}
is proved using the marginal distribution of each $\hat{N}(Q|\mathbf{y})$
alone, and not their \emph{joint} distribution. It should also be
noted that the second argument of $\mathbf{S}(\tilde{Q}_{Y};\cdot,\cdot)$
in \eqref{eq: S bold function} is a \emph{function} of the joint
type $Q$, and the third argument is a set of joint types. Finally,
since the types are dense in the subspace of the simplex of all the
type satisfying $Q_{Y}=\tilde{Q}_{Y}$, then the exclusion of a \emph{single}
type form the intersection in \eqref{eq: exponential of intersection}
does not change the result of the lemma.
\end{rem}

\begin{rem}
\label{rem: J function convex}As $Q_{X}=P_{X}$ the minimization
in \eqref{eq: S bold function} is in fact over the variables $\{Q_{Y|X}(y|x)\}_{x\in{\cal X},y\in{\cal Y}}$.
Thus, whenever $J(Q)$ is convex in $Q_{Y|X}$, then 
\begin{align}
\mathbf{S}(\tilde{Q}_{Y};J,{\cal Q}) & =\min_{Q\in{\cal Q}:\, Q{}_{Y}=\tilde{Q}_{Y}}\max_{0\leq\lambda\leq1}\left[I(Q)+\lambda J(Q)\right]\\
 & \trre[=,a]\max_{0\leq\lambda\leq1}\min_{Q\in{\cal Q}:\, Q{}_{Y}=\tilde{Q}_{Y}}\left[I(Q)+\lambda J(Q)\right]\label{eq: S bold J convex}
\end{align}
where $(a)$ is by the minimax theorem \cite{Sion1958minimax}, as
both $I(Q)$ and $J(Q)$ are convex in $Q_{Y|X}$ and the minimization
set involves only linear constraints and thus convex. This dual form
is simpler to compute than \eqref{eq: S bold function}, since the
inner minimization in \eqref{eq: S bold J convex} is a convex optimization
problem \cite{Boyd}, and the outer maximization problem requires
only a simple line-search. Note that the function $\mathbf{s}(\tilde{Q}_{Y};\gamma)$
is a specific instance of $\mathbf{S}(\tilde{Q}_{Y};\cdot,\cdot)$
defined in \eqref{eq: S bold function} with ${\cal Q}={\cal Q}_{W}$
and $J(Q)=-\alpha-f_{W}(Q)+\gamma$ which is convex in $Q_{Y|X}$
(in fact, linear). 

We are now ready to prove Theorem \ref{thm: RC bound type-enumeration}. \end{rem}
\begin{IEEEproof}[Proof of Theorem \ref{thm: RC bound type-enumeration}]
 We begin by analyzing the FA exponent. Assume, without loss of generality,
that the first message is transmitted. Let us condition on the event
$\mathbf{X}_{1}=\mathbf{x}_{1}$ and $\mathbf{Y}=\mathbf{y}$, and
analyze the average over the ensemble of fixed composition codes of
type $P_{X}$. For brevity, we will denote $\tilde{Q}=\hat{Q}_{\mathbf{x}_{1}\mathbf{y}}$.
The average conditional FA probability for the decoder $\phi'$ with
parameter $\alpha$ is given by
\begin{align}
\overline{P_{\st[FA]}}(\mathbf{x}_{1},\mathbf{y}) & \dfn\P\left(\mathbf{y}\in{\cal R}_{0}'|\mathbf{X}_{1}=\mathbf{x}_{1},\mathbf{Y}=\mathbf{y}\right)\\
 & \trre[=,a]\P\left(W(\mathbf{y}|\mathbf{x}_{1})+\sum_{m=2}^{M}W(\mathbf{y}|\mathbf{X}_{m})\leq e^{-n\alpha}\cdot V(\mathbf{y}|\mathbf{x}_{1})+e^{-n\alpha}\cdot\sum_{m=2}^{M}V(\mathbf{y}|\mathbf{X}_{m})\right)\\
 & \trre[\doteq,UR]\P\left(W(\mathbf{y}|\mathbf{x}_{1})+\sum_{m=2}^{M}W(\mathbf{y}|\mathbf{X}_{m})\leq e^{-n\alpha}\cdot V(\mathbf{y}|\mathbf{x}_{1})\right)\nonumber \\
 & \hphantom{=}+\P\left(W(\mathbf{y}|\mathbf{x}_{1})+\sum_{m=2}^{M}W(\mathbf{y}|\mathbf{X}_{m})\leq e^{-n\alpha}\cdot\sum_{m=2}^{M}V(\mathbf{y}|\mathbf{X}_{m})\right)\\
 & \trre[\doteq,IR]\P\left(\sum_{m=2}^{M}W(\mathbf{y}|\mathbf{X}_{m})\leq e^{-n\alpha}\cdot V(\mathbf{y}|\mathbf{x}_{1})\right)\cdot\I\left\{ W(\mathbf{y}|\mathbf{x}_{1})\leq e^{-n\alpha}\cdot V(\mathbf{y}|\mathbf{x}_{1})\right\} \nonumber \\
 & \hphantom{=}+\P\left(W(\mathbf{y}|\mathbf{x}_{1})+\sum_{m=2}^{M}W(\mathbf{y}|\mathbf{X}_{m})\leq e^{-n\alpha}\cdot\sum_{m=2}^{M}V(\mathbf{y}|\mathbf{X}_{m})\right)\\
 & =\P\left(\sum_{Q}N(Q|\mathbf{y})e^{nf_{W}(Q)}\leq e^{-n\alpha}\cdot e^{nf_{V}(\tilde{Q})}\right)\cdot\I\left\{ f_{W}(\tilde{Q})\leq-\alpha+f_{V}(\tilde{Q})\right\} \nonumber \\
 & \hphantom{=}+\P\left(e^{nf_{W}(\tilde{Q})}+\sum_{Q}N(Q|\mathbf{y})e^{nf_{W}(Q)}\leq e^{-n\alpha}\cdot\sum_{Q}N(Q|\mathbf{y})e^{nf_{V}(Q)}\right)\\
 & \dfn A(\tilde{Q})+B(\tilde{Q})\\
 & \doteq\max\left\{ A(\tilde{Q}),B(\tilde{Q})\right\} ,\label{eq: FA intial derivation}
\end{align}
where $A(\tilde{Q})$ and $B(\tilde{Q})$ were implicitly defined,
and $(a)$ is because $\{\mathbf{X}_{m}\}_{m=2}^{M}$ are chosen independently
of $(\mathbf{X}_{1},\mathbf{Y})$. For the first term, 
\begin{align}
A(\tilde{Q}) & \trre[\doteq,IR]\P\left(\bigcap_{Q:\, f_{W}(Q)>-\infty}\left\{ N(Q|\mathbf{y})<e^{n\left[-\alpha+f_{V}(\tilde{Q})-f_{W}(Q)\right]}\right\} \right)\cdot\I\left\{ f_{W}(\tilde{Q})\leq-\alpha+f_{V}(\tilde{Q})\right\} \\
 & \trre[\doteq,a]\I\left\{ \mathbf{S}(\tilde{Q}_{Y};-\alpha+f_{V}(\tilde{Q})-f_{W}(Q),{\cal Q}_{W})>R\right\} \cdot\I\left\{ f_{W}(\tilde{Q})\leq-\alpha+f_{V}(\tilde{Q})\right\} ,
\end{align}
where $(a)$ is by Lemma \ref{lem: behavior of intersection} . Upon
averaging over $(\mathbf{X}_{1},\mathbf{Y})$, we obtain the exponent
$E_{A}$ of \eqref{eq: EA}, when utilizing the definition in \eqref{eq: s bold lower case}.
Moving on to the second term, we first assume that $e^{nf_{W}(\tilde{Q})}>0$.
Then,\textbf{ }
\begin{align}
B(\tilde{Q}) & \trre[\doteq,UR]\sum_{\overline{Q}}\P\left(e^{nf_{W}(\tilde{Q})}+\sum_{Q}N(Q|\mathbf{y})e^{nf_{W}(Q)}\leq e^{-n\alpha}\cdot N(\overline{Q}|\mathbf{y})e^{nf_{V}(\overline{Q})}\right)\\
 & \trre[\doteq,IR]\sum_{\overline{Q}}\P\left(\bigcap_{Q\neq\overline{Q}}\left\{ N(Q|\mathbf{y})e^{nf_{W}(Q)}\leq e^{-n\alpha}\cdot N(\overline{Q}|\mathbf{y})e^{nf_{V}(\overline{Q})}\right\} \cap\right.\nonumber \\
 & \hphantom{\doteq\sum_{\overline{Q}}\P}\left.\vphantom{\bigcap_{Q\neq\overline{Q}}\left\{ N(Q|\mathbf{y})e^{nf_{W}(Q)}\leq e^{-n\alpha}\cdot N(\overline{Q}|\mathbf{y})e^{nf_{V}(\overline{Q})}\right\} \cap}\left\{ N(\overline{Q}|\mathbf{y})e^{nf_{W}(\overline{Q})}\leq e^{-n\alpha}\cdot N(\overline{Q}|\mathbf{y})e^{nf_{V}(\overline{Q})}\right\} \cap\left\{ e^{nf_{W}(\tilde{Q})}\leq e^{-n\alpha}\cdot N(\overline{Q}|\mathbf{y})e^{nf_{V}(\overline{Q})}\right\} \right)\\
 & \trre[=,a]\sum_{\overline{Q}:\, f_{W}(\overline{Q})\leq-\alpha+f_{V}(\overline{Q})}\P\left(\bigcap_{Q\neq\overline{Q}}\left\{ N(Q|\mathbf{y})e^{nf_{W}(Q)}\leq e^{-n\alpha}\cdot N(\overline{Q}|\mathbf{y})e^{nf_{V}(\overline{Q})}\right\} \right.\nonumber \\
 & \hphantom{\doteq\sum_{\overline{Q}:\, f_{W}(\overline{Q})\leq-\alpha+f_{V}(\overline{Q})}\P}\left.\vphantom{\bigcap_{Q\neq\overline{Q}}\left\{ N(Q|\mathbf{y})e^{nf_{W}(Q)}\leq e^{-n\alpha}\cdot N(\overline{Q}|\mathbf{y})e^{nf_{V}(\overline{Q})}\right\} \cap}\left\{ e^{nf_{W}(\tilde{Q})}\leq e^{-n\alpha}\cdot N(\overline{Q}|\mathbf{y})e^{nf_{V}(\overline{Q})}\right\} \right)\\
 & \trre[=,b]\sum_{\overline{Q}:\, f_{W}(\overline{Q})\leq-\alpha+f_{V}(\overline{Q})}\P\left(\bigcap_{Q\neq\overline{Q}:\, f_{W}(Q)>-\infty}\left\{ N(Q|\mathbf{y})\leq e^{n\left[-\alpha+f_{V}(\overline{Q})-f_{W}(Q)\right]}\cdot N(\overline{Q}|\mathbf{y})\right\} \right.\nonumber \\
 & \hphantom{\doteq\sum_{\overline{Q}:\, f_{W}(\overline{Q})\leq-\alpha+f_{V}(\overline{Q})}\P}\left.\vphantom{\bigcap_{Q\neq\overline{Q}}\left\{ N(Q|\mathbf{y})e^{nf_{W}(Q)}\leq e^{-n\alpha}\cdot N(\overline{Q}|\mathbf{y})e^{nf_{V}(\overline{Q})}\right\} \cap}\left\{ 1\leq e^{n\left[-\alpha+f_{V}(\overline{Q})-f_{W}(\tilde{Q})\right]}\cdot N(\overline{Q}|\mathbf{y})\right\} \right)\\
 & \dfn\sum_{\overline{Q}:\, f_{W}(\overline{Q})\leq-\alpha+f_{V}(\overline{Q})}\zeta(\overline{Q}),\label{eq: B term}
\end{align}
where $(a)$ is since when $f_{W}(\overline{Q})>-\alpha+f_{V}(\overline{Q})$
the second event in the intersection implies $N(\overline{Q}|\mathbf{y})=0$,
but this implies that the third event does not occur, and in $(b)$
we have rearranged the terms. To continue the analysis of the exponential
behavior of $B(\tilde{Q})$, we split the analysis into three cases:

\uline{Case 1:} $0<I(\overline{Q})\leq R$. For any $0<\epsilon<R-I(\overline{Q})$
let 
\[
{\cal G}_{n}\dfn\left\{ e^{n\left[R-I(\overline{Q})-\epsilon\right]}\leq N(\overline{Q}|\mathbf{y})\leq e^{n\left[R-I(\overline{Q})+\epsilon\right]}\right\} ,
\]
which satisfies $\P\left[{\cal G}_{n}\right]\doteq1$. Thus, 
\begin{align}
\zeta(\overline{Q}) & =\P\left(\bigcap_{Q\neq\overline{Q}:\, f_{W}(Q)>-\infty}\left\{ N(Q|\mathbf{y})\leq e^{n\left[-\alpha+f_{V}(\overline{Q})-f_{W}(Q)\right]}\cdot N(\overline{Q}|\mathbf{y})\right\} \cap\right.\nonumber \\
 & \hphantom{\doteq\P}\left.\vphantom{\bigcap_{Q\neq\overline{Q}}\left\{ N(Q|\mathbf{y})e^{nf_{W}(Q)}\leq e^{-n\alpha}\cdot N(\overline{Q}|\mathbf{y})e^{nf_{V}(\overline{Q})}\right\} \cap}\left\{ 1\leq e^{n\left[-\alpha+f_{V}(\overline{Q})-f_{W}(\tilde{Q})\right]}\cdot N(\overline{Q}|\mathbf{y})\right\} \right)\\
 & \leq\P\left(\bigcap_{Q\neq\overline{Q}:\, f_{W}(Q)>-\infty}\left\{ N(Q|\mathbf{y})\leq e^{n\left[-\alpha+f_{V}(\overline{Q})-f_{W}(Q)\right]}\cdot N(\overline{Q}|\mathbf{y})\right\} \cap\right.\nonumber \\
 & \hphantom{\doteq\P}\left.\vphantom{\bigcap_{Q\neq\overline{Q}}\left\{ N(Q|\mathbf{y})e^{nf_{W}(Q)}\leq e^{-n\alpha}\cdot N(\overline{Q}|\mathbf{y})e^{nf_{V}(\overline{Q})}\right\} \cap}\left\{ 1\leq e^{n\left[-\alpha+f_{V}(\overline{Q})-f_{W}(\tilde{Q})\right]}\cdot N(\overline{Q}|\mathbf{y})\right\} \right)\P({\cal G}_{n})+\P(\overline{{\cal G}_{n}})\\
 & \doteq\P\left(\bigcap_{Q\neq\overline{Q}:\, f_{W}(Q)>-\infty}\left\{ N(Q|\mathbf{y})\leq e^{n\left[-\alpha+f_{V}(\overline{Q})-f_{W}(Q)\right]}\cdot N(\overline{Q}|\mathbf{y})\right\} \cap\right.\nonumber \\
 & \hphantom{\doteq\P}\left.\vphantom{\bigcap_{Q\neq\overline{Q}:\, f_{W}(Q)>-\infty}\left\{ N(Q|\mathbf{y})\leq e^{n\left[-\alpha+f_{V}(\overline{Q})-f_{W}(Q)\right]}\cdot N(\overline{Q}|\mathbf{y})\right\} \cap}\left\{ 1\leq e^{n\left[-\alpha+f_{V}(\overline{Q})-f_{W}(\tilde{Q})\right]}\cdot N(\overline{Q}|\mathbf{y})\right\} |{\cal G}_{n}\right)\\
 & \dot{\leq}\P\left(\bigcap_{Q\neq\overline{Q}:\, f_{W}(Q)>-\infty}\left\{ N(Q|\mathbf{y})\leq e^{n\left[-\alpha+f_{V}(\overline{Q})-f_{W}(Q)+R-I(\overline{Q})+\epsilon\right]}\right\} \cap\right.\nonumber \\
 & \hphantom{\doteq\P}\left.\vphantom{\bigcap_{Q\neq\overline{Q}:\, f_{W}(Q)>-\infty}\left\{ N(Q|\mathbf{y})\leq e^{n\left[-\alpha+f_{V}(\overline{Q})-f_{W}(Q)\right]}\cdot N(\overline{Q}|\mathbf{y})\right\} \cap}\left\{ 1\leq e^{n\left[-\alpha+f_{V}(\overline{Q})-f_{W}(\tilde{Q})+R-I(\overline{Q})+\epsilon\right]}\right\} |{\cal G}_{n}\right)\\
 & \trre[\doteq,a]\I\left\{ \mathbf{S}(\tilde{Q}_{Y};-\alpha+f_{V}(\overline{Q})-f_{W}(Q)+R-I(\overline{Q})+\epsilon,{\cal Q}_{W})>R\right\} \times\nonumber \\
 & \hphantom{\doteq\I}\I\left\{ -\alpha+f_{V}(\overline{Q})-f_{W}(\tilde{Q})+R-I(\overline{Q})+\epsilon\geq0\right\} ,
\end{align}
where $(a)$ is since conditioned on ${\cal G}_{n}$, $N(Q|\mathbf{y})$
is a binomial random variable with probability of success $\doteq e^{-nI(Q)}$
(see Lemma \ref{lem: conditional Ber}), and more than $e^{nR}-e^{n\left[R-I(\overline{Q})-\epsilon\right]}\doteq e^{nR}$
trials (whenever $Q_{Y}=\overline{Q}_{Y}$, and $N(Q|\mathbf{y})=0$
otherwise), and by using Lemma \ref{lem: behavior of intersection}
and Remark \ref{rem: continuity of S}.%
\footnote{We have also implicitly used the following obvious monotonicity property:
If $N_{1}$ and $N_{2}$ are two binomial random variables pertaining
to the same probability of success but the number of trials of $N_{1}$
is larger than the number of trials of $N_{2}$ then $\P\left(N_{1}\leq L\right)\leq\P\left(N_{2}\leq L\right)$.%
} Similarly,
\begin{align}
\zeta(\overline{Q}) & =\P\left(\bigcap_{Q\neq\overline{Q}:\, f_{W}(Q)>-\infty}\left\{ N(Q|\mathbf{y})\leq e^{n\left[-\alpha+f_{V}(\overline{Q})-f_{W}(Q)\right]}\cdot N(\overline{Q}|\mathbf{y})\right\} \right.\nonumber \\
 & \hphantom{\doteq\P}\left.\vphantom{\bigcap_{Q\neq\overline{Q}}\left\{ N(Q|\mathbf{y})e^{nf_{W}(Q)}\leq e^{-n\alpha}\cdot N(\overline{Q}|\mathbf{y})e^{nf_{V}(\overline{Q})}\right\} \cap}\left\{ 1\leq e^{n\left[-\alpha+f_{V}(\overline{Q})-f_{W}(\tilde{Q})\right]}\cdot N(\overline{Q}|\mathbf{y})\right\} \right)\\
 & \geq\P\left(\bigcap_{Q\neq\overline{Q}:\, f_{W}(Q)>-\infty}\left\{ N(Q|\mathbf{y})\leq e^{n\left[-\alpha+f_{V}(\overline{Q})-f_{W}(Q)\right]}\cdot N(\overline{Q}|\mathbf{y})\right\} \cap\right.\nonumber \\
 & \hphantom{\doteq\P}\left.\vphantom{\bigcap_{Q\neq\overline{Q}}\left\{ N(Q|\mathbf{y})e^{nf_{W}(Q)}\leq e^{-n\alpha}\cdot N(\overline{Q}|\mathbf{y})e^{nf_{V}(\overline{Q})}\right\} \cap}\left\{ 1\leq e^{n\left[-\alpha+f_{V}(\overline{Q})-f_{W}(\tilde{Q})\right]}\cdot N(\overline{Q}|\mathbf{y})\right\} |{\cal G}_{n}\right)\P({\cal G}_{n})\\
 & \doteq\P\left(\bigcap_{Q\neq\overline{Q}:\, f_{W}(Q)>-\infty}\left\{ N(Q|\mathbf{y})\leq e^{n\left[-\alpha+f_{V}(\overline{Q})-f_{W}(Q)\right]}\cdot N(\overline{Q}|\mathbf{y})\right\} \cap\right.\nonumber \\
 & \hphantom{\doteq\P}\left.\vphantom{\bigcap_{Q\neq\overline{Q}:\, f_{W}(Q)>-\infty}\left\{ N(Q|\mathbf{y})\leq e^{n\left[-\alpha+f_{V}(\overline{Q})-f_{W}(Q)\right]}\cdot N(\overline{Q}|\mathbf{y})\right\} \cap}\left\{ 1\leq e^{n\left[-\alpha+f_{V}(\overline{Q})-f_{W}(\tilde{Q})\right]}\cdot N(\overline{Q}|\mathbf{y})\right\} |{\cal G}_{n}\right)\\
 & \dot{\geq}\P\left(\bigcap_{Q\neq\overline{Q}:\, f_{W}(Q)>-\infty}\left\{ N(Q|\mathbf{y})\leq e^{n\left[-\alpha+f_{V}(\overline{Q})-f_{W}(Q)+R-I(\overline{Q})-\epsilon\right]}\right\} \cap\right.\nonumber \\
 & \hphantom{\doteq\P}\left.\vphantom{\bigcap_{Q\neq\overline{Q}:\, f_{W}(Q)>-\infty}\left\{ N(Q|\mathbf{y})\leq e^{n\left[-\alpha+f_{V}(\overline{Q})-f_{W}(Q)\right]}\cdot N(\overline{Q}|\mathbf{y})\right\} \cap}\left\{ 1\leq e^{n\left[-\alpha+f_{V}(\overline{Q})-f_{W}(\tilde{Q})+R-I(\overline{Q})-\epsilon\right]}\right\} |{\cal G}_{n}\right)\\
 & \trre[\doteq,a]\I\left\{ \mathbf{S}(\tilde{Q}_{Y};-\alpha+f_{V}(\overline{Q})-f_{W}(Q)+R-I(\overline{Q})-\epsilon,{\cal Q}_{W})>R\right\} \times\nonumber \\
 & \hphantom{\doteq\I}\I\left\{ -\alpha+f_{V}(\overline{Q})-f_{W}(\tilde{Q})+R-I(\overline{Q})-\epsilon\geq0\right\} ,
\end{align}
where $(a)$ is now since conditioned on ${\cal G}_{n}$, $N(Q|\mathbf{y})$
is a binomial random variable, with probability of success $\doteq e^{-nI(Q)}$
(see Lemma \ref{lem: conditional Ber}), and less than $e^{nR}$ trials
(whenever $Q_{Y}=\overline{Q}_{Y}$, and $N(Q|\mathbf{y})=0$ otherwise),
and by utilizing again Lemma \ref{lem: behavior of intersection}
and Remark \ref{rem: continuity of S}. As $\epsilon>0$ is arbitrary,
\begin{align}
\zeta(\overline{Q}) & \doteq\I\left\{ \mathbf{S}(\tilde{Q}_{Y};-\alpha+f_{V}(\overline{Q})-f_{W}(Q)+R-I(\overline{Q}),{\cal Q}_{W})>R\right\} \times\nonumber \\
 & \hphantom{\doteq}\I\left\{ -\alpha+f_{V}(\overline{Q})-f_{W}(\tilde{Q})+R-I(\overline{Q})>0\right\} 
\end{align}
\uline{Case 2:}\emph{ }Assume that $I(\overline{Q})=0$. This case
is not significantly different from Case 1. Indeed, for any $0<\epsilon<R$,
let 
\[
{\cal G}_{n}\dfn\left\{ e^{n(R-\epsilon)}\leq N(\overline{Q}|\mathbf{y})\leq\frac{1}{2}e^{nR}\right\} ,
\]
then $\P\left[{\cal G}_{n}\right]\doteq1$. To see this, we note that
for $\mathbf{X}_{l}$ drawn uniformly within ${\cal T}(P_{X})$. 
\begin{align}
\E\left[N(\overline{Q}|\mathbf{y})\right] & =e^{nR}\cdot\P\left(\hat{Q}_{\mathbf{X}_{l}\mathbf{y}}=\overline{Q}\right)\\
 & \trre[\leq,a]\frac{1}{4}e^{nR}
\end{align}
for all $n$ sufficiently large, where $(a)$ is since $\P\left(\hat{Q}_{\mathbf{X}_{l}\mathbf{y}}=Q\right)\to0$
as $n\to\infty$. So, by Markov inequality 
\begin{align}
\P\left\{ N(\overline{Q}|\mathbf{y})\leq\frac{1}{2}e^{nR}\right\}  & \geq\P\left\{ N(\overline{Q}|\mathbf{y})\leq2\E\left[N(\overline{Q}|\mathbf{y})\right]\right\} \geq\frac{1}{2}.
\end{align}
Since, as before $\P\left\{ e^{n(R-\epsilon)}\leq N(\overline{Q}|\mathbf{y})\right\} \doteq1$,
and the intersection of two high probability sets also has high probability,
we obtain $\P\left[{\cal G}_{n}\right]\doteq1$. The rest of the analysis
follows as in Case 1, and the result is the same, when setting $I(\overline{Q})=0$.

\uline{Case 3:} Assume that $I(\overline{Q})>R$. Then, for any
$\epsilon>0$ 
\begin{align}
\zeta(\overline{Q}) & =\P\left(\bigcap_{Q\neq\overline{Q}:\, f_{W}(Q)>-\infty}\left\{ N(Q|\mathbf{y})\leq e^{n\left[-\alpha+f_{V}(\overline{Q})-f_{W}(Q)\right]}\cdot N(\overline{Q}|\mathbf{y})\right\} \cap\right.\nonumber \\
 & \hphantom{\doteq\P}\left.\vphantom{\bigcap_{Q\neq\overline{Q}:\, f_{W}(Q)>-\infty}\left\{ N(Q|\mathbf{y})\leq e^{n\left[-\alpha+f_{V}(\overline{Q})-f_{W}(Q)\right]}\cdot N(\overline{Q}|\mathbf{y})\right\} \cap}\left\{ 1\leq e^{n\left[-\alpha+f_{V}(\overline{Q})-f_{W}(\tilde{Q})\right]}\cdot N(\overline{Q}|\mathbf{y})\right\} \right)\\
 & \trre[\doteq,a]\P\left(\bigcap_{Q\neq\overline{Q}:\, f_{W}(Q)>-\infty}\left\{ N(Q|\mathbf{y})\leq e^{n\left[-\alpha+f_{V}(\overline{Q})-f_{W}(Q)\right]}\cdot N(\overline{Q}|\mathbf{y})\right\} \cap\right.\nonumber \\
 & \hphantom{\doteq\P}\left.\vphantom{\bigcap_{Q\neq\overline{Q}}\left\{ N(Q|\mathbf{y})e^{nf_{W}(Q)}\leq e^{-n\alpha}\cdot N(\overline{Q}|\mathbf{y})e^{nf_{V}(\overline{Q})}\right\} \cap}\left\{ 1\leq e^{n\left[-\alpha+f_{V}(\overline{Q})-f_{W}(\tilde{Q})\right]}\cdot N(\overline{Q}|\mathbf{y})\right\} |1\leq N(\overline{Q}|\mathbf{y})\leq e^{n\epsilon}\right)\P\left(1\leq N(\overline{Q}|\mathbf{y})\leq e^{n\epsilon}\right)\\
 & \trre[\dot{\geq},b]\P\left(\bigcap_{Q\neq\overline{Q}:\, f_{W}(Q)>-\infty}\left\{ N(Q|\mathbf{y})\leq e^{n\left[-\alpha+f_{V}(\overline{Q})-f_{W}(Q)\right]}\right\} \cap\right.\nonumber \\
 & \hphantom{\doteq\P}\left.\vphantom{\bigcap_{Q\neq\overline{Q}}\left\{ N(Q|\mathbf{y})e^{nf_{W}(Q)}\leq e^{-n\alpha}\cdot N(\overline{Q}|\mathbf{y})e^{nf_{V}(\overline{Q})}\right\} \cap}\left\{ 1\leq e^{n\left[-\alpha+f_{V}(\overline{Q})-f_{W}(\tilde{Q})\right]}\right\} |1\leq N(\overline{Q}|\mathbf{y})\leq e^{n\epsilon}\right)e^{-n\left(I(\overline{Q})-R\right)}\\
 & \trre[\doteq,c]\I\left\{ \mathbf{S}(\tilde{Q}_{Y};-\alpha+f_{V}(\overline{Q})-f_{W}(Q),{\cal Q}_{W})>R\right\} \I\left\{ -\alpha+f_{V}(\overline{Q})-f_{W}(\tilde{Q})\geq0\right\} e^{-n\left(I(\overline{Q})-R\right)},
\end{align}
where $(a)$ is since conditioned on $N(\overline{Q}|\mathbf{y})=0$
the probability of the event is $0$, and 
\[
\P\left[N(\overline{Q}|\mathbf{y})\geq e^{n\epsilon}\right]\doteq0,
\]
$(b)$ is since
\begin{align}
\P\left(1\leq N(\overline{Q}|\mathbf{y})\leq e^{n\epsilon}\right) & \geq\P\left(N(\overline{Q}|\mathbf{y})=1\right)\\
 & \doteq e^{-n\left(I(\overline{Q})-R\right)},
\end{align}
and $(c)$ is since conditioned on $1\leq N(\overline{Q}|\mathbf{y})\leq e^{n\epsilon}$,
$N(Q|\mathbf{y})$ is a binomial random variable, with probability
of success $\doteq e^{-nI(Q)}$ (see Lemma \ref{lem: conditional Ber}),
and $\doteq e^{nR}$ trials (whenever $Q_{Y}=\overline{Q}_{Y}$, and
$N(Q|\mathbf{y})=0$ otherwise), and by utilizing once again Lemma
\ref{lem: behavior of intersection} and Remark \ref{rem: continuity of S}.
Similarly, using 
\[
\P\left(1\leq N(\overline{Q}|\mathbf{y})\leq e^{n\epsilon}\right)\leq e^{n\epsilon}\P\left(N(\overline{Q}|\mathbf{y})=1\right)\doteq e^{-n\left(I(\overline{Q})-R-\epsilon\right)},
\]
the same analysis as in the previous case, shows a reversed inequality.
As $\epsilon>0$ is arbitrary, then
\[
\zeta(\overline{Q})\doteq\I\left\{ \mathbf{S}(\tilde{Q}_{Y};-\alpha+f_{V}(\overline{Q})-f_{W}(Q),{\cal Q}_{W})>R\right\} \I\left\{ -\alpha+f_{V}(\overline{Q})-f_{W}(\tilde{Q})>0\right\} e^{-n\left(I(\overline{Q})-R\right)}.
\]
Returning to \eqref{eq: B term}, we obtain that $B(\tilde{Q})$ is
exponentially equal to the maximum between
\[
\max_{\overline{Q}:\, f_{W}(\overline{Q})<-\alpha+f_{V}(\overline{Q}),\, I(\overline{Q})\leq R,\, f_{V}(\overline{Q})>\alpha+f_{W}(\tilde{Q})-R+I(\overline{Q})}\I\left\{ \mathbf{S}(\tilde{Q}_{Y};-\alpha+f_{V}(\overline{Q})-f_{W}(Q)+R-I(\overline{Q}),{\cal Q}_{W})>R\right\} ,
\]
and
\[
\max_{\overline{Q}:\, f_{W}(\overline{Q})<-\alpha+f_{V}(\overline{Q}),\, I(\overline{Q})>R,\, f_{V}(\overline{Q})>\alpha+f_{W}(\tilde{Q})}\I\left\{ \mathbf{S}(\tilde{Q}_{Y};-\alpha+f_{V}(\overline{Q})-f_{W}(Q),{\cal Q}_{W})>R\right\} e^{-n\left(I(\overline{Q})-R\right)},
\]
or, more succinctly,
\[
B(\tilde{Q})=\max_{\overline{Q}}\I\left\{ \mathbf{S}(\tilde{Q}_{Y};-\alpha+f_{V}(\overline{Q})-f_{W}(Q)+\left[R-I(\overline{Q})\right]_{+},{\cal Q}_{W})>R\right\} e^{-n\left[I(\overline{Q})-R\right]_{+}}
\]
where the maximization is over
\[
\left\{ \overline{Q}:\, f_{W}(\overline{Q})<-\alpha+f_{V}(\overline{Q}),\, f_{V}(\overline{Q})>\alpha+f_{W}(\tilde{Q})-\left[R-I(\overline{Q})\right]_{+}\right\} .
\]
Now, in the evaluation of $B(\tilde{Q})$ we have assumed that $e^{nf_{W}(\tilde{Q})}>0$.
However, there is no need to analyze the case $e^{nf_{W}(\tilde{Q})}=0$
since as
\[
f_{W}(\tilde{Q})=-D(\tilde{Q}_{Y|X}||W|P_{X})-H_{\tilde{Q}}(Y|X)
\]
and $H_{\tilde{Q}}(Y|X)\leq\log|{\cal Y}|<\infty$, then $e^{nf_{W}(\tilde{Q})}=0$
implies $\P(\hat{Q}_{\mathbf{x}_{1}\mathbf{y}}=\tilde{Q})\doteq\exp\left[-nD(\tilde{Q}_{Y|X}||W|P_{X})\right]\doteq e^{-n\infty}$.
Thus, upon averaging over $(\mathbf{X}_{1},\mathbf{Y})$ we obtain
the exponent $E_{B}$ of \eqref{eq: EB}, utilizing \eqref{eq: s bold lower case}.
Then, we obtain the required result from \eqref{eq: FA intial derivation}.

Next, for the MD exponent, we observe that as $E_{\st[TE]}^{\st[RC]}\left(R,\alpha,P_{X},W,V\right)$
is continuous in $\alpha$, Fact \ref{fact: alternative expression for MD exponent}
above implies that the MD exponent will be also continuous in $\alpha$.
So, Proposition \ref{prop: exponents balance} implies that when the
codewords are drawn from a fixed composition ensemble with distribution
$P_{X}$,
\[
\lim_{l\to\infty}-\frac{1}{n_{l}}\log\overline{P_{\st[MD]}}({\cal C}_{n_{l}},\phi^{*})=E_{\st[TE]}^{\st[RC]}\left(R,\alpha,P_{X},W,V\right)-\alpha.
\]

Finally, the continuity of $E_{\st[TE]}^{\st[RC]}\left(R,\alpha,P_{X},W,V\right)$
in $P_{X}$ implies that for all sufficiently large $n$, one can
find a distribution $P_{X}'$ close enough to $P_{X}$ such that \eqref{eq: FA exponent RC type-enumeration thm}
and \eqref{eq: MD exponent RC type-enumaration thm} hold, which completes
the proof of the theorem.
\end{IEEEproof}
To keep the flow of the proof, we have omitted a technical point which
we now address.
\begin{rem}
\label{rem:uniform continuity}The ensemble average FA probability
should be obtained by averaging $\overline{P_{\st[FA]}}(\mathbf{X}_{1},\mathbf{Y})$
w.r.t. $(\mathbf{X}_{1},\mathbf{Y})$. However, we have averaged its
asymptotic equivalence in the exponential scale, resulting from analyzing
the terms $A(\tilde{Q})$ and $B(\tilde{Q})$. Thus, in a sense, we
have interchanged the expectation and limit order. This is possible
due to the fact that all the asymptotic equivalence relations become
tight for $n$ sufficiently large, which \emph{does not depend on
$\tilde{Q}$} (i.e. on $(\mathbf{X}_{1},\mathbf{Y})$). Indeed, the
union and intersection rules add a negligible term to the exponent.
This term depends only on the number of types, which is polynomial
in $n$, independent of the specific type $\tilde{Q}$. The asymptotic
equivalence relations that stem from Lemma \ref{lem: behavior of intersection}
do not depend on $\tilde{Q}$, as functions of $\tilde{Q}$ only play
the role of bounds on the sums of weighted type enumerators. Indeed,
it is evident from the proof of Lemma \ref{lem: behavior of intersection}
that the required blocklength $n$ to approach convergence of the
probability does not depend on $J(Q)$.
\end{rem}

\subsection{Expurgated Exponents\label{sub:An-Expurgated-Bound}}

We begin again with several definitions. Throughout, $P_{X\tilde{X}}$
will represent a joint type of a pair of codewords. Let us define
the Chernoff distance%
\footnote{When $s$ is maximized, then the result is the Chernoff information
\cite[Section 11.9]{Cover:2006:EIT:1146355}. For $s=\frac{1}{2}$
this is the Bhattacharyya distance.%
}
\begin{equation}
d_{s}(x,\tilde{x})\dfn-\log\left(\sum_{y\in{\cal Y}}W^{1-s}(y|x)V^{s}(y|\tilde{x})\right)\label{eq: Chernoff distance}
\end{equation}
and the set
\begin{equation}
{\cal L}\dfn\left\{ P_{X\tilde{X}}:\; P_{\tilde{X}}=P_{X},\, I(P_{X\tilde{X}})\leq R\right\} .\label{eq: L defintion (for expurgated)}
\end{equation}
In addition, let us define the \emph{type-enumeration detection expurgated
exponent }as
\begin{equation}
E_{\st[TE]}^{\st[EX]}\left(R,\alpha,P_{X},W,V\right)\dfn\max_{0\leq s\leq1}\min_{P_{X\tilde{X}}\in{\cal L}}\left\{ \alpha s+\E\left[d_{s}(X,\tilde{X})\right]+I(P_{X\tilde{X}})-R\right\} .\label{eq: Detection exponent EX enumeration}
\end{equation}

\begin{thm}
\label{thm: EX bound type-enumeration}Let a distribution $P_{X}$
and a parameter $\alpha\in\mathbb{R}$ be given. Then, there exists
a sequence of codes ${\cal C}=\{{\cal C}_{n}\}_{n=1}^{\infty}$ of
rate $R$ such that for any $\delta>0$
\begin{equation}
E_{\st[FA]}\left({\cal C},\phi^{*}\right)\geq E_{\st[TE]}^{\st[EX]}\left(R,\alpha,P_{X},W,V\right)-\delta,\label{eq: FA exponent EX type-enumeration thm}
\end{equation}
\begin{equation}
E_{\st[MD]}\left({\cal C},\phi^{*}\right)\geq E_{\st[TE]}^{\st[EX]}\left(R,\alpha,P_{X},W,V\right)-\alpha-\delta.\label{eq: MD exponent EX type-enumaration thm}
\end{equation}

\end{thm}
The proof can be found in Appendix \ref{sec:Proof-of-Theorem EX TE}.
\begin{rem}
H\"{o}lder inequality shows that $d_{s}(x,\tilde{x})\geq0$. In \eqref{eq: Detection exponent EX enumeration},
there is freedom to maximize over $0\leq s\leq1$, and naturally,
$s=\frac{1}{2}$ is a valid choice. Due to the symmetry of $d_{s}(x,\tilde{x})$
in $s$ around $s=\frac{1}{2}$ when $W=V$, for the ordinary decoding
exponent, the optimal choice is $s=\frac{1}{2}$ (as also manifested
at $R=0$ by the Shannon-Gallager-Berlekamp upper bound \cite[Theorem 4]{shannon1967lowerII}),
but here, no such symmetry exists.
\end{rem}

\begin{rem}
In Theorem \ref{thm: EX bound type-enumeration} we have assumed a
fixed composition code of type $P_{X}$. As discussed in \cite[Problem 10.23 (b)]{csiszar2011information},
for ordinary decoding, the exponent \eqref{eq: Detection exponent EX enumeration}
is at least as large as the corresponding exponent using Gallager's
approach to expurgation \cite[Section 5.7]{gallager1968information},
and for the maximizing $P_{X}$, the two bounds coincide. Thus, for
ordinary decoding, the exponent bound \eqref{eq: Detection exponent EX enumeration}
offers an improvement over Gallager's approach when the input type
$P_{X}$ is constrained. For joint detection/decoding, there is an
additional source of possible improvement - the input type $P_{X}$
which best suits channel coding is not necessarily the best input
type for the detection problem. We also mention that for $R=0$, an
improvement at any given $P_{X}$ can be obtained by taking the \emph{upper
concave envelope} of \eqref{eq: Detection exponent EX enumeration}
(see \cite[Problem 10.22]{csiszar2011information} and the discussion
in \cite[Section II]{merhav2014zero_rate}).
\end{rem}

\begin{rem}
This expurgation technique can be used also for continuous alphabet
channels, and specifically, for AWGN channels, see \cite[Section 4]{Merhav13_List}.
\end{rem}

\subsection{Exact Random Coding Exponents of Simplified Detectors/Decoders\label{sub:Exponents-Simplified-Decoders}}

We now discuss the random coding exponents achieved by the simplified
detectors/decoders $\phi_{\st[L]}$ and $\phi_{\st[H]}$ introduced
in Subsection \ref{sub:Simplified-Detector/Decoders}. We begin with
$\phi_{\st[L]}$. For $\gamma\in\mathbb{R}$, let us define
\begin{equation}
\mathbf{t}(\tilde{Q}_{Y},\gamma)\dfn\min_{Q\in{\cal Q}_{W}:\, Q=\tilde{Q}_{Y},-\alpha-f_{W}(Q)+\gamma\leq0}I(Q),\label{eq: t bold lower case}
\end{equation}
the sets ${\cal J}_{1,\st[L]}\dfn{\cal J}_{1}$ and 
\[
{\cal J}_{2,\st[L]}\dfn\left\{ \tilde{Q}:\;\mathbf{t}\left(\tilde{Q}_{Y},f_{V}(\tilde{Q})\right)\geq R\right\} ,
\]
the exponent
\begin{equation}
E_{A,\st[L]}\dfn\min_{\tilde{Q}\cap_{i=1}^{2}{\cal J}_{i,\st[L]}}D(\tilde{Q}{}_{Y|X}\|W|P_{X}),\label{eq: EA low rates}
\end{equation}
the sets ${\cal K}_{1,\st[L]}\dfn{\cal K}_{1}$, ${\cal K}_{2,\st[L]}\dfn{\cal K}_{2}$%
\footnote{It can be noticed that the only difference between ${\cal K}_{3,\st[L]},{\cal K}_{4,\st[L]}$
and ${\cal K}_{3},{\cal K}_{4}$ are the exclusion of $I(Q)-R$ terms
and replacing $\mathbf{s}(\tilde{Q}_{Y},\gamma)$ with $\mathbf{t}(\tilde{Q}_{Y},\gamma)$.%
}
\[
{\cal K}_{3,\st[L]}\dfn\left\{ (\tilde{Q},\overline{Q}):\; f_{V}(\overline{Q})\geq\alpha+f_{W}(\tilde{Q})\right\} ,
\]
\[
{\cal K}_{4,\st[L]}\dfn\left\{ (\tilde{Q},\overline{Q}):\;\mathbf{t}\left(\tilde{Q}_{Y},f_{V}(\overline{Q})\right)\geq R\right\} ,
\]
and the exponent
\begin{equation}
E_{B,\st[L]}\dfn\min_{(\tilde{Q},\overline{Q})\in\cap_{i=1}^{4}{\cal K}_{i,\st[L]}}D(\tilde{Q}{}_{Y|X}\|W|P_{X})+\left[I(\overline{Q})-R\right]_{+}.\label{eq: EB low rates}
\end{equation}
In addition, let us define the \emph{low-rate detection random coding
exponent} as
\begin{equation}
E_{\st[L]}^{\st[RC]}\left(R,\alpha,P_{X},W,V\right)\dfn\min\left\{ E_{A,\st[L]},E_{B,\st[L]}\right\} .\label{eq: Detection exponent RC enumeration low rates}
\end{equation}

\begin{thm}
\label{thm: RC bound type-enumeration simplified low}Let a distribution
$P_{X}$ and a parameter $\alpha\geq0$ be given. Then, there exists
a sequence of codes ${\cal C}=\{{\cal C}_{n}\}_{n=1}^{\infty}$ of
rate $R$ such that for any $\delta>0$
\begin{equation}
E_{\st[FA]}\left({\cal C},\phi^{*}\right)\geq E_{\st[L]}^{\st[RC]}\left(R,\alpha,P_{X},W,V\right)-\delta,\label{eq: FA exponent RC type-enumeration thm low rates}
\end{equation}
\begin{equation}
E_{\st[MD]}\left({\cal C},\phi^{*}\right)\geq E_{\st[L]}^{\st[RC]}\left(R,-\alpha,P_{X},V,W\right)-\delta.\label{eq: MD exponent RC type-enumaration thm low rates}
\end{equation}

\end{thm}
The proof can be found in Appendix \ref{sec:Proof-of-Theorem RC low TE}.

Next, we discuss the random coding exponents of $\phi_{\st[H]}$.
As this is a simple hypothesis testing between two memoryless sources
$\tilde{W}$ and $\tilde{V}$, the standard analysis \cite{blahut1974hypothesis}
and \cite[Section 11.7]{Cover:2006:EIT:1146355} is applicable verbatim.
For given $0\leq\mu\leq1$, let
\[
Q_{\mu}(y)\dfn\frac{\tilde{W}^{\mu}(y)\tilde{V}^{1-\mu}(y)}{\sum_{y'\in{\cal Y}}\tilde{W}^{\mu}(y')\tilde{V}^{1-\mu}(y')}
\]
for all $x\in{\cal X}$, and let us define the high-rate detection
random coding exponent as 
\[
E_{\st[H]}^{\st[RC]}\left(R,\alpha,P_{X},W,V\right)\dfn D(Q_{\mu(\alpha)}||\tilde{W}),
\]
where $\mu(\alpha)$ is chosen so that
\[
D(Q_{\mu(\alpha)}||\tilde{W})-D(Q_{\mu(\alpha)}||\tilde{V})=-\alpha.
\]

\begin{thm}
\label{thm: RC bound type-enumeration simplified high}Let a distribution
$P_{X}$ and a parameter $\alpha\geq0$ be given. Then, there exists
a sequence of codes ${\cal C}=\{{\cal C}_{n}\}_{n=1}^{\infty}$ of
rate $R$ such that for any $\delta>0$
\begin{equation}
E_{\st[FA]}\left({\cal C},\phi^{*}\right)\geq E_{\st[H]}^{\st[RC]}\left(R,\alpha,P_{X},W,V\right)-\delta,\label{eq: FA exponent RC type-enumeration thm high rate}
\end{equation}
\begin{equation}
E_{\st[MD]}\left({\cal C},\phi^{*}\right)\geq E_{\st[H]}^{\st[RC]}\left(R,\alpha,P_{X},W,V\right)-\alpha-\delta.\label{eq: MD exponent RC type-enumaration thm high rate}
\end{equation}
\end{thm}
\begin{IEEEproof}
The proof follows the standard analysis in \cite[Section 11.7]{Cover:2006:EIT:1146355}.\end{IEEEproof}
\begin{rem}
The decoder $\phi_{\st[H]}$ and its random coding exponents do not
depend on the rate $R$.
\end{rem}

\subsection{Gallager/Forney-Style Exponents\label{sub:Gallager-Forney-Style-Bounds}}

Next, we derive achievable exponents using the classical Gallager/Forney
technique.

\subsubsection{Random Coding Exponents}

For a given distribution $\{P_{X}(x)\}_{x\in{\cal X}}$, and parameters
$s,\rho$, define
\begin{equation}
E_{0}'(s,\rho)\dfn-\log\left[\sum_{y\in{\cal Y}}\left(\sum_{x\in{\cal X}}P_{X}(x)W^{\nicefrac{(1-s)}{\rho}}(y|x)V^{\nicefrac{s}{\rho}}(y|x)\right)^{\rho}\right],\label{eq: E0'}
\end{equation}
and
\begin{equation}
E_{0}''(s,\rho)\dfn-\log\left[\sum_{y\in{\cal Y}}\left(\sum_{x\in{\cal X}}P_{X}(x)W^{\nicefrac{(1-s)}{\rho}}(y|x)\right)^{\rho}\left(\sum_{x\in{\cal X}}P_{X}(x)V^{\nicefrac{s}{\rho}}(y|x)\right)^{\rho}\right],\label{eq: E0''}
\end{equation}
and let the \emph{Gallager/Forney detection random coding exponent}
be defined as
\begin{align}
E_{\st[GF]}^{\st[RC]}\left(R,\alpha,P_{X},W,V\right) & \dfn\max_{0\leq s\leq1,\max\left\{ s,1-s\right\} \leq\rho\leq1}\min\left\{ \alpha s+E_{0}'(s,\rho)-(\rho-1)R,\right.\nonumber \\
 & \hphantom{\dfn\max_{0\leq s\leq1,\max\left\{ s,1-s\right\} \leq\rho\leq1}\min\{}\left.\alpha s+E_{0}''(s,\rho)-(2\rho-1)R\right\} .\label{eq: Detection exponent RC Gallager}
\end{align}

\begin{thm}
\label{thm: RC bound Gallager}Let a distribution $P_{X}$ and a parameter
$\alpha\in\mathbb{R}$ be given. Then, there exists a sequence of
codes ${\cal C}=\{{\cal C}_{n}\}_{n=1}^{\infty}$ of rate $R$ such
that for any $\delta>0$
\begin{equation}
E_{\st[FA]}\left({\cal C},\phi^{*}\right)\geq E_{\st[GF]}^{\st[RC]}\left(R,\alpha,P_{X},W,V\right)-\delta,\label{eq: FA exponent RC Gallager thm}
\end{equation}
\begin{equation}
E_{\st[MD]}\left({\cal C},\phi^{*}\right)\geq E_{\st[GF]}^{\st[RC]}\left(R,\alpha,P_{X},W,V\right)-\alpha-\delta.\label{eq: MD exponent RC Gallager thm}
\end{equation}

\end{thm}
The proof can be found in Appendix \ref{sec:Proof-of-Theorem RC GF}.

\subsubsection{Expurgated Exponents}

For a given distribution $\{P_{X}(x)\}_{x\in{\cal X}}$ and parameters
$s,\rho$, define
\begin{equation}
E_{x}'(s)\dfn-\log\left[\sum_{x\in{\cal X}}P_{X}(x)\sum_{y\in{\cal Y}}W^{1-s}(y|x)V^{s}(y|x)\right],\label{eq: Ex'}
\end{equation}
and
\begin{equation}
E_{x}''(s)\dfn-\log\left[\sum_{y\in{\cal Y}}\left(\sum_{x\in{\cal X}}P_{X}(x)W^{1-s}(y|x)\right)\left(\sum_{x\in{\cal X}}P_{X}(x)V^{s}(y|x)\right)\right],\label{eq: Ex''}
\end{equation}
and let the \emph{Gallager/Forney detection expurgated exponent} be
defined as
\begin{equation}
E_{\st[GF]}^{\st[EX]}\left(R,\alpha,P_{X},W,V\right)\dfn\sup_{0\leq s\leq1,\rho\geq1}\min\left\{ s\alpha+E_{x}'(s),s\alpha+E_{x}''(s)-\rho R\right\} .\label{eq: Detection exponent EX Gallager}
\end{equation}

\begin{thm}
\label{thm: Expurgated bound Gallager}Let a distribution $P_{X}$
and a parameter $\alpha\in\mathbb{R}$ be given. Then, there exists
a sequence of codes ${\cal C}=\{{\cal C}_{n}\}_{n=1}^{\infty}$ of
rate $R$ such that for any $\delta>0$
\begin{equation}
E_{\st[FA]}\left({\cal C},\phi^{*}\right)\geq E_{\st[GF]}^{\st[EX]}\left(R,\alpha,P_{X},W,V\right)-\delta,\label{eq: FA exponent EX Gallager thm}
\end{equation}
\begin{equation}
E_{\st[MD]}\left({\cal C},\phi^{*}\right)\geq E_{\st[GF]}^{\st[EX]}\left(R,\alpha,P_{X},W,V\right)-\alpha-\delta.\label{eq: MD exponent EX Gallager thm}
\end{equation}

\end{thm}
The proof can be found in Appendix \ref{sec:Proof-of-Theorem EX GF}.

\subsection{Discussion}

We summarize this section with the following discussion.
\begin{enumerate}
\item \emph{Monotonicity in the rate}:\emph{ }The ordinary random coding
exponents are decreasing with the rate $R$, and vanish at $I(P_{X}\times W)$.
By contrast, the detection exponents are not necessarily so. Indeed,
the exponent $E_{A}$ of \eqref{eq: EA} is increasing with the rate.
For the exponent $E_{B}$ of \eqref{eq: EB}, as $R$ increases, the
objective function decreases and ${\cal K}_{3}$ expands, but the
set ${\cal K}_{4}$ diminishes%
\footnote{As its r.h.s. always increases, but its l.h.s. does not.%
}, and so no monotonicity is assured for $E_{B}$, and as a results,
also for $E_{\st[TE]}^{\st[RC]}\left(R,\alpha,P_{X},W,V\right)$.
The same holds for $\phi_{\st[L]}$, whereas $\phi_{\st[H]}$ does
not depend on $R$ at all. The expurgated exponent $E_{\st[TE]}^{\st[EX]}\left(R,\alpha,P_{X},W,V\right)$
of \eqref{eq: Detection exponent EX enumeration} decreases in $R$.
To gain intuition, recall from \eqref{eq: large deviations of type enumerator},
that when $I(Q)<R$ the type enumerator $N(Q|\mathbf{y})$ concentrates
double-exponentially rapidly around its average $\doteq\exp\left[n(R-I(Q))\right]$.
Thus, for any given $\mathbf{y}$, an increase of the rate will introduce
codewords having a joint type that was not typically seen at lower
rates, and this new joint type might dominate one of the likelihoods.
However, it is not clear to which direction this new type will tip
the scale in the likelihoods comparison, and so the rate increase
does not necessarily imply an increase or a decrease of one of the
exponents. In addition, the above discussion and \eqref{eq: IE probability is minimum betweeen ordinary and FA}
imply that the largest achievable rate such that $P_{\st[IE]}\to0$
as $n\to\infty$, may still be the mutual information $I(P_{X}\times W)$,
or, in other words, the detection does not cause a rate loss.
\item \emph{Computation of the exponents}:\emph{ }Unfortunately, the optimization
problems involved in computing the exact exponents of Subsections
\ref{sub:A-Tight-Random} and \ref{sub:Exponents-Simplified-Decoders}
are usually not convex, and might be complex to solve when the alphabets
are large. For example, for the exact exponents, computing $E_{A}$
of \eqref{eq: EA} is not a convex optimization problem since ${\cal J}_{2}$
is not a convex set of $\tilde{Q}$, and computing $E_{B}$ of \eqref{eq: EB}
is not a convex optimization problem since ${\cal K}_{3}$ and ${\cal K}_{4}$
are not convex sets of $(\tilde{Q},\overline{Q})$, and not even of
$(\tilde{Q}_{Y|X},\overline{Q}_{Y|X})$. An efficient algorithm their
efficient computation is an important open problem. However, the expurgated
exponent \eqref{eq: Detection exponent EX enumeration} is concave%
\footnote{The second derivative w.r.t. $s$ of $d_{s}(x,\tilde{x})$ is the
variance of $\log\frac{V(y|\tilde{x})}{W(y|x)}$ w.r.t. the distribution
$P_{Y}$ which satisfies $P_{Y}(y)\propto W^{1-s}(y|x)V^{s}(y|\tilde{x})$.%
} in $s$ and convex in $P_{X\tilde{X}}$. This promotes the importance
of the lower bounds derived in Subsection \ref{sub:Gallager-Forney-Style-Bounds},
which only require two-dimensional optimization problems, irrespective
of the alphabet sizes.
\item \emph{Choice of input distribution}: Thus far, the input distribution
$P_{X}$ was assumed fixed, but it can obviously be optimized. Nonetheless,
there might be a tension between the optimal choice for channel coding
versus the optimal choice for detection. For example, consider the
detection problem between $W$, a Z-channel, i.e. $W(0|0)=1,W(0|1)=w$
for some $0\leq w\leq1$, and $V$, an S-channel, i.e. $V(1|0)=v,V(1|1)=1$
for some $0\leq v\leq1$. Choosing $P_{X}(0)=1$ will result an infinite
FA and MD exponents (upon appropriate choice of $\alpha$), but is
useless from the channel coding perspective. One possible remedy is
to define a Lagrangian that weighs, e.g. the FA and ordinary decoding
exponents with some weight, and optimize it over the input type. However,
still, the resulting optimization might be non-tractable.
\item \emph{Simplified decoders}: Intuitively, the low-rate simplified detector/decoder
$\phi_{\st[L]}$ has worse FA-MD trade-off than the optimal detector/decoder
$\phi'$ since the effect of a non-typical codeword may be averaged
out in $\frac{1}{M}\sum_{m=1}^{M}W(\mathbf{y}|\mathbf{x}_{m})$, but
may totally change $\max_{1\leq m\leq M}W(\mathbf{y}|\mathbf{x}_{m}).$
However, there exists a \emph{critical rate} $R_{\st[cr]}$ such that
for all $R\leq R_{\st[cr]}$ the exponents of the two detectors/decoders
coincide, when using the same parameter $\alpha$. To see this, first
let 
\[
\tilde{Q}_{A}\dfn\argmin_{\tilde{Q}\in{\cal J}_{1}}D(\tilde{Q}{}_{Y|X}\|W|P_{X}),
\]
i.e. the exponent $E_{A}$ for $R=0$, and in fact, for all rates
satisfying
\[
R\leq\mathbf{s}\left(\tilde{Q}_{Y};f_{V}(\tilde{Q}_{A})\right)\dfn R_{\st[cr,A]}.
\]
Since from Remark \ref{rem: J function convex indicator} (Appendix
\ref{sec:Proof-of-Theorem RC low TE}) 
\[
\mathbf{s}(\tilde{Q}_{Y},\gamma)\leq\mathbf{t}(\tilde{Q}_{Y},\gamma)
\]
this is also the exponent $E_{A,\st[L]}$. Now, letting $R=0$ in
$\{{\cal K}_{i}\}_{i=3}^{4}$ and then solving
\[
(\tilde{Q}_{B},\overline{Q}_{B})\dfn\argmin{}_{(\tilde{Q},\overline{Q})\in\cap_{i=1}^{4}{\cal K}_{i}}\left\{ D(\tilde{Q}{}_{Y|X}\|W|P_{X})+I(\overline{Q})\right\} 
\]
we get the exponent $E_{B}$ for $R=0$, and in fact, for all rates
satisfying 
\[
R\leq\min\left\{ I(\overline{Q}_{B}),\mathbf{s}\left(\tilde{Q}_{Y};f_{V}(\overline{Q}_{B})\right)\right\} \dfn R_{\st[cr,B]}.
\]
Similarly, this is also the exponent $E_{B,\st[L]}$. In conclusion,
for all $R\leq R_{\st[cr]}\dfn\min\left\{ R_{\st[cr,A]},R_{\st[cr,B]}\right\} $
it is assured that the FA exponents of $\phi'$ and $\phi_{\st[L]}$
are exactly the same. In the same manner, a critical rate can be found
for the MD exponent. For the the high-rate simplified detector/decoder
$\phi_{\st[H]}$ we only remark that in some cases,  the output distributions
$\tilde{W}$ and $\tilde{V}$ may be equal, and so this detector/decoder
is useless, even though $\phi'$ achieves strictly positive exponents
(cf. the example in Section \ref{sec:Examples}).
\item \emph{Continuous alphabet channels}: As previously mentioned, one
of the advantages of the Gallager/Forney-Style bounds is their simple
generalization to continuous channels with input constraints. We briefly
describe this well known technique \cite[Chapter 7]{gallager1968information}.
For concreteness, let us focus on the power constraint $\E[X^{2}]\leq1$.
In this technique a one-dimensional input distribution is chosen,
say with density $f_{X}(x)$, which satisfies the input constraint.
Then, an $n$-dimensional distribution is defined as follows
\[
f_{n}(\mathbf{x})=\psi^{-1}\I\left\{ n-\delta\leq\sum_{i=1}^{n}x_{m,i}^{2}\leq n\right\} \prod_{i=1}^{n}P_{X}(x_{i}),
\]
where $\psi$ is a normalization factor. This distribution corresponds
to a uniform distribution over a thin $n$-dimensional spherical shell,
which is the surface of the $n$-dimensional `ball' of sequences which
satisfy the input constraint. While this input distribution is not
memoryless, it is easily upper bounded by a memoryless distribution:
by introducing a parameter $r\geq0$, and using 
\[
\I\left\{ n-\delta\leq\sum_{i=1}^{n}x_{m,i}\leq n\right\} \leq\exp\left[r\cdot\left(\sum_{i=1}^{n}x_{m,i}^{2}-n+\delta\right)\right]
\]
we get 
\begin{equation}
f_{n}(\mathbf{x})\leq\psi^{-1}e^{r\delta}\prod_{i=1}^{n}P_{X}(x_{i})e^{r\left[x_{i}^{2}-1\right]}.\label{eq: upper bound on pdf-1}
\end{equation}
Now, e.g., in the derivation in \eqref{eq: Psi def} we may use 
\begin{align}
\E\left[W^{\nicefrac{(1-s)}{\rho}}(\mathbf{y}|\mathbf{X}_{m})V^{\nicefrac{s}{\rho}}(\mathbf{y}|\mathbf{X}_{m})\right] & =\int_{\mathbf{x}}f_{n}(\mathbf{x})W^{\nicefrac{(1-s)}{\rho}}(\mathbf{y}|\mathbf{X}_{m})V^{\nicefrac{s}{\rho}}(\mathbf{y}|\mathbf{X}_{m})d\mathbf{x}\\
 & \leq\psi^{-1}e^{r\delta}\left[\int_{x}f_{X}(x)e^{r\left[x^{2}-1\right]}W^{\nicefrac{(1-s)}{\rho}}(y_{i}|x)V^{\nicefrac{s}{\rho}}(y_{i}|x)dx\right]^{n}.
\end{align}
As discussed in \cite[p. 341]{gallager1968information}, the term
$\psi^{-1}e^{r\delta}$ is sub-exponential, and can be disregarded.
Now, the resulting exponential functions can be modified. For example,
for a pair of power constrained AWGN channels $W$ and $V$, we may
define%
\footnote{Since the additive noise has a density, the probability distributions
in the bounds of subsection \ref{sub:Gallager-Forney-Style-Bounds}
can be simply replaced by densities, and the summations can be replaced
by integrals.%
} 
\begin{equation}
E_{0}'(s,\rho,r)\dfn-\log\int_{-\infty}^{\infty}\left(\int_{-\infty}^{\infty}f_{X}(x)e^{r\left[x^{2}-1\right]}W^{\nicefrac{(1-s)}{\rho}}(y|x)V^{\nicefrac{s}{\rho}}(y|x)dx\right)^{\rho}dy,\label{eq: E0' AWGN}
\end{equation}
where the dependence in $r$ was made explicit, and similarly, 
\begin{equation}
E_{0}''(s,\rho,r_{1},r_{2})\dfn-\log\int_{-\infty}^{\infty}\left(\int_{-\infty}^{\infty}f_{X}(x)e^{r_{1}\left[x^{2}-1\right]}W^{\nicefrac{(1-s)}{\rho}}(y|x)dx\right)^{\rho}\left(\int_{-\infty}^{\infty}f_{X}(x)e^{r_{2}\left[x^{2}-1\right]}V^{\nicefrac{s}{\rho}}(y|x)dx\right)^{\rho}dy,\label{eq: E0'' AWGN}
\end{equation}
which requires two new parameters $r_{1},r_{2}$. Then, the exponent
in \eqref{eq: Detection exponent RC Gallager} can be computed exactly
in the same way, with additional maximization over non-negative $r,r_{1},r_{2}$.
To obtain an explicit bound, it is required to choose an input distribution.
The natural choice is the Gaussian distribution, which is appropriate
from the channel coding perspective%
\footnote{Nevertheless, it should be recalled that Gaussian input is optimal
at high rates (above some critical rate). At low rates, the optimal
input distribution is not known, even for pure channel coding.%
}, and also enables to obtain analytic bounds. Of course, it might
be very far from being optimal for the purpose of pure detection.
Then, the integrals in \eqref{eq: E0' AWGN} can be solved by `completing
the square' in the exponent of Gaussian distributions%
\footnote{Namely, the identities $\int_{t=-\infty}^{\infty}\exp\left[-at^{2}-bt\right]dt=\sqrt{\frac{\pi}{a}}\cdot e^{\frac{b^{2}}{4a}}$
and $\int_{t=-\infty}^{\infty}\exp\left[-a\frac{t^{2}}{2}\right]dt=\sqrt{\frac{2\pi}{a}}$. %
}, and the optimal values of $r$ and $\rho$ can be found analytically
\cite[Section 7.4]{gallager1968information}. Here, since two channels
are involved, and we also need to optimize over $s$, we have not
been able to obtain simple expressions%
\footnote{Nonetheless, for a given $s$, the expression for $E_{0}'(s,\rho,r)$
is rather similar to the ordinary decoding exponent $E_{0}(\rho,r)$
and so the optimal $\rho$ and $r$ can be analytically found.%
}. Nonetheless, the required optimization problem is only four-dimensional,
and can be easily solved by an exhaustive search. Finally, it can
be noticed that the computing the expurgated bounds is a similar problem
as 
\[
E_{x}'(s,r)=E_{0}'(s,\rho=1,r)
\]
and 
\[
E_{x}''(s,r)=E_{0}''(s,\rho=1,r).
\]

\item \emph{Comparison with }\cite{CoN}: As mentioned in the introduction
(Section \ref{sec:Introduction}), the problem studied here is a generalization
of \cite{CoN}. Indeed, when the channel $V$ does not depend on the
input, i.e. $V(\mathbf{y}|\mathbf{x})=Q_{0}(\mathbf{y})$, then the
problem studied in \cite{CoN} is obtained%
\footnote{The meaning of FA and MD here is opposite to their respective meaning
in \cite{CoN}, as sanctioned by the motivating applications.%
}. Of course, the detectors derived in Section \ref{sec:Joint-Detector/Decoders}
can be used directly for this special case. Moreover, the exponent
expressions can be slightly simplified as follows. A joint type $\tilde{Q}$
is feasible if and only if $f_{W}(P_{X}\times\tilde{Q}_{Y})\leq-\alpha+f_{V}(P_{X}\times\tilde{Q}_{Y})$,
both in $E_{A}$ of \eqref{eq: EA} and $E_{B}$ of \eqref{eq: EB},
as otherwise, the sets ${\cal J}_{2}$ and ${\cal K}_{4}$ are empty.
For any such $\tilde{Q}$ which satisfies this condition, when utilizing
the fact that $f_{V}(\overline{Q})$ depends only on $\overline{Q}_{Y}=\tilde{Q}_{Y}$,
the optimal choice for $E_{B}$ is $\overline{Q}=P_{X}\times\tilde{Q}_{Y}$,
since it results $I(\overline{Q})=0$. Under this choice, we get ${\cal J}_{1}\subset{\cal K}_{3}$
and ${\cal J}_{2}\subset{\cal K}_{4}$ and so $E_{A}\geq E_{B}$.
Thus, from \eqref{eq: Detection exponent RC enumeration}
\[
E_{\st[TE]}^{\st[RC]}\left(R,\alpha,P_{X},W,V\right)=\min_{\tilde{Q}\in\cap_{i=3}^{4}{\cal M}_{i}}D(\tilde{Q}{}_{Y|X}\|W|P_{X})
\]
where
\[
{\cal M}_{3}\dfn\left\{ \tilde{Q}:\; f_{V}(\tilde{Q})\geq\alpha+f_{W}(\tilde{Q})-R\right\} ,
\]
replaces ${\cal K}_{3}$, and 
\[
{\cal M}_{4}\dfn\left\{ \tilde{Q}:\;\mathbf{s}\left(\tilde{Q}_{Y},f_{V}(\tilde{Q}_{Y})+R\right)\geq R\right\} ,
\]
 replaces ${\cal K}_{4}$. Thus, the minimization in the exponent
is only on $\tilde{Q}$.
\end{enumerate}

\section{Composite Detection\label{sec:Composite-Detection}}

Up until now, we have assumed that detection is performed between
two simple hypotheses, namely $W$ and $V$. In this section, we briefly
discuss the generalization of the random coding analysis to composite
hypotheses, to wit, a detection between a channel $W\in{\cal W}$
and a channel $V\in{\cal V}$, where ${\cal W}$ and ${\cal V}$ are
disjoint. Due to the nature of the problems outlined in the introduction
(Section \ref{sec:Introduction}), we adopt a \emph{worst case} approach\emph{.}
For a codebook ${\cal C}_{n}$ and a given detector/decoder $\phi$,
we generalize the FA probability to 

\begin{equation}
P_{\st[FA]}({\cal C}_{n},\phi)\dfn\max_{W\in{\cal W}}\frac{1}{M}\sum_{m=1}^{M}W({\cal R}_{0}|\mathbf{x}_{m}),\label{eq: FA probability universal}
\end{equation}
and analogously, the MD and IE probabilities are obtained by maximizing
over $V\in{\cal V}$ and $W\in{\cal W}$, respectively. Then, the
trade-off between the IE probability and the FA and MD probabilities
in \eqref{eq: IE- FA and MD tradeoff} is defined exactly the same
way. 

Just as we have seen in \eqref{eq: IE exponent is minimum between ordinary and FA}
(proof of Proposition \ref{prop:Full tension decoder is asym optimal}),
for any sequence of codebooks ${\cal C}_{n}$ and decoder $\phi$
\[
E_{\st[IE]}({\cal C}_{n},\phi)=\min\left\{ E_{\st[O]}({\cal C}_{n},\phi),E_{\st[FA]}({\cal C}_{n},\phi)\right\} 
\]
where here, $E_{\st[O]}({\cal C}_{n},\phi)$ is the exponent achieved
by an ordinary decoder, which is not aware of $W$. Thus, the asymptotic
separation principle holds here too, in the sense that the optimal
detector/decoder may first use a detector which achieves the optimal
trade-off between the FA and MD exponents, and then a decoder which
achieves the optimal ordinary exponent.

We next discuss the achievable random coding exponents. %
\footnote{In universal decoding, typically only the random coding exponents
are attempted to be achieved, cf. Remark \ref{rem: universal - only random coding}. %
} As is well known, the \emph{maximum mutual information} \cite{goppa1975nonprobabilistic},
\cite[Chapter 10, p. 147]{csiszar2011information} universally achieves
the random for ordinary decoding. So, as in the simple hypotheses
case, it remains to focus on the optimal trade-off between the FA
and MD exponents, namely, solve
\begin{alignat}{1}
 & \mbox{minimize}~~~P_{\st[FA]}\nonumber \\
 & \mbox{subject to}~~P_{\st[MD]}\le e^{-n\overline{E}_{\st[MD]}}\label{eq: FA-MD tradeoff universal}
\end{alignat}
for some given exponent $\overline{E}_{\st[MD]}>0$. The next Lemma
shows that the following \emph{universal} detector/decoder $\phi^{\st[U]}$,
whose rejection region is 
\begin{equation}
{\cal R}_{0}^{\st[U]}\dfn\left\{ \mathbf{y}:~e^{n\alpha}\cdot\sum_{m=1}^{M}\max_{W\in{\cal W}}W(\mathbf{y}|\mathbf{x}_{m})\leq\sum_{m=1}^{M}\max_{V\in{\cal V}}V(\mathbf{y}|\mathbf{x}_{m})\right\} ,\label{eq: R0 optimal universal}
\end{equation}
solves \eqref{eq: FA-MD tradeoff universal}. The universality here
is in the sense of \eqref{eq: FA-MD tradeoff universal}, i.e., achieving
the best worst-case (over $W$) FA exponent, under a worst case constraint
(over $V$) on the MD exponent. There might be, however, a loss in
exponents compared to a detector which is aware of the actual pair
$(W,V)$ (cf. Corollary \ref{cor: universal exponent}). 
\begin{lem}
\label{lem: Optimal universal detector/decoder}Let ${\cal C}=\{{\cal C}_{n}\}$
be a given sequence of codebooks, let $\phi^{\st[U]}$ be as above,
and let $\phi$ be any other partition of ${\cal Y}^{n}$ into $M+1$
regions. Then, if $E_{\st[FA]}({\cal C},\phi)\geq E_{\st[FA]}({\cal C},\phi^{*})$
then $E_{\st[MD]}({\cal C},\phi)\leq E_{\st[MD]}({\cal C}_{n},\phi^{*})$.\end{lem}
\begin{IEEEproof}
The idea is that the maximum in \eqref{eq: FA probability universal}
can be interchanged with the sum without affecting the exponential
behavior. Specifically, let us define the sets of channels which maximize
$f_{W}(Q)$ for some $Q$
\[
{\cal W}_{\st[U]}\dfn\left\{ W\in{\cal W}:\;\exists Q\mbox{\,\ such that }W=\argmax_{W'\in{\cal W}}f_{W'}(Q)\right\} .
\]
Clearly, since $f_{W}(Q)$ is only a function of the joint type, the
cardinality of the sets ${\cal W}_{\st[U]}$ is not larger than the
number of different joint types, and so their cardinality increases
only polynomially with $n$. Then, 
\begin{align}
P_{\st[FA]}({\cal C}_{n},\phi) & =\max_{W\in{\cal W}}\sum_{\mathbf{y}\in{\cal R}_{0}}\frac{1}{M}\sum_{m=1}^{M}W(\mathbf{y}|\mathbf{x}_{m})\\
 & \leq\sum_{\mathbf{y}\in{\cal R}_{0}}\frac{1}{M}\sum_{m=1}^{M}\max_{W\in{\cal W}}W(\mathbf{y}|\mathbf{x}_{m})\\
 & =\sum_{\mathbf{y}\in{\cal R}_{0}}\frac{1}{M}\sum_{m=1}^{M}\max_{W\in{\cal W}_{\st[U]}}W(\mathbf{y}|\mathbf{x}_{m})\\
 & \dfn\sum_{\mathbf{y}\in{\cal R}_{0}}g(\mathbf{y})\\
 & \leq\sum_{\mathbf{y}\in{\cal R}_{0}}\frac{1}{M}\sum_{m=1}^{M}\sum_{W\in{\cal W}_{\st[U]}}W(\mathbf{y}|\mathbf{x}_{m})\\
 & =\sum_{W\in{\cal W}_{\st[U]}}\frac{1}{M}\sum_{m=1}^{M}\sum_{\mathbf{y}\in{\cal R}_{0}}W(\mathbf{y}|\mathbf{x}_{m})\\
 & \doteq\max_{W\in{\cal W}_{\st[U]}}\frac{1}{M}\sum_{m=1}^{M}\sum_{\mathbf{y}\in{\cal R}_{0}}W(\mathbf{y}|\mathbf{x}_{m})\\
 & \leq\max_{W\in{\cal W}}\frac{1}{M}\sum_{m=1}^{M}\sum_{\mathbf{y}\in{\cal R}_{0}}W(\mathbf{y}|\mathbf{x}_{m})\\
 & =P_{\st[FA]}({\cal C}_{n},\phi)
\end{align}
where the measure $g(\mathbf{y})$ was implicitly defined. Thus, up
to a sub-exponential term which does not affect exponents, 
\[
P_{\st[FA]}({\cal C}_{n},\phi)\doteq\sum_{\mathbf{y}\in{\cal R}_{0}}g(\mathbf{y}).
\]
Similarly, defining the measure
\[
h(\mathbf{y})\dfn\frac{1}{M}\sum_{m=1}^{M}\max_{V\in{\cal V}}V(\mathbf{y}|\mathbf{x}_{m})
\]
we get
\[
P_{\st[MD]}({\cal C}_{n},\phi)=\sum_{\mathbf{y}\in\overline{{\cal R}_{0}}}h(\mathbf{y}).
\]
Now, the ordinary Neyman-Pearson lemma \cite[Theorem 11.7.1]{Cover:2006:EIT:1146355}
can be invoked%
\footnote{Note that the Neyman-Pearson lemma is also valid for general positive
measures, not just for probability distributions. This can also be
seen from the Lagrange formulation \eqref{eq: Lagrange formulation}.%
} to show that the optimal detector is of the form \eqref{eq: R0 optimal universal},
which completes the theorem. 
\end{IEEEproof}
It now remains to evaluate, for a given pair of channels $(W,V)\in{\cal W}\times{\cal V}$,
the resulting random coding exponents when $\phi^{\st[U]}$ is used.
Fortunately, this is an easy task given Theorem \ref{thm: RC bound type-enumeration}.
Let us define the generalized normalized log-likelihood ratio of the
set of channels ${\cal W}$ as 
\begin{equation}
f_{{\cal W}}(Q)\dfn\max_{W\in{\cal W}}\sum_{x\in{\cal X},y\in{\cal Y}}Q(x,y)\log W(y|x).\label{eq: generalized log likelihood definition}
\end{equation}
The following is easily verified.
\begin{cor}[to Theorem \ref{thm: RC bound type-enumeration}]
\label{cor: universal exponent}Let a distribution $P_{X}$ and a
parameter $\alpha\in\mathbb{R}$ be given. Then, there exists a sequence
of codes ${\cal C}=\{{\cal C}_{n}\}_{n=1}^{\infty}$ of rate $R$,
such that for any $\delta>0$
\begin{equation}
E_{\st[FA]}\left({\cal C},\phi^{\st[U]}\right)\geq E_{\st[TE,U]}^{\st[RC]}\left(R,\alpha,P_{X},W,V\right)-\delta,\label{eq: FA exponent RC type-enumeration thm universal}
\end{equation}
\begin{equation}
E_{\st[MD]}\left({\cal C},\phi^{\st[U]}\right)\geq E_{\st[TE,U]}^{\st[RC]}\left(R,\alpha,P_{X},W,V\right)-\alpha-\delta\label{eq: MD exponent RC type-enumaration thm universal}
\end{equation}
where $E_{\st[TE,U]}^{\st[RC]}\left(R,\alpha,P_{X},W,V\right)$ is
defined as $E_{\st[TE]}^{\st[RC]}\left(R,\alpha,P_{X},W,V\right)$
of \eqref{eq: Detection exponent RC enumeration}, but replacing $f_{W}(Q)$
with $f_{{\cal W}}(Q)$ and $f_{V}(Q)$ with $f_{{\cal V}}(Q)$ in
all the definitions preceding Theorem \ref{thm: RC bound type-enumeration}.
\end{cor}
We conclude with a few remarks.
\begin{rem}
The function $f_{{\cal W}}(Q)$ is a convex function of $Q$ (as a
pointwise maximum of linear functions), but not a linear function.
This may harden the optimization problems involved in computing the
exponents. Also, we implicitly assume that the set of channels ${\cal W}$
is sufficiently `regular', so that $f_{{\cal W}}(Q)$ is a continuous
function of $Q$.
\end{rem}

\begin{rem}
\label{rem: universal - only random coding}The same technique works
for the simplified low-rate detector/decoder. Unfortunately, since
the bound \eqref{eq: Gallager bound on false alarm for a specific code}
(Appendix \ref{sec:Proof-of-Theorem EX TE}) utilizes the structure
of the optimal detector/decoder, it is difficult to generalize the
bounds which rely on it, namely, the expurgated exponents and the
Gallager/Forney-style bounds. This is common to many other problem
in universal decoding - for a non-exhaustive list of examples, see
\cite{csiszar1977new,Ahlswede_permutations,Ziv_finite_state,Neri_universal_Gaussians,universal_memory}.
\end{rem}

\begin{rem}
A different approach to composite hypothesis testing is the competitive
minimax approach \cite{competitive_minimax}. In this approach, a
detector/decoder is sought which achieves the largest fraction of
the error exponents achieved for a detection of only a pair of channels
$(W,V)$, uniformly over all possible pairs of channels $(W,V)$.
The application of this method on generalized decoders was exemplified
for Forney's erasure/list decoder \cite{Forney68} in \cite{MF07,universal_erasure},
and the same techniques can work for this problem.
\end{rem}

\section{An Example: A Detection of a Pair Binary Symmetric Channels\label{sec:Examples}}

Let $W$ and $V$ be a pair of BSCs with crossover probabilities $w\in(0,1)$
and $v\in(0,1)$, respectively. In this case the exponent bounds of
Section \ref{sec:Achievable-Error-Exponents } can be greatly simplified,
if the input distribution is uniform, i.e. $P_{X}=(\frac{1}{2},\frac{1}{2})$.
Indeed, in Appendix \ref{sec:Simplified-Expressions-for-BSC} we provide
simplified expressions for the type-enumeration based exponents. Interestingly,
while this input distribution is optimal from the channel coding perspective,
the two output distributions $\tilde{W}$ and $\tilde{V}$ it induces
are also uniform, and so the simple decoder which only uses the output
statistics, namely $\phi_{\st[H]}$ of Subsection \ref{sub:Simplified-Detector/Decoders},
is utterly useless. However, the optimal decoder $\phi'$ can produce
strictly positive exponents. 

\begin{figure}
\centering{}\includegraphics[scale=0.4]{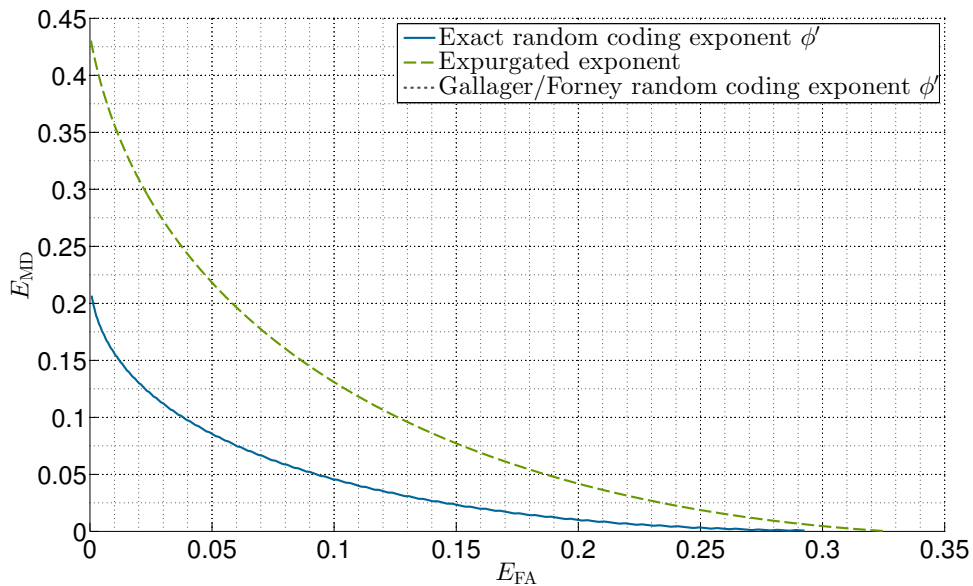} \protect\caption{The trade-off between the FA exponent and the MD exponent at $R=0$,
for the detection of a BSC $W$ with crossover probability $0.1$,
from a BSC $V$ with crossover probability $0.4$, when using the
optimal detector $\phi'$. The solid line corresponds to the exact
random coding exponent, and also to the Gallager/Forney-style random
coding exponent. The dashed line corresponds to the expurgated exponent.
\label{fig: zero rate}}
\end{figure}

\begin{figure}
\centering{}\includegraphics[scale=0.4]{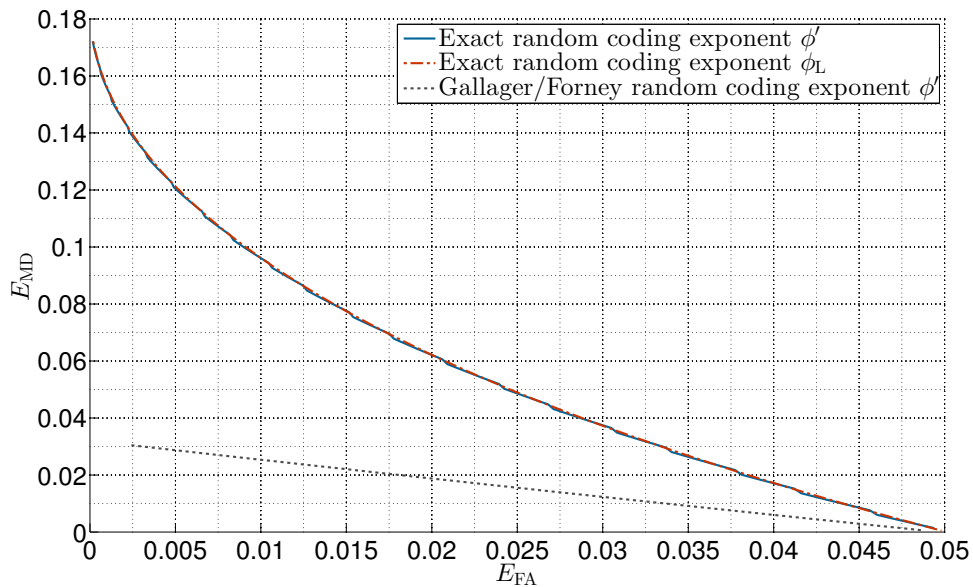} \protect\caption{The trade-off between the FA exponent and the MD exponent at $R=0.5\cdot C_{W}$,
for the detection of a BSC $W$ with crossover probability $0.1$,
from a BSC $V$ with crossover probability $0.4$. The solid line
corresponds to the exact random coding exponent of $\phi'$, and also
to the exact random coding exponent of $\phi_{\st[L]}$. The dotted
line corresponds Gallager/Forney-style random coding exponent of $\phi'$.\label{fig: medium rate}}
\end{figure}

We have plotted the FA exponent versus the MD exponent for the detection
between two BSCs with $w=0.1$ and $v=0.4$. We have assumed the uniform
input distribution $P_{X}=(\frac{1}{2},\frac{1}{2})$, which results
the capacity $C_{W}\dfn I(P_{X}\times W)\approx0.37\mbox{ (nats)}$.
Figure \ref{fig: zero rate} shows that at zero rate, the expurgated
bound which is based on type-enumeration significantly improves the
random coding bound. In addition, the Gallager/Forney-style random
coding exponent coincides with the exact exponent. By contrast, the
Gallager/Forney-style expurgated exponent offers no improvement over
the ordinary random coding bound (and thus not displayed). Figure
\ref{fig: medium rate} shows that at $R=0.5\cdot C_{W}$, the simplified
low-rate detector/decoder $\phi_{\st[L]}$ still performs as well
as the optimal detector/decoder $\phi'$. This, in fact continues
to hold for all rates less than $R\approx0.8\cdot C_{W}$. In addition,
it is evident that the Gallager/Forney-style random coding exponent
is a poor bound, which exemplifies the importance of the ensemble-tight
bounding technique of the type enumeration method.

\appendices{\numberwithin{equation}{section}}

\section{Proof of Theorem \ref{thm: EX bound type-enumeration}\label{sec:Proof-of-Theorem EX TE}}

Before getting into the proof, we derive a standard bound on the FA
probability, which will also be used in Appendices \ref{sec:Proof-of-Theorem RC GF}
and \ref{sec:Proof-of-Theorem EX GF}. For any given code and $s\geq0$
\begin{align}
P_{\st[FA]}({\cal C}_{n},\phi') & =\sum_{\mathbf{y}\in{\cal R}_{0}'}\frac{1}{M}\sum_{m=1}^{M}W(\mathbf{y}|\mathbf{x}_{m})\\
 & =\sum_{\mathbf{y}\in{\cal R}_{0}'}\left[\frac{1}{M}\sum_{m=1}^{M}W(\mathbf{y}|\mathbf{x}_{m})\right]^{1-s}\left[\frac{1}{M}\sum_{m=1}^{M}W(\mathbf{y}|\mathbf{x}_{m})\right]^{s}\\
 & \trre[\leq,a]e^{-n\alpha s}\sum_{\mathbf{y}\in{\cal R}_{0}'}\left[\frac{1}{M}\sum_{m=1}^{M}W(\mathbf{y}|\mathbf{x}_{m})\right]^{1-s}\left[\frac{1}{M}\sum_{m=1}^{M}V(\mathbf{y}|\mathbf{x}_{m})\right]^{s}\\
 & \leq e^{-n\alpha s}\sum_{\mathbf{y}\in{\cal Y}^{n}}\left[\frac{1}{M}\sum_{m=1}^{M}W(\mathbf{y}|\mathbf{x}_{m})\right]^{1-s}\left[\frac{1}{M}\sum_{m=1}^{M}V(\mathbf{y}|\mathbf{x}_{m})\right]^{s},\label{eq: Gallager bound on false alarm for a specific code}
\end{align}
where $(a)$ is from \eqref{eq: asymptotical rejection region full tension}.
\begin{IEEEproof}[Proof of Theorem \ref{thm: EX bound type-enumeration}]
For a given code ${\cal C}_{n}$, a codeword $1\leq m\leq M$, and
a joint type $P_{X\tilde{X}}$, define the \emph{type class enumerator}
\begin{equation}
\acute{N}_{m}(P_{X\tilde{X}},{\cal C}_{n})\dfn\left|\left\{ \mathbf{x}\in{\cal C}_{n}\backslash\mathbf{x}_{m}:\;\hat{Q}_{\mathbf{x}_{m}\mathbf{x}}=P_{X\tilde{X}}\right\} \right|.\label{eq: type class enumerator expurgated}
\end{equation}
Upon restricting $0\leq s\leq1$ in \eqref{eq: Gallager bound on false alarm for a specific code},
we obtain the bound
\begin{align}
P_{\st[FA]}({\cal C}_{n},\phi') & \leq e^{-n\alpha s}\sum_{\mathbf{y}\in{\cal Y}^{n}}\left[\frac{1}{M}\sum_{m=1}^{M}W(\mathbf{y}|\mathbf{x}_{m})\right]^{1-s}\left[\frac{1}{M}\sum_{m=1}^{M}V(\mathbf{y}|\mathbf{x}_{m})\right]^{s}\\
 & \trre[\leq,a]e^{-n\alpha s}\frac{1}{M}\sum_{m=1}^{M}\sum_{k=1}^{M}\sum_{\mathbf{y}\in{\cal Y}^{n}}W^{1-s}(\mathbf{y}|\mathbf{x}_{m})V^{s}(\mathbf{y}|\mathbf{x}_{k})\\
 & \trre[=,b]e^{-n\alpha s}\frac{1}{M}\sum_{m=1}^{M}\sum_{P_{X\tilde{X}}}\acute{N}_{m}(P_{X\tilde{X}},{\cal C}_{n})\exp\left[-n\left(\E_{P_{X\tilde{X}}}\left[d_{s}(X,\tilde{X})\right]\right)\right],
\end{align}
where $(a)$ follows from $\sum_{i}a_{i}^{\nu}\geq\left(\sum_{i}a_{i}\right)^{\nu}$
for $\nu\leq1$, and $(b)$ is using \eqref{eq: type class enumerator expurgated}
and \eqref{eq: Chernoff distance}. Now, the packing lemma \cite[Problem 10.2]{csiszar2011information}
essentially shows (see also \cite[Appendix]{Merhav13_List}) that
for any $\delta>0$, there exists a code ${\cal C}_{n}^{*}$ (of rate
$R$) such that
\[
\acute{N}_{m}(P_{X\tilde{X}},{\cal C}_{n}^{*})\leq\begin{cases}
\exp\left[n\left(R+\delta-I(P_{X\tilde{X}})\right)\right], & I(P_{X\tilde{X}})\leq R+\delta\\
0, & I(P_{X\tilde{X}})>R+\delta
\end{cases}
\]
for all $1\leq m\leq M$ and $P_{X\tilde{X}}$. This, along with Proposition
\ref{prop: exponents balance} completes the proof of the theorem.
\end{IEEEproof}

\section{Proof of Theorem \ref{thm: RC bound type-enumeration simplified low}\label{sec:Proof-of-Theorem RC low TE}}

The proof is very similar to the proof of Theorem \ref{thm: RC bound type-enumeration}.
We will use the following lemma, which is analogous to Lemma \ref{lem: behavior of intersection}.
\begin{lem}
\label{lem: behavior of intersection indicator}Under the conditions
of Lemma \ref{lem: behavior of intersection}, 
\begin{equation}
\P\left(\bigcap_{Q\in{\cal Q}:\, Q_{Y}=\tilde{Q}_{Y}}\left\{ \I\left\{ \hat{N}(Q|\mathbf{y})\geq1\right\} <e^{nJ(Q)}\right\} \right)\;\begin{cases}
=1-o(n), & \mathbf{T}(\tilde{Q}_{Y};J,{\cal Q})>R\\
\doteq e^{-n\infty}, & \mbox{otherwise}
\end{cases},\label{eq: exponential of intersection indicator}
\end{equation}
where $\mathbf{y}\in{\cal T}(\tilde{Q}_{Y})$, and 
\begin{equation}
\mathbf{T}(\tilde{Q}_{Y};J,{\cal Q})\dfn\min_{Q\in{\cal Q}:\, Q=\tilde{Q}_{Y},J(Q)\leq0}I(Q).\label{eq: T bold function}
\end{equation}
\end{lem}
\begin{IEEEproof}
We have
\[
\P\left(\bigcap_{Q\in{\cal Q}:\, Q_{Y}=\tilde{Q}_{Y}}\left\{ \I\left\{ N(Q|\mathbf{y})\geq1\right\} <e^{nJ(Q)}\right\} \right)=\P\left(\bigcap_{Q\in{\cal Q}:\, Q_{Y}=\tilde{Q}_{Y},J(Q)\leq0}\left\{ \I\left\{ N(Q|\mathbf{y})=0\right\} \right\} \right).
\]
From this point onward, the proof follows the same lines of the proof
of Lemma \ref{lem: behavior of intersection}.\end{IEEEproof}
\begin{rem}
\label{rem: J function convex indicator}Remarks \ref{rem: continuity of S}
and \ref{rem: J function convex} are also valid here. If $J(Q)$
is convex in $Q_{Y|X}$ then Lagrange duality \cite[Chapter 5]{Boyd}
implies
\begin{align}
\mathbf{T}(\tilde{Q}_{Y};J,{\cal Q}) & =\min_{Q\in{\cal Q}:\, Q=\tilde{Q}_{Y}}\max_{\lambda\geq0}\left[I(Q)+\lambda J(Q)\right]\\
 & =\max_{\lambda\geq0}\min_{Q\in{\cal Q}:\, Q=\tilde{Q}_{Y}}\left[I(Q)+\lambda J(Q)\right].\label{eq: T bold J convex}
\end{align}
The only difference from $\mathbf{S}(\tilde{Q}_{Y};J,{\cal Q})$ of
\eqref{eq: S bold J convex} in this case is the maximization domain
for $\lambda$. Note that the function $\mathbf{t}(\tilde{Q}_{Y};\gamma)$
of \eqref{eq: t bold lower case} is a specific instance of $\mathbf{T}(\tilde{Q}_{Y};\cdot,\cdot)$
defined in \eqref{eq: T bold function} with ${\cal Q}={\cal Q}_{W}$
and $J(Q)=-\alpha-f_{W}(Q)+\gamma$ which is convex in $Q_{Y|X}$
(in fact, linear).\end{rem}
\begin{IEEEproof}[Proof of Theorem \ref{thm: RC bound type-enumeration simplified low}]
In general, since
\[
\sum_{m=2}^{M}W(\mathbf{y}|\mathbf{x}_{m})=\sum_{Q}N(Q|\mathbf{y})e^{nf_{W}(Q)}
\]
but 
\begin{align}
\max_{2\leq m\leq M}W(\mathbf{y}|\mathbf{x}_{m}) & =\max_{Q}\I\left\{ N(Q|\mathbf{y})\geq1\right\} e^{nf_{W}(Q)}\\
 & \doteq\sum_{Q}\I\left\{ N(Q|\mathbf{y})\geq1\right\} e^{nf_{W}(Q)},
\end{align}
then the analysis of the FA exponent of $\phi_{\st[L]}$ follows the
same lines as the analysis in the proof of Theorem \ref{thm: RC bound type-enumeration},
when replacing $N(Q|\mathbf{y})$ with $\I\left\{ N(Q|\mathbf{y})\geq1\right\} $.
Thus, in the following we only highlight the main changes. Just as
in the derivations leading to \eqref{eq: FA intial derivation}, 
\begin{align}
\overline{P_{\st[FA]}}(\mathbf{x}_{1},\mathbf{y}) & \dfn\P\left(\mathbf{y}\in{\cal R}_{0,\st[L]}|\mathbf{X}_{1}=\mathbf{x}_{1},\mathbf{Y}=\mathbf{y}\right)\\
 & \doteq\max\left\{ A_{\st[L]}(\tilde{Q}),B_{\st[L]}(\tilde{Q})\right\} ,\label{eq: FA initial derivation low}
\end{align}
where
\[
A_{\st[L]}(\tilde{Q})\dfn\P\left(\sum_{Q}\I\left\{ N(Q|\mathbf{y})\geq1\right\} e^{nf_{W}(Q)}\leq e^{-n\alpha}\cdot e^{nf_{V}(\tilde{Q})}\right)\cdot\I\left\{ f_{W}(\tilde{Q})\leq-\alpha+f_{V}(\tilde{Q})\right\} 
\]
and
\[
B_{\st[L]}(\tilde{Q})\dfn\P\left(e^{nf_{W}(\tilde{Q})}+\max_{Q}\I\left\{ N(Q|\mathbf{y})\geq1\right\} e^{nf_{W}(Q)}\leq e^{-n\alpha}\cdot\max_{Q}\I\left\{ N(Q|\mathbf{y})\geq1\right\} e^{nf_{V}(Q)}\right).
\]

For the first term, 
\begin{align}
A_{\st[L]}(\tilde{Q}) & \trre[\doteq,IR]\P\left(\bigcap_{Q:\, f_{W}(Q)>-\infty}\left\{ \I\left\{ N(Q|\mathbf{y})\geq1\right\} <e^{n\left[-\alpha+f_{V}(\tilde{Q})-f_{W}(Q)\right]}\right\} \right)\cdot\I\left\{ f_{W}(\tilde{Q})\leq-\alpha+f_{V}(\tilde{Q})\right\} \\
 & \trre[\doteq,a]\I\left\{ \mathbf{T}(\tilde{Q}_{Y};-\alpha+f_{V}(\tilde{Q})-f_{W}(Q),{\cal Q}_{W})>R\right\} \cdot\I\left\{ f_{W}(\tilde{Q})\leq-\alpha+f_{V}(\tilde{Q})\right\} ,
\end{align}
where $(a)$ is by Lemma \ref{lem: behavior of intersection indicator}.
Upon averaging over $(\mathbf{X}_{1},\mathbf{Y})$, we obtain the
exponent $E_{A,\st[L]}$ of \eqref{eq: EA low rates} (utilizing the
definition \eqref{eq: t bold lower case}).

Moving on to the second term, similarly as in the analysis leading
to \eqref{eq: B term}
\begin{align}
B_{\st[L]}(\tilde{Q}) & \doteq\sum_{\overline{Q}:\, f_{W}(\overline{Q})\leq-\alpha+f_{V}(\overline{Q})}\nonumber \\
 & \P\left(\bigcap_{Q\neq\overline{Q}:\, f_{W}(Q)>-\infty}\left\{ \I\left\{ N(Q|\mathbf{y})\geq1\right\} \leq e^{n\left[-\alpha+f_{V}(\overline{Q})-f_{W}(Q)\right]}\cdot\I\left\{ N(\overline{Q}|\mathbf{y})\geq1\right\} \right\} \cap\right.\nonumber \\
 & \left.\vphantom{\bigcap_{Q\neq\overline{Q}}\left\{ N(Q|\mathbf{y})e^{nf_{W}(Q)}\leq e^{-n\alpha}\cdot N(\overline{Q}|\mathbf{y})e^{nf_{V}(\overline{Q})}\right\} \cap}\left\{ 1\leq e^{n\left[-\alpha+f_{V}(\overline{Q})-f_{W}(\tilde{Q})\right]}\cdot\I\left\{ N(\overline{Q}|\mathbf{y})\geq1\right\} \right\} \right)\\
 & \dfn\sum_{\overline{Q}:\, f_{W}(\overline{Q})\leq-\alpha+f_{V}(\overline{Q})}\zeta_{\st[L]}(\overline{Q}).\label{eq: B term low}
\end{align}
We now split the analysis into three cases:

\uline{Cases 1 and 2:} Assume $0\leq I(\overline{Q})<R$. An analysis
similar to cases 1 and 2 in the proof of Theorem \ref{thm: RC bound type-enumeration}
shows that 
\[
\zeta_{\st[L]}(\overline{Q})\doteq\I\left\{ \mathbf{T}(\tilde{Q}_{Y};-\alpha+f_{V}(\overline{Q})-f_{W}(Q),{\cal Q}_{W})>R\right\} \I\left\{ -\alpha+f_{V}(\overline{Q})-f_{W}(\tilde{Q})>0\right\} .
\]

\uline{Case 3:} Assume that $I(\overline{Q})>R$. An analysis similar
to case 3 in the proof of Theorem \ref{thm: RC bound type-enumeration}
shows that the inner probability in \eqref{eq: B term low} is exponentially
equal to
\[
\zeta_{\st[L]}(\overline{Q})\doteq\I\left\{ \mathbf{T}(\tilde{Q}_{Y};-\alpha+f_{V}(\overline{Q})-f_{W}(Q),{\cal Q}_{W})>R\right\} \I\left\{ -\alpha+f_{V}(\overline{Q})-f_{W}(\tilde{Q})>0\right\} e^{-n\left(I(\overline{Q})-R\right)}.
\]
Returning to \eqref{eq: B term low} we obtain that $B_{\st[L]}(\tilde{Q})$
is exponentially equal to the maximum between
\[
\max_{\overline{Q}:\, f_{W}(\overline{Q})<-\alpha+f_{V}(\overline{Q}),\, I(\overline{Q})<R,\, f_{V}(\overline{Q})>\alpha+f_{W}(\tilde{Q})}\I\left\{ \mathbf{T}(\tilde{Q}_{Y};-\alpha+f_{V}(\overline{Q})-f_{W}(Q),{\cal Q}_{W})>R\right\} ,
\]
and
\[
\max_{\overline{Q}:\, f_{W}(\overline{Q})<-\alpha+f_{V}(\overline{Q}),\, I(\overline{Q})\geq R,\, f_{V}(\overline{Q})>\alpha+f_{W}(\tilde{Q})}\I\left\{ \mathbf{T}(\tilde{Q}_{Y};-\alpha+f_{V}(\overline{Q})-f_{W}(Q),{\cal Q}_{W})>R\right\} e^{-n\left(I(\overline{Q})-R\right)},
\]
or, more succinctly, 
\[
B(\tilde{Q})=\max_{\overline{Q}}\I\left\{ \mathbf{T}(\tilde{Q}_{Y};-\alpha+f_{V}(\overline{Q})-f_{W}(Q),{\cal Q}_{W})>R\right\} e^{-n\left[I(\overline{Q})-R\right]_{+}}
\]
where the maximization is over
\[
\left\{ \overline{Q}:\; f_{W}(\overline{Q})<-\alpha+f_{V}(\overline{Q}),\, f_{V}(\overline{Q})\geq\alpha+f_{W}(\tilde{Q})\right\} .
\]
Upon averaging over $(\mathbf{X}_{1},\mathbf{Y})$, we obtain the
exponent $E_{B,\st[L]}$ of \eqref{eq: EB low rates} (utilizing again
\eqref{eq: t bold lower case}), and the proof of the FA exponent
\eqref{eq: FA exponent RC type-enumeration thm low rates} is proved
using \eqref{eq: FA initial derivation low}. 

For the MD expression, since $\phi_{\st[L]}$ is not necessarily the
optimal detector in the Neyman-Pearson sense, we cannot use Proposition
\ref{prop: exponents balance}. However, due to the symmetry in ${\cal R}_{0,\st[L]}$
of $W$ and $V$, a similar observation as in Fact \ref{fact: alternative expression for MD exponent}
holds, which leads directly to \eqref{eq: MD exponent RC type-enumaration thm low rates}.
The rest of the proof follows the same lines as the proof of theorem
\ref{thm: RC bound type-enumeration}. 
\end{IEEEproof}

\section{Proof of Theorem \ref{thm: RC bound Gallager}\label{sec:Proof-of-Theorem RC GF}}
\begin{IEEEproof}[Proof of Theorem \ref{thm: RC bound Gallager}]
As in the proof of Theorem \ref{thm: RC bound type-enumeration},
we only need to upper bound the FA probability as the MD probability
can be easily evaluated from the FA bound, using Proposition \ref{prop: exponents balance}.
It remains to derive an upper bound on the average FA error probability.
We assume the ensemble of randomly selected codes of size $M=\left\lceil e^{nR}\right\rceil $,
where each codeword is selected independently at random, with i.i.d.
components from the distribution $P_{X}$. Introducing a parameter
$\rho\geq\max\left\{ s,1-s\right\} $, we continue the bound \eqref{eq: Gallager bound on false alarm for a specific code}
as follows:
\begin{align}
P_{\st[FA]}({\cal C}_{n},\phi') & \leq e^{-n(\alpha s+R)}\sum_{\mathbf{y}\in{\cal Y}^{n}}\left[\sum_{m=1}^{M}W(\mathbf{y}|\mathbf{x}_{m})\right]^{\nicefrac{\rho(1-s)}{\rho}}\left[\sum_{m=1}^{M}V(\mathbf{y}|\mathbf{x}_{m})\right]^{\nicefrac{\rho s}{\rho}}\\
 & \trre[\leq,a]e^{-n(\alpha s+R)}\sum_{\mathbf{y}\in{\cal Y}^{n}}\left[\sum_{m=1}^{M}W^{\nicefrac{(1-s)}{\rho}}(\mathbf{y}|\mathbf{x}_{m})\right]^{\rho}\left[\sum_{m=1}^{M}V^{\nicefrac{s}{\rho}}(\mathbf{y}|\mathbf{x}_{m})\right]^{\rho}\\
 & =e^{-n(\alpha s+R)}\sum_{\mathbf{y}\in{\cal Y}^{n}}\left[\sum_{m=1}^{M}\sum_{k=1}^{M}W^{\nicefrac{(1-s)}{\rho}}(\mathbf{y}|\mathbf{x}_{m})V^{\nicefrac{s}{\rho}}(\mathbf{y}|\mathbf{x}_{k})\right]^{\rho},
\end{align}
where $(a)$ follows from $\left(\sum_{i}a_{i}\right)^{\nu}\leq\sum_{i}a_{i}^{\nu}$
for $\nu\leq1$. Using now the fact that the codewords are selected
at random, we obtain
\begin{align}
\overline{P_{\st[FA]}}({\cal C}_{n},\phi') & \leq e^{-n(\alpha s+R)}\sum_{\mathbf{y}\in{\cal Y}^{n}}\E\left\{ \left[\sum_{m=1}^{M}\sum_{k=1}^{M}W^{\nicefrac{(1-s)}{\rho}}(\mathbf{y}|\mathbf{X}_{m})V^{\nicefrac{s}{\rho}}(\mathbf{y}|\mathbf{X}_{k})\right]^{\rho}\right\} \\
 & \trre[\leq,a]e^{-n(\alpha s+R)}\sum_{\mathbf{y}\in{\cal Y}^{n}}\left\{ \sum_{m=1}^{M}\sum_{k=1}^{M}\E\left[W^{\nicefrac{(1-s)}{\rho}}(\mathbf{y}|\mathbf{X}_{m})V^{\nicefrac{s}{\rho}}(\mathbf{y}|\mathbf{X}_{k})\right]\right\} ^{\rho},\label{eq: averaging over the ensemble}
\end{align}
where $(a)$ is by restricting $\rho\leq1$ and using Jensen Inequality.
For a given $\mathbf{y}$, let us focus on the inner expectation.
If $m=k$ then
\begin{align}
\E\left[W^{\nicefrac{(1-s)}{\rho}}(\mathbf{y}|\mathbf{X}_{m})V^{\nicefrac{s}{\rho}}(\mathbf{y}|\mathbf{X}_{m})\right] & =\E\left[\prod_{i=1}^{n}W^{\nicefrac{(1-s)}{\rho}}(y_{i}|X_{m,i})V^{\nicefrac{s}{\rho}}(y_{i}|X_{m,i})\right]\\
 & =\prod_{i=1}^{n}\E\left[W^{\nicefrac{(1-s)}{\rho}}(y_{i}|X_{m,i})V^{\nicefrac{s}{\rho}}(y_{i}|X_{m,i})\right]\\
 & =\prod_{i=1}^{n}\left(\sum_{x\in{\cal X}}P_{X}(x)W^{\nicefrac{(1-s)}{\rho}}(y_{i}|x)V^{\nicefrac{s}{\rho}}(y_{i}|x)\right)\\
 & \dfn\Psi_{s,\rho}(\mathbf{y}).\label{eq: Psi def}
\end{align}
Otherwise, if $m\neq k$, then since the codewords are selected independently
\begin{align}
\E\left[W^{\nicefrac{(1-s)}{\rho}}(\mathbf{y}|\mathbf{X}_{m})V^{\nicefrac{s}{\rho}}(\mathbf{y}|\mathbf{X}_{k})\right] & =\E\left[W^{\nicefrac{(1-s)}{\rho}}(\mathbf{y}|\mathbf{X}_{m})\right]\E\left[V^{\nicefrac{s}{\rho}}(\mathbf{y}|\mathbf{X}_{k})\right]\\
 & =\E\left[\prod_{i=1}^{n}W^{\nicefrac{(1-s)}{\rho}}(y_{i}|X_{m,i})\right]\E\left[\prod_{i=1}^{n}V^{\nicefrac{s}{\rho}}(y_{i}|X_{k,i})\right]\\
 & =\prod_{i=1}^{n}\E\left[W^{\nicefrac{(1-s)}{\rho}}(y_{i}|X_{m,i})\right]\E\left[V^{\nicefrac{s}{\rho}}(y_{i}|X_{k,i})\right]\\
 & =\prod_{i=1}^{n}\left(\sum_{x\in{\cal X}}P_{X}(x)W^{\nicefrac{(1-s)}{\rho}}(y_{i}|x)\right)\left(\sum_{x\in{\cal X}}P_{X}(x)V^{\nicefrac{s}{\rho}}(y_{i}|x)\right)\\
 & \dfn\Gamma_{s,\rho}(\mathbf{y}).\label{eq: Gamma def}
\end{align}
So, the double inner summand in \eqref{eq: averaging over the ensemble}
is bounded as
\begin{align}
\left\{ \sum_{m=1}^{M}\sum_{k=1}^{M}\E\left[W^{\nicefrac{(1-s)}{\rho}}(\mathbf{y}|\mathbf{X}_{m})V^{\nicefrac{s}{\rho}}(\mathbf{y}|\mathbf{X}_{k})\right]\right\} ^{\rho} & =\left\{ M\Psi_{s,\rho}(\mathbf{y})+M(M-1)\Gamma_{s,\rho}(\mathbf{y})\right\} ^{\rho}\\
 & \leq2^{\rho}\max\left\{ M^{\rho}\Psi_{s,\rho}^{\rho}(\mathbf{y}),M^{2\rho}\Gamma_{s,\rho}^{\rho}(\mathbf{y})\right\} ,
\end{align}
using $\left\{ c+d\right\} ^{\rho}\leq\left[2\max\{c,d\}\right]^{\rho}$
for any $c,d\geq0$. Thus, we may continue the bound of \eqref{eq: averaging over the ensemble}
as
\[
\overline{P_{\st[FA]}}({\cal C}_{n},\phi')\leq e^{-n(\alpha s+R)}2^{\rho}\max\left\{ \sum_{\mathbf{y}\in{\cal Y}^{n}}M^{\rho}\Psi_{s,\rho}^{\rho}(\mathbf{y}),\sum_{\mathbf{y}\in{\cal Y}^{n}}M^{2\rho}\Gamma_{s,\rho}^{\rho}(\mathbf{y})\right\} .
\]
The first term in the above maximization is given by
\begin{align}
\hphantom{=} & e^{-n\left(\alpha s-(\rho-1)R-\frac{\rho\log2}{n}\right)}\sum_{\mathbf{y}\in{\cal Y}^{n}}\prod_{i=1}^{n}\left(\sum_{x\in{\cal X}}P_{X}(x)W^{\nicefrac{(1-s)}{\rho}}(y_{i}|x)V^{\nicefrac{s}{\rho}}(y_{i}|x)\right)^{\rho}\\
= & e^{-n\left(\alpha s-(\rho-1)R-\frac{\rho\log2}{n}\right)}\prod_{i=1}^{n}\sum_{y\in{\cal Y}}\left(\sum_{x\in{\cal X}}P_{X}(x)W^{\nicefrac{(1-s)}{\rho}}(y|x)V^{\nicefrac{s}{\rho}}(y|x)\right)^{\rho}\\
= & e^{-n\left(\alpha s-(\rho-1)R-\frac{\rho\log2}{n}\right)}\left[\sum_{y\in{\cal Y}}\left(\sum_{x\in{\cal X}}P_{X}(x)W^{\nicefrac{(1-s)}{\rho}}(y|x)V^{\nicefrac{s}{\rho}}(y|x)\right)^{\rho}\right]^{n}\\
= & \exp\left[-n\cdot\left(\alpha s+E_{0}'(s,\rho)-(\rho-1)R-\frac{\rho\log2}{n}\right)\right]
\end{align}
where $E_{0}'(s,\rho)$ was defined in \eqref{eq: E0'}. In a similar
manner, the second term in the maximization is given by
\begin{align}
 & \hphantom{=}e^{-n\left(\alpha s-(2\rho-1)R-\frac{\rho\log2}{n}\right)}\sum_{\mathbf{y}\in{\cal Y}^{n}}\prod_{i=1}^{n}\left(\sum_{x\in{\cal X}}P_{X}(x)W^{\nicefrac{(1-s)}{\rho}}(y_{i}|x)\right)^{\rho}\left(\sum_{x\in{\cal X}}P_{X}(x)V^{\nicefrac{s}{\rho}}(y_{i}|x)\right)^{\rho}\\
 & \leq e^{-n\left(\alpha s-(2\rho-1)R-\frac{\rho\log2}{n}\right)}\sum_{\mathbf{y}\in{\cal Y}^{n}}\prod_{i=1}^{n}\left(\sum_{x\in{\cal X}}P_{X}(x)W^{\nicefrac{(1-s)}{\rho}}(y_{i}|x)\right)^{\rho}\left(\sum_{x\in{\cal X}}P_{X}(x)V^{\nicefrac{s}{\rho}}(y_{i}|x)\right)^{\rho}\\
 & =e^{-n\left(\alpha s-(2\rho-1)R-\frac{\rho\log2}{n}\right)}\left[\sum_{y\in{\cal Y}}\left(\sum_{x\in{\cal X}}P_{X}(x)W^{\nicefrac{(1-s)}{\rho}}(y|x)\right)^{\rho}\left(\sum_{x\in{\cal X}}P_{X}(x)V^{\nicefrac{s}{\rho}}(y|x)\right)^{\rho}\right]^{n}\\
 & =\exp\left[-n\cdot\left(\alpha s+E_{0}''(s,\rho)-(2\rho-1)R-\frac{\rho\log2}{n}\right)\right]
\end{align}
where $E_{0}''(s,\rho)$ was defined in \eqref{eq: E0''}. Definition
\eqref{eq: Detection exponent RC Gallager} then implies the achievability
in \eqref{eq: FA exponent RC Gallager thm}. 
\end{IEEEproof}

\section{Proof of Theorem \ref{thm: Expurgated bound Gallager}\label{sec:Proof-of-Theorem EX GF}}
\begin{IEEEproof}[Proof of Theorem \ref{thm: Expurgated bound Gallager}]
Let us begin with the FA probability. We start again from the bound
\eqref{eq: Gallager bound on false alarm for a specific code} and
restrict $s\leq1$
\begin{align}
P_{\st[FA]}({\cal C}_{n},\phi') & \leq e^{-n\alpha s}\frac{1}{M}\sum_{\mathbf{y}\in{\cal Y}^{n}}\left[\sum_{m=1}^{M}W(\mathbf{y}|\mathbf{x}_{m})\right]^{1-s}\left[\sum_{m=1}^{M}V(\mathbf{y}|\mathbf{x}_{m})\right]^{s}\\
 & \trre[\leq,a]e^{-n\alpha s}\frac{1}{M}\sum_{m=1}^{M}\sum_{k=1}^{M}\sum_{\mathbf{y}\in{\cal Y}^{n}}W^{1-s}(\mathbf{y}|\mathbf{x}_{m})V^{s}(\mathbf{y}|\mathbf{x}_{k})\label{eq: FA probability expurgated basic inequality}
\end{align}
where $(a)$ follows from $\sum_{i}a_{i}^{\nu}\geq\left(\sum_{i}a_{i}\right)^{\nu}$
for $\nu\leq1$. Let us denote the random variable 
\begin{equation}
Z_{m}\dfn\sum_{k=1}^{M}\sum_{\mathbf{y}\in{\cal Y}^{n}}W^{1-s}(\mathbf{y}|\mathbf{X}_{m})V^{s}(\mathbf{y}|\mathbf{X}_{k})\label{eq: Zm definition}
\end{equation}
over a random choice of codewords from i.i.d. distribution $P_{X}$.
Introducing a parameter $\rho\geq1$, for any given $B>0$, we may
use the classical variation of the Markov inequality, as e.g. in \cite[Eqs. (96)-(98)]{Forney68},
\begin{align}
\P\left(Z_{m}\geq B\right) & \leq\E\left[\sum_{k=1}^{M}\frac{\left[\sum_{\mathbf{y}\in{\cal Y}^{n}}W^{1-s}(\mathbf{y}|\mathbf{X}_{m})V^{s}(\mathbf{y}|\mathbf{X}_{k})\right]^{\nicefrac{1}{\rho}}}{B^{\nicefrac{1}{\rho}}}\right]\\
 & =B^{-\nicefrac{1}{\rho}}\sum_{k=1}^{M}\E\left\{ \left[\sum_{\mathbf{y}\in{\cal Y}^{n}}W^{1-s}(\mathbf{y}|\mathbf{X}_{m})V^{s}(\mathbf{y}|\mathbf{X}_{k})\right]^{\nicefrac{1}{\rho}}\right\} \\
 & \trre[\leq,a]B^{-\nicefrac{1}{\rho}}\sum_{k=1}^{M}\left\{ \sum_{\mathbf{y}\in{\cal Y}^{n}}\E\left[W^{1-s}(\mathbf{y}|\mathbf{X}_{m})V^{s}(\mathbf{y}|\mathbf{X}_{k})\right]\right\} ^{\nicefrac{1}{\rho}}\\
 & =B^{-\nicefrac{1}{\rho}}\left\{ \left[\sum_{\mathbf{y}\in{\cal Y}^{n}}\Psi_{s,1}(\mathbf{y})\right]^{\nicefrac{1}{\rho}}+\left(M-1\right)\left[\sum_{\mathbf{y}\in{\cal Y}^{n}}\Gamma_{s,1}(\mathbf{y})\right]^{\nicefrac{1}{\rho}}\right\} \\
 & <B^{-\nicefrac{1}{\rho}}2\cdot\max\left\{ \left[\sum_{\mathbf{y}\in{\cal Y}^{n}}\Psi_{s,1}(\mathbf{y})\right]^{\nicefrac{1}{\rho}},M\left[\sum_{\mathbf{y}\in{\cal Y}^{n}}\Gamma_{s,1}(\mathbf{y})\right]^{\nicefrac{1}{\rho}}\right\} ,\label{eq: expurgation probability}
\end{align}
where $(a)$ follows from Jensen inequality, and we have used the
definitions of $\Gamma_{s,\rho}(\mathbf{y})$ and $\Psi_{s,\rho}(\mathbf{y})$
from \eqref{eq: Gamma def} and \eqref{eq: Psi def}. Now, as 
\begin{align}
\sum_{\mathbf{y}\in{\cal Y}^{n}}\Psi_{s,1}(\mathbf{y}) & =\sum_{\mathbf{y}\in{\cal Y}^{n}}\prod_{i=1}^{n}\left(\sum_{x\in{\cal X}}P_{X}(x)W^{1-s}(y_{i}|x)V^{s}(y_{i}|x)\right)\\
 & =\left[\sum_{x\in{\cal X}}P_{X}(x)\sum_{y\in{\cal Y}}W^{1-s}(y|x)V^{s}(y|x)\right]^{n},
\end{align}
and
\begin{align}
\sum_{\mathbf{y}\in{\cal Y}^{n}}\Gamma_{s,1}(\mathbf{y}) & =\sum_{\mathbf{y}\in{\cal Y}^{n}}\prod_{i=1}^{n}\left(\sum_{x\in{\cal X}}P_{X}(x)W^{1-s}(y_{i}|x)\right)\left(\sum_{x\in{\cal X}}P_{X}(x)V^{s}(y_{i}|x)\right)\\
 & =\left[\sum_{y\in{\cal Y}}\left(\sum_{x\in{\cal X}}P_{X}(x)W^{1-s}(y|x)\right)\left(\sum_{x\in{\cal X}}P_{X}(x)V^{s}(y|x)\right)\right]^{n},
\end{align}
then using the definition of $E_{x}'(s)$ and $E_{x}''(s)$ in \eqref{eq: Ex'}
and \eqref{eq: Ex''}, respectively, as well as
\[
F_{x}(s,\rho,\alpha,P_{X})\dfn\min\left\{ \frac{1}{\rho}E_{x}'(s),\frac{1}{\rho}E_{x}''(s)-R\right\} ,
\]
we get that \eqref{eq: expurgation probability} is
\begin{equation}
\P\left(Z_{m}\geq B\right)\leq2B^{-\nicefrac{1}{\rho}}\cdot\exp\left[-n\cdot F_{x}(s,\rho,\alpha)\right].\label{eq: expurgation probability 2}
\end{equation}
For any given $\delta>0$ let us choose 
\[
B^{*}=e^{\nicefrac{n\delta}{2}}4^{\rho}\exp\left[-n\cdot\rho F_{x}(s,\rho,\alpha)\right]
\]
we obtain
\begin{equation}
\P\left(Z_{m}\geq B^{*}\right)<\frac{1}{2}e^{-\nicefrac{n\delta}{2\rho}}.\label{eq: probability of large Z}
\end{equation}
So, if we expurgate $\frac{1}{2}$ of the bad codewords in a randomly
chosen codebook, then\textbf{ 
\[
\P\left(\bigcup_{m=1}^{M}\left\{ Z_{m}\geq B^{*}\right\} \right)<e^{-\nicefrac{n\delta}{2\rho}}
\]
}where the probability is over the random codebooks (note also that
this expurgation only causes the sum over $k$ in \eqref{eq: Zm definition}
to decrease). Indeed, to see this, define $\frak{C}_{n}$ as the set
of `bad' codes which have $\{Z_{m}>B^{*}\}$ for more than half of
the codewords. Assume by contradiction, that the probability of a
`bad' code is larger than $e^{-\frac{n\delta}{2\rho}}$. Hence, from
the symmetry of the codewords
\begin{align}
\P\left(Z_{m}\geq B^{*}\right) & =\sum_{{\cal C}_{n}}\P\left({\cal C}_{n}\right)\I\left\{ Z_{m}>B^{*}\right\} \\
 & =\sum_{{\cal C}_{n}}\P\left({\cal C}_{n}\right)\frac{1}{M}\sum_{m=1}^{m}\I\left\{ Z_{m}>B^{*}\right\} \\
 & \geq\sum_{{\cal C}_{n}\in\frak{C}_{n}}\P\left({\cal C}\right)\frac{1}{2}\\
 & \geq\frac{1}{2}e^{-\nicefrac{n\delta}{2\rho}},
\end{align}
which contradicts \eqref{eq: probability of large Z}. Namely, if
we expurgate $\frac{1}{2}$ of the bad codewords of each codebook,
then 
\[
\overline{P_{\st[FA]}}({\cal C}_{n},\phi')\leq\exp\left[-n\cdot\left(E_{\st[GF]}^{\st[EX]}\left(R,\alpha,P_{X},W,V\right)-\delta\right)\right]
\]
for all sufficiently large $n$, with probability tending exponentially
fast to $1$ (over the random ensemble). Then, Proposition \ref{prop: exponents balance}
implies that also
\[
\overline{P_{\st[MD]}}({\cal C}_{n},\phi')\leq\exp\left[-n\cdot\left(E_{\st[GF]}^{\st[EX]}\left(R,\alpha,P_{X},W,V\right)-\alpha-\delta\right)\right].
\]
Thus, one can find a \emph{single} sequence of codebooks, of size
larger than $\frac{M}{2}$ which simultaneously achieves both upper
bounds above.
\end{IEEEproof}

\section{Simplified Expressions for BSC\label{sec:Simplified-Expressions-for-BSC}}

In Subsection \ref{sub:A-Tight-Random} (respectively, \ref{sub:Exponents-Simplified-Decoders}),
the exponents \eqref{eq: EA} and \eqref{eq: EB} (respectively, \eqref{eq: EA low rates}
and \eqref{eq: EB low rates}) are given as minimization problems
over the joint types $\tilde{Q},\overline{Q}$, and also over $Q$,
via $\mathbf{s}(\tilde{Q}_{Y},\gamma)$ (respectively, $\mathbf{t}(\tilde{Q}_{Y},\gamma)$).
These joint types are constrained to $\tilde{Q}_{X}=\overline{Q}_{X}=Q_{X}=P_{X}$
and $\tilde{Q}_{Y}=\overline{Q}_{Y}=Q_{Y}$. To obtain simplified
expressions, we will show that the optimal joint types are symmetric,
to wit, they result from an input distributed according to $P_{X}$
which undergoes a BSC. Thus, as both the input and output distributions
for such symmetric joint types are uniform, it is only remains to
optimize over the crossover probabilities $\tilde{q},\overline{q},q$.

To prove the above claim, we introduce some new notation of previously
defined quantities, but specified for the binary symmetric case. For
$q,q_{1},q_{2}\in[0,1]$, the \emph{binary normalized log likelihood
ratio }is defined as
\begin{align}
f_{w,\st[B]}(q) & \dfn\frac{1}{n}\log\left[w^{qn}(1-w)^{(1-q)n}\right]\\
 & =\log(1-w)-q\rho_{w},
\end{align}
where $\rho_{w}\dfn\log\frac{1-w}{w}$, the\emph{ binary entropy}
is denoted by
\[
h_{\st[B]}(q)\dfn-q\log q-(1-q)\log(1-q),
\]
and the \emph{binary information divergence }is denoted by\emph{
\[
D_{\st[B]}(q_{1}||q_{2})\dfn q_{1}\log\frac{q_{1}}{q_{2}}+(1-q_{1})\log\frac{(1-q_{1})}{(1-q_{2})}.
\]
}For a given type $Q$, let us define the \emph{average crossover
probability} 
\[
\hat{q}(Q)\dfn\frac{1}{2}[Q_{Y|X}(0|1)+Q_{Y|X}(1|0)],
\]
and let ${\cal Q}$ be a set of joint types, for which the inclusion
of $Q$ in ${\cal Q}$ depends on $Q$ only via $\hat{q}(Q)$. It
is easy to verify the following facts:
\begin{enumerate}
\item The information divergence satisfies 
\[
\min_{Q_{Y|X}\in{\cal Q}}D(Q_{Y|X}||W|P_{X})=\min_{0\leq q\leq1}D_{\st[B]}(q||w).
\]
from the convexity of the information divergence in $Q_{Y|X}$ and
symmetry of $P_{X}$ and $W$.
\item The normalized log likelihood ratio $f_{W}(Q)$ depends on $Q$ only
via $\hat{q}(Q)$, and so
\begin{align}
f_{W}(Q) & =\sum_{x\in{\cal X},y\in{\cal Y}}Q(x,y)\log W(y|x)\\
 & =\left(1-\hat{q}(Q)\right)\log(1-w)+\hat{q}(Q)\log(w)\\
 & =f_{w,\st[B]}\left(\hat{q}(Q)\right).
\end{align}

\item Let $L(q)$ be a linear function of $q$. Then 
\[
\max_{\tilde{Q}_{Y}}\min_{Q:\, Q_{Y}=\tilde{Q}_{Y}}\left\{ I(Q)+L\left[\hat{q}(Q)\right]\right\} =\min_{0\leq q\leq1}\left\{ \log2-h_{\st[B]}(q)+L(q)\right\} .
\]
To see this, note that $I(Q)$ is concave in $\tilde{Q}_{Y}$ (as
the input distribution to the reverse channel $Q_{X|Y}$), and $L\left[\hat{q}(Q)\right]$
is linear in $\tilde{Q}_{Y}$. So, 
\[
\min_{Q:\, Q_{Y}=\tilde{Q}_{Y}}\left\{ I(Q)+L(\hat{q}(Q))\right\} =\min_{Q_{X|Y}}\left\{ I(\tilde{Q}_{Y}\times Q_{X|Y})+L\left[\hat{q}(\tilde{Q}_{Y}\times Q_{X|Y})\right]\right\} 
\]
is a pointwise minimum of concave functions in $\tilde{Q}_{Y}$ and
thus a concave function. Moreover, it is symmetric in the sense that
if $\tilde{Q}_{Y}(0)$ is replaced with $\tilde{Q}_{Y}(1)$, and $Q_{X|Y}(\cdot|0)$
is replaced with $Q_{X|Y}(\cdot|1)$, then the same value for the
objective function is obtained. This fact along with convexity implies
that the maximizing $\tilde{Q}_{Y}$ is uniform. Since $P_{X}$ is
also uniform, the minimizing $Q_{X|Y}$ is also symmetric.
\end{enumerate}
We are now ready to provide the various bounds for detection of two
BSCs under uniform input using the facts above.

\subsection{Exact Random Coding Exponents}

Let us begin with $E_{A}$ of \eqref{eq: EA}. Assume by contradiction
that the optimal $\tilde{Q}^{*}$ is not symmetric. Fact 1 implies
that if the inputs are permuted, $\tilde{Q}^{*}(\cdot|0)\leftrightarrow\tilde{Q}^{*}(\cdot|1)$
and this joint type is averaged with $\tilde{Q}^{*}$ with weight
$\frac{1}{2}$ to result a new type $\tilde{Q}^{**}$ then
\[
D(\tilde{Q}_{Y|X}^{**}||W|P_{X})\leq D(\tilde{Q}_{Y|X}^{*}||W|P_{X}).
\]
Also, Fact 2 implies that $\tilde{Q}^{**}\in{\cal J}_{1}$. In addition,
since the function $J(Q)\dfn-\alpha+f_{V}(\tilde{Q})-f_{W}(Q)$ is
linear in $Q$ and depends on $Q$ only via $\hat{q}(Q)$, then Remark
\ref{rem: J function convex} and Fact 3 above implies that $\tilde{Q}^{**}\in{\cal J}_{2}$.
Consequently, the optimal $\tilde{Q}^{*}$ must be symmetric, and
the minimization problem involved in computing $E_{A}$ \eqref{eq: EA}
may be reduced to optimizing only over crossover probabilities, rather
than joint types. The result is as follows. Let $\gamma_{wv}\dfn\log\frac{1-v}{1-w}$.
Then,
\begin{align}
{\cal J}_{1,\st[B]} & \dfn\left\{ \tilde{q}:\; f_{w,\st[B]}(\tilde{q})+\alpha-f_{v,\st[B]}(\tilde{q})\leq0\right\} \\
 & =\left\{ \tilde{q}:\;\tilde{q}(\rho_{v}-\rho_{w})\leq-\alpha+\gamma_{wv}\right\} 
\end{align}
and
\begin{align}
{\cal J}_{2,\st[B]} & \dfn\left\{ \tilde{q}:\;\max_{0\leq\lambda\leq1}\min_{0\leq q\leq1}\left\{ \log2-h_{\st[B]}(q)+\lambda\left[-\alpha+f_{v,\st[B]}(\tilde{q})-f_{w,\st[B]}(q)\right]\right\} >R\right\} \\
 & \trre[=,a]\left\{ \tilde{q}:\;\max_{0\leq\lambda\leq1}\left\{ \log2-h_{\st[B]}(q^{*})+\lambda\left[-\alpha+f_{v,\st[B]}(\tilde{q})-f_{w,\st[B]}(q^{*})\right]\right\} >R\right\} 
\end{align}
where $(a)$ is obtained by simple differentiation and $q^{*}=\frac{w^{\lambda}}{(1-w)^{\lambda}+w^{\lambda}}$.
Then,
\[
E_{A,\st[B]}\dfn\min_{\tilde{q}\in\cap_{i=1}^{2}{\cal J}_{i,\st[B]}}D_{\st[B]}(\tilde{q}\|w).
\]
Let us now inspect $E_{B}$ of \eqref{eq: EB}. The same reasoning
as above shows that the optimal $(\tilde{Q},\overline{Q})$ must be
symmetric. Now, let
\[
{\cal K}_{2,\st[B]}\dfn\left\{ (\tilde{q},\overline{q}):\;\overline{q}(\rho_{v}-\rho_{w})\leq-\alpha+\gamma_{wv}\right\} 
\]
\[
{\cal K}_{3,\st[B]}\dfn\left\{ (\tilde{q},\overline{q}):\; f_{v,\st[B]}(\overline{q})\geq\alpha+f_{w,\st[B]}(\tilde{q})-\left[R-\log2+h_{\st[B]}(\overline{q})\right]_{+}\right\} 
\]
and
\begin{align}
{\cal K}_{4,\st[B]} & \dfn\left\{ (\tilde{q},\overline{q}):\;\max_{0\leq\lambda\leq1}\min_{0\leq q\leq1}\left\{ \log2-h_{\st[B]}(q)+\lambda\left[-\alpha+f_{v,\st[B]}(\overline{q})-f_{w,\st[B]}(q)+\left[R-\log2+h_{\st[B]}(\overline{q})\right]_{+}\right]\right\} >R\right\} \\
 & =\left\{ (\tilde{q},\overline{q}):\;\max_{0\leq\lambda\leq1}\left\{ \log2-h_{\st[B]}(q^{*})+\lambda\left[-\alpha+f_{v,\st[B]}(\overline{q})-f_{w,\st[B]}(q^{*})+\left[R-\log2+h_{\st[B]}(\overline{q})\right]_{+}\right]\right\} >R\right\} 
\end{align}
we obtain
\[
E_{B,\st[B]}\dfn\min_{(\tilde{q},\overline{q})\in\cap_{i=2}^{4}{\cal K}_{i,\st[B]}}D_{\st[B]}(\tilde{q}\|w)+\left[\log2-h_{\st[B]}(\overline{q})-R\right]_{+}.
\]
The most difficult optimization problem to solve, namely $E_{B,\st[B]}$,
is only two-dimensional.

\subsection{Expurgated Exponents}

The Chernoff distance \eqref{eq: Chernoff distance} for a pair of
BSCs with crossover probabilities $w$ and $v$ is
\[
d_{s}(x,\tilde{x})=\begin{cases}
-\log\left[(1-w)^{s}v^{1-s}+w^{s}(1-v)^{1-s}\right], & x\neq\tilde{x}\\
-\log\left[(1-w)^{s}(1-v)^{1-s}+w^{s}v^{1-s}\right], & x=\tilde{x}
\end{cases}.
\]
Now, let us analyze \eqref{eq: Detection exponent EX enumeration}.
Since $P_{X}$ is uniform, then the definition of the set ${\cal L}$
in \eqref{eq: L defintion (for expurgated)} implies that $P_{X\tilde{X}}$
is symmetric. So, 
\begin{align}
E_{\st[TE]}^{\st[EX]}\left(R,\alpha,P_{X},W,V\right) & =\max_{0\leq s\leq1}\min_{q:\,\log2-h_{\st[B]}(q)\leq R}\left\{ \alpha s+(1-q)d_{s}(1,0)+qd_{s}(0,0)+\log2-h_{\st[B]}(q)-R\right\} \\
 & =\max_{0\leq s\leq1}\left\{ \alpha s+(1-q^{*})d_{s}(1,0)+q^{*}d_{s}(0,0)+\log2-h_{\st[B]}(q^{*})-R\right\} 
\end{align}
where
\[
q^{*}=\frac{\exp\left[\frac{1}{\mu}\left(d_{s}(1,0)-d_{s}(0,0)\right)\right]}{1+\exp\left[\frac{1}{\mu}\left(d_{s}(1,0)-d_{s}(0,0)\right)\right]}
\]
and $\mu\geq1$ is either chosen to satisfy $h_{\st[B]}(q^{*})=\log2-R$
or $\mu=1$.

\subsection{Exact Random Coding Exponents of Simplified Detectors/Decoders}

As was previously mentioned, the simplified detector/decoder for high
rates is useless in this case. For the simplified detector/decoder
for low rates, we may use the same reasoning as for the optimal detector/decoder.
Let ${\cal J}_{1,\st[L,B]}\dfn{\cal J}_{1,\st[B]}$ and

\begin{align}
{\cal J}_{2,\st[L,B]} & \dfn\left\{ \tilde{q}:\;\max_{\lambda\geq0}\min_{0\leq q\leq1}\left\{ \log2-h_{\st[B]}(q)+\lambda\left[-\alpha+f_{v,\st[B]}(\tilde{q})-f_{w,\st[B]}(q)\right]\right\} >R\right\} \\
 & =\left\{ \tilde{q}:\;\max_{\lambda\geq0}\left\{ \log2-h_{\st[B]}(q^{*})+\lambda\left[-\alpha+f_{v,\st[B]}(\tilde{q})-f_{w,\st[B]}(q^{*})\right]\right\} >R\right\} 
\end{align}
where $q^{*}=\frac{w^{\lambda}}{(1-w)^{\lambda}+w^{\lambda}}$. Then,
\[
E_{A,\st[L,B]}\dfn\min_{\tilde{q}\in\cap_{i=1}^{2}{\cal J}_{i,\st[L,B]}}D_{\st[B]}(\tilde{q}\|w).
\]
Let ${\cal K}_{2,\st[L,B]}\dfn{\cal K}_{2,\st[B]}$ and
\[
{\cal K}_{3,\st[L,B]}\dfn\left\{ (\tilde{q},\overline{q}):\; f_{v,\st[B]}(\overline{q})\geq\alpha+f_{w,\st[B]}(\tilde{q})\right\} ,
\]
and
\begin{align}
{\cal K}_{4,\st[L,B]} & \dfn\left\{ (\tilde{q},\overline{q}):\;\max_{\lambda\geq0}\min_{0\leq q\leq1}\left\{ \log2-h_{\st[B]}(q)+\lambda\left[-\alpha+f_{v,\st[B]}(\overline{q})-f_{w,\st[B]}(q)\right]\right\} >R\right\} \\
 & =\left\{ (\tilde{q},\overline{q}):\;\max_{\lambda\geq0}\left\{ \log2-h_{\st[B]}(q^{*})+\lambda\left[-\alpha+f_{v,\st[B]}(\overline{q})-f_{w,\st[B]}(q^{*})\right]\right\} >R\right\} ,
\end{align}
then
\[
E_{B,\st[L,B]}\dfn\min_{(\tilde{q},\overline{q})\in\cap_{i=2}^{4}{\cal K}_{i,\st[L,B]}}D_{\st[B]}(\tilde{q}\|w)+\left[\log2-h_{\st[B]}(\overline{q})-R\right]_{+}.
\]

\bibliographystyle{plain}
\bibliography{Joint_Identification_Decoding}

\end{document}